\documentclass[trackchanges, twocolumn,authoryear]{aastex701}

\usepackage[version=4]{mhchem}
\usepackage{graphicx,epsf}
\usepackage{subfigure}
\usepackage{graphicx}	
\usepackage{amsmath}	
\usepackage{amssymb}	
\usepackage{newtxtext,newtxmath}
\usepackage{siunitx}
\usepackage{footnote}
\usepackage[para]{footmisc}
\usepackage{natbib}
\usepackage{array}
\newcolumntype{C}[1]{>{\centering\arraybackslash}p{#1}}

\newcommand{\Msun}{{\rm M}_\odot}
\newcommand{\Rsun}{{\rm R}_\odot}

\newcommand{\kms}{\textrm{km}\,\textrm{s}^{-1}}

\newcommand{\mdot}{M$_{\odot}$~yr$^{-1}$}

\def\arcsec{\hbox{$^{\prime\prime}$}}

\DeclareRobustCommand{\ion}[2]{\relax\ifmmode\ifx\testbx\f@series{\mathbf{#1\,\mathsc{#2}}}\else{\mathrm{#1\,\mathsc{#2}}}\fi\else\textup{#1\,{\mdseries\textsc{#2}}}\fi}

\usepackage[scale=0.94]{tgbonum}
\usepackage[mode=text]{siunitx}

\makeatletter
\DeclareTextCompositeCommand{\r}{OT1}{A}{%
  \leavevmode\vbox{%
    \offinterlineskip
    \ialign{\hfil##\hfil\cr\char23\cr\noalign{\kern-1.15ex}A\cr}%
  }%
}
\makeatother

\begin{document}

\title{SN 2024iss: A Multi-Wavelength Exposé of a Type IIb Supernova with an Early-Time Ultraviolet Spectrum and Shock Breakout Constraints}

\author[0009-0005-3569-9944]{R.~J.~Yete}
\affiliation{Department of Physics and Astronomy, University of North Carolina at Chapel Hill, Chapel Hill, NC 27599-3255, USA}
\email[show]{rjyete@unc.edu}

\author[0000-0002-3934-2644]{W.~V.~Jacobson-Gal\'{a}n}
\altaffiliation{NASA Hubble Fellow}
\affiliation{Cahill Center for Astrophysics, California Institute of Technology, MC 249-17, 1216 E California Boulevard, Pasadena, CA 91125, USA}
\email{wynnjg@caltech.edu}

\author[0009-0008-7581-3096]{Ferdinand}
\affiliation{Department of Astronomy, University of Illinois at Urbana-Champaign, 1002 W. Green St., IL 61801, USA}
\email{}

\author[0000-0003-0599-8407]{L.~Dessart}
\affiliation{Institut d’Astrophysique de Paris, CNRS-Sorbonne Université, 98 bis boulevard Arago, F-75014 Paris, France}
\email{}

\author[0000-0002-5619-4938]{M.~M.~Kasliwal}
\affiliation{Cahill Center for Astrophysics, California Institute of Technology, MC 249-17, 1216 E California Boulevard, Pasadena, CA 91125, USA}
\email{}

\author[0000-0002-5680-4660]{K.~W.~Davis}
\affiliation{Department of Astronomy and Astrophysics, University of California, Santa Cruz, CA 95064, USA}
\email{}

\author[0000-0001-6806-0673]{A.~L.~Piro}
\affiliation{The Observatories of the Carnegie Institute for Science, 813 Santa Barbara St., Pasadena, CA 91101, USA}
\email{}

\author[0000-0002-1125-9187]{V.~A.~Villar}
\affiliation{Center for Astrophysics \textbar{} Harvard \& Smithsonian, 60 Garden Street, Cambridge, MA 02138-1516, USA}
\newcommand{\IAIFI}{\affiliation{The NSF AI Institute for Artificial Intelligence and Fundamental Interactions}}
\email{}


\author[0000-0002-8977-1498]{I.~Andreoni}
\affiliation{Department of Physics and Astronomy, University of North Carolina at Chapel Hill, Chapel Hill, NC 27599-3255, USA}
\email{}

\author[0000-0002-4449-9152]{K.~Auchettl}
\affiliation{Department of Astronomy and Astrophysics, University of California, Santa Cruz, CA 95064, USA}
\affiliation{OzGrav, School of Physics, The University of Melbourne, Parkville, VIC, Australia}
\email{}

\author[0000-0001-6965-7789]{K.~C.~Chambers}
\affiliation{Institute for Astronomy, University of Hawaii, 2680 Woodlawn Drive, Honolulu, HI 96822, USA}
\email{}

\author[0000-0002-7706-5668]{R.~Chornock}
\affiliation{Department of Astronomy, University of California, Berkeley, CA 94720-3411, USA} 
\affiliation{Berkeley Center for Multi-messenger Research on Astrophysical Transients and Outreach (Multi-RAPTOR), University of California, Berkeley, CA 94720-3411, USA}
\email{}

\author[0000-0002-8262-2924]{M.~W.~Coughlin}
\affiliation{School of Physics and Astronomy, University of Minnesota, Minneapolis, MN 55455, USA}
\email{}

\author[0000-0003-4263-2228]{D.~A.~Coulter}
\affiliation{William H. Miller III Department of Physics \& Astronomy, Johns Hopkins University, 3400 N Charles St, Baltimore, MD 21218, USA}
\affiliation{Space Telescope Science Institute, Baltimore, MD 21218, USA}
\email{}

\author[0000-0002-5884-7867]{R.~Dekany}
\affiliation{Caltech Optical Observatories, California Institute of Technology, Pasadena, CA  91125, USA}
\email{}

\author[0000-0002-2445-5275]{R.~J.~Foley}
\affiliation{Department of Astronomy and Astrophysics, University of California, Santa Cruz, CA 95064, USA}
\email{}

\author[0000-0003-2238-1572]{O.~D.~Fox}
\affiliation{Space Telescope Science Institute, Baltimore, MD 21218, USA}
\email{}

\author[0000-0002-1296-6887]{L.~Galbany}
\affiliation{Institute of Space Sciences (ICE-CSIC), Campus UAB, Carrer de Can Magrans, s/n, E-08193 Barcelona, Spain.}
\affiliation{Institut d'Estudis Espacials de Catalunya (IEEC), 08860 Castelldefels (Barcelona), Spain}
\email{}

\author[0000-0002-8526-3963]{C.~Gall}
\affiliation{DARK, Niels Bohr Institute, University of Copenhagen, Jagtvej 128, 2200 Copenhagen, Denmark}
\email{}

\author[0000-0002-3884-5637]{A.~Gangopadhyay}
\affiliation{The Oskar Klein Centre, Department of Astronomy, Stockholm University, AlbaNova, SE-10691 Stockholm, Sweden}
\email{}

\author[0000-0003-1012-3031]{J.~A.~Goldberg}
\affiliation{Department of Physics and Astronomy, Michigan State University, East Lansing, MI 48824, USA}
\email{}

\author[0000-0003-1196-3761]{G.~Govindaraj}
\affiliation{Department of Physics, Drexel University, Philadelphia, PA 19104, USA}
\email{}

\author[0009-0002-9727-8326]{X.~Guo}
\affiliation{Department of Astronomy, University of California, Berkeley, CA 94720-3411, USA}
\email{}

\author[0000-0003-1059-9603]{M.~E.~Huber}
\affiliation{Institute for Astronomy, University of Hawaii, 2680 Woodlawn Drive, Honolulu, HI 96822, USA}
\email{}

\author[0000-0002-0987-3372]{J.~Castaneda Jaimes}
\affiliation{Division of Physics, Mathematics and Astronomy, California Institute of Technology, 1200 E. California Blvd, Pasadena, CA 91125, USA}
\email{}

\author[0009-0000-3122-8321]{J.~Karcheski}
\affiliation{Department of Astronomy and Astrophysics, University of California, Santa Cruz, CA 95064, USA}
\email{}

\author[0009-0005-1871-7856]{R.~Kaur}
\affiliation{Department of Astronomy and Astrophysics, University of California, Santa Cruz, CA 95064, USA}
\email{}

\author[0000-0002-5740-7747]{C.~D.~Kilpatrick}
\affiliation{Center for Interdisciplinary Exploration and Research in Astrophysics (CIERA), Northwestern University, 1800 Sherman Ave., Evanston, IL 60201, USA}
\affiliation{Department of Physics and Astronomy, Northwestern University, Evanston, IL 60208, USA}
\email{}

\author[0000-0003-2451-5482]{R.~R.~Laher}
\affiliation{IPAC, California Institute of Technology, 1200 E. California Blvd, Pasadena, CA 91125, USA}
\email{}

\author[0000-0002-2249-0595]{N.~LeBaron}
\affiliation{Department of Astronomy, University of California, Berkeley, CA 94720-3411, USA}
\affiliation{Berkeley Center for Multi-messenger Research on Astrophysical Transients and Outreach (Multi-RAPTOR), University of California, Berkeley, CA 94720-3411, USA}
\email{}

\author[0000-0002-7272-5129]{C.~-C.~Lin}
\affiliation{Institute for Astronomy, University of Hawaii, 2680 Woodlawn Drive, Honolulu, HI 96822, USA}
\email{}

\author[0000-0003-4768-7586]{R.~Margutti}
\affiliation{Department of Astronomy, University of California, Berkeley, CA 94720-3411, USA}
\affiliation{Department of Physics, University of California, Berkeley, CA 94720-7300, USA}
\affiliation{Berkeley Center for Multi-messenger Research on Astrophysical Transients and Outreach (Multi-RAPTOR), University of California, Berkeley, CA 94720-3411, USA}
\email{}

\author[0000-0001-8415-6720]{Y.-C.~Pan}
\affiliation{Graduate Institute of Astronomy, National Central University, 300 Zhongda Road, Zhongli, Taoyuan 32001, Taiwan}
\email{}

\author[0000-0002-1092-6806]{K.~C.~Patra}
\affiliation{Department of Astronomy and Astrophysics, University of California, Santa Cruz, CA 95064, USA}
\email{}

\author[0009-0000-5561-9116]{H.~M.~L.~Perkins}
\affiliation{Department of Astronomy, University of Illinois at Urbana-Champaign, 1002 W. Green St., IL 61801, USA}
\affiliation{Center for Astrophysical Surveys, National Center for Supercomputing Applications, Urbana, IL 61801, USA}
\affiliation{Illinois Center for Advanced Studies of the Universe, University of Illinois Urbana-Champaign, Urbana, IL 61801}
\email{}

\author[0000-0002-8041-8559]{P.~J.~Pessi}
\affiliation{Astrophysics Division, National Centre for Nuclear Research, Pasteura 7, 02-093 Warsaw, Poland}
\email{}

\author[0009-0007-5449-4041]{O.~Pyshna}
\affiliation{Cahill Center for Astrophysics, California Institute of Technology, MC 249-17, 1216 E California Boulevard, Pasadena, CA 91125, USA}
\email{}

\author[0000-0002-4410-5387]{A.~Rest}
\affiliation{William H. Miller III Department of Physics \& Astronomy, Johns Hopkins University, 3400 N Charles St, Baltimore, MD 21218, USA}
\affiliation{Space Telescope Science Institute, Baltimore, MD 21218, USA}
\email{}

\author[]{S.~Romagnoli}
\affiliation{School of Physics, The University of Melbourne, VIC 3010, Australia}
\email{}

\author[0000-0001-6797-1889]{S.~Schulze}
\affiliation{Center for Interdisciplinary Exploration and Research in Astrophysics (CIERA), Northwestern University, 1800 Sherman Ave., Evanston, IL 60201, USA}
\affiliation{Department of Particle Physics and Astrophysics, Weizmann Institute of Science, 234 Herzl St, 76100 Rehovot, Israel}
\email{}

\author[0000-0001-8023-4912]{H.~Sears}
\affiliation{Department of Physics and Astronomy, Rutgers, the State University of New Jersey, 136 Frelinghuysen Road, Piscataway, NJ 08854-8019, USA}
\email{}

\author[0000-0002-9158-750X]{A.~Sedgewick}
\affiliation{DARK, Niels Bohr Institute, University of Copenhagen, Jagtvej 128, 2200 Copenhagen, Denmark}
\email{}

\author[0000-0003-2445-3891]{M.~R.~Siebert}
\affiliation{Space Telescope Science Institute, Baltimore, MD 21218, USA}
\email{}

\author[0000-0003-2091-622X]{A.~Singh}
\affiliation{The Oskar Klein Centre, Department of Astronomy, Stockholm University, AlbaNova, SE-10691 Stockholm, Sweden}
\email{}

\author[0000-0003-1546-6615]{J.~Sollerman}
\affiliation{The Oskar Klein Centre, Department of Astronomy, Stockholm University, AlbaNova, SE-10691 Stockholm, Sweden}
\email{}

\author[]{N.~Sravan}
\affiliation{Department of Physics, Drexel University, Philadelphia, PA 19104, USA}
\email{}

\author[0000-0002-1481-4676]{S.~Tinyanont}
\affiliation{National Astronomical Research Institute of Thailand, 260 Moo 4, Donkaew, Maerim, Chiang Mai, 50180, Thailand}
\email{}

\author[0000-0002-1341-0952]{R.~J.~Wainscoat}
\affiliation{Institute for Astronomy, University of Hawaii, 2680 Woodlawn Drive, Honolulu, HI 96822, USA}
\email{}

\author[0000-0001-5233-6989]{Q.~Wang}
\affiliation{Department of Physics and Kavli Institute for Astrophysics and Space Research, Massachusetts Institute of Technology, 77 Massachusetts Avenue, Cambridge, MA 02139, USA}
\email{}

\author[]{D.~Warshofsky}
\affiliation{School of Physics and Astronomy, University of Minnesota, Minneapolis, MN 55455, USA}
\email{}

\author[0000-0001-6747-8509]{Y.~Yao}
\affiliation{Department of Astronomy, University of California, Berkeley, CA 94720-3411, USA} 
\affiliation{Miller Institute for Basic Research in Science, 206B Stanley Hall, Berkeley, CA 94720, USA}
\affiliation{Berkeley Center for Multi-messenger Research on Astrophysical Transients and Outreach (Multi-RAPTOR), University of California, Berkeley, CA 94720-3411, USA}
\email{}

\author[0000-0002-0632-8897]{Y.~Zenati}
\affiliation{Astrophysics Research Center of the Open University (ARCO), Department of Natural Sciences, Ra’anana 4353701, Israel}
\affiliation{William H. Miller III Department of Physics \& Astronomy, Johns Hopkins University, 3400 N Charles St, Baltimore, MD 21218, USA}
\email{}



\begin{abstract}

We present multi-wavelength observations and analysis of SN2024iss, a Type IIb supernova (SN IIb) located at $\sim$14 Mpc. The first ZTF detection of SN2024iss is $\sim$40 minutes after first light (one of the earliest detections of a SN IIb to date) and a {\it Hubble Space Telescope} ultraviolet spectrum was obtained at 7 days (the earliest \textit{HST} UV SN IIb spectrum to date). We estimate an ejecta mass range of $1.1-3.3 \Msun$ and $^{56}\textrm{Ni}$ mass of $0.11 \pm 0.01 \Msun$ from bolometric light curve modeling and He-star model comparisons. We fit shock-cooling emission models to estimate a progenitor radius of $100-320\ \Rsun$ and envelope mass of $0.07-0.46\ \Msun$. We compare optical/UV spectra to binary progenitor models, which indicate a stripped hydrogen-rich envelope mass of $0.19-0.28\ \Msun$. We use early-time X-ray detections to calculate CSM densities that are consistent with a progenitor mass-loss rate of $5\times10^{-4}$ \mdot ($v_w=100 \kms$) corresponding to 2-5 years pre-explosion. In the UV, we observe strong \ion{Mg}{ii} emission extending to 15000 km/s as well as weak P-Cygni profiles of iron-group elements (e.g., Fe, Ti, Al, Ni) present in outer SN ejecta. The overall spectroscopic evolution of SN2024iss is comparable to SNe IIb, but increased brightness at peak is influenced by SN ejecta-CSM interaction. Nebular spectroscopy of SN2024iss at $\sim 260-412$ days reveals multi-peaked forbidden line profiles of \ion{O}{i} and \ion{Mg}{i}] indicative of inner ejecta asymmetry and/or clumping. We demonstrate the utility of a rich, multi-wavelength dataset for constraining progenitor systems and explosion dynamics of SNe IIb.

\end{abstract}

\keywords{\uat{Supernovae}{1668} --- \uat{Type II supernovae}{1731} ---\uat{Ultraviolet spectroscopy}{2284} --- \uat{High Energy astrophysics}{739} --- \uat{Circumstellar matter}{241} -- \uat{X-ray astronomy}{1810} } 


\section{Introduction} 

Type II supernovae (SNe II) occur from the core collapse of massive stars ($\gtrsim8 \ \Msun$), and are characterized by hydrogen features in their spectra. SNe IIb are a transitional class of SNe II that display hydrogen features in early time spectra, but evolve to be dominated by helium (\ion{He}{i}) features in later times as typically seen in Type Ib SNe \citep{filippenko_optical_1997}. These features are often attributed to the progenitor possessing a low-mass hydrogen envelope \citep{woosley_sn_1994, woosley_physics_2005, hachinger_how_2012}, indicating high mass-loss rates from the progenitor leading up to its explosion \citep{Maund04, smith_observed_2011, ben-ami_ultraviolet_2015}. While not fully understood, this mass-loss can be explained by interactions of the progenitor with a companion star in a binary system \citep{1992ApJ...391..246P, Maund04, Yoon10, yoon_type_2017, Ercolino24}. Several SNe IIb progenitors have been identified as yellow supergiant stars through direct imaging of the progenitor (e.g., \citealt{Aldering94, Maund04, bersten_type_2012, van_dyk_type_2014, reguitti_sn_2025}). Previous studies have suggested that yellow supergiant progenitors began their final stages as red supergiants, but evolved to yellow supergiants by the time of explosion due to high mass-loss rates \citep{georgy_yellow_2012}. 

Surface composition and mass-loss of the progenitor can be explored with early time observations of the SN \citep{gal-yam_wolf-rayet-like_2014}. Chemical elements ejected by the progenitor during mass-loss remain in the medium surrounding the star, and are ionized by the shock wave radiation resulting from the collision of SN ejecta and circumstellar material (CSM; \citealt{Yaron17, 2017A&A...605A..83D}). Evidence of this photoionization can be seen through narrow emission features in ultraviolet spectra of the supernova until $\sim~7$ days after first light \citep{zimmerman_complex_2024}. Previous studies of early-time UV spectra of SNe II have been successful in tracing progenitor atmospheres, mass-loss, CSM abundance and density, as well as the structure and temperature of the supernova ejecta \citep[e.g.,][]{vasylyev_early-time_2022, vasylyev_early-time_2023, bostroem_sn_2023, zimmerman_complex_2024}. 

\begin{figure*}[h!t]
\centering
\includegraphics[width=\textwidth]{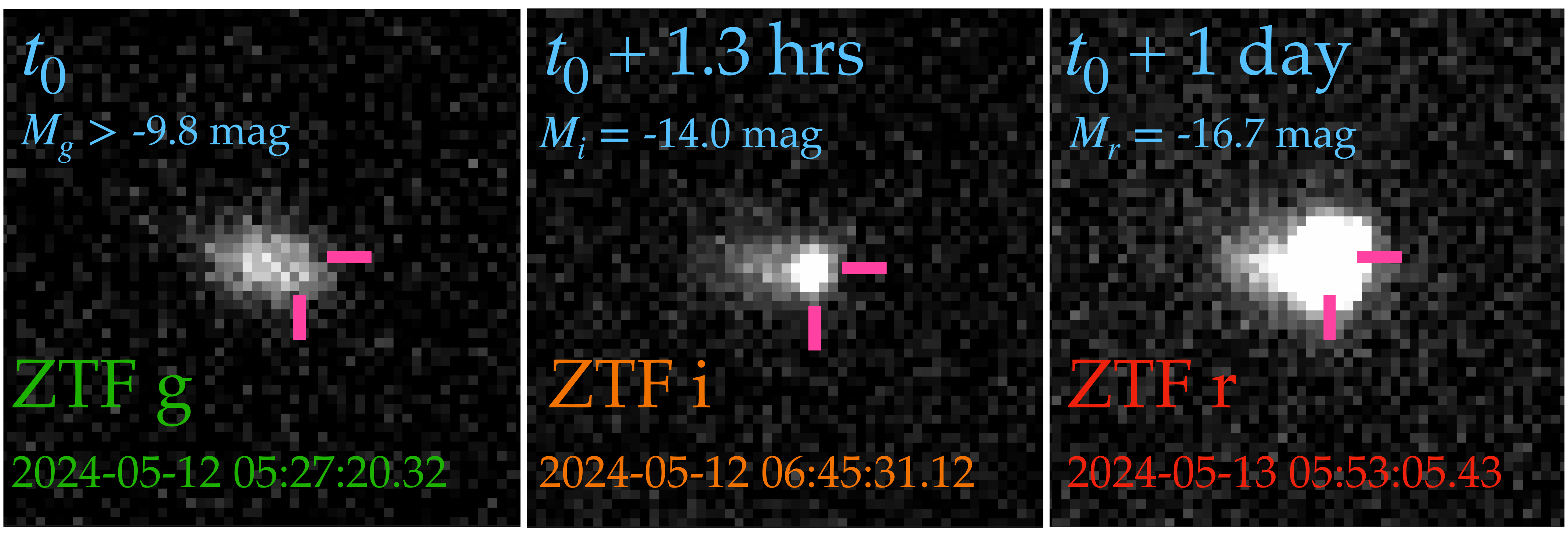}
\caption{{\it Left:} Last pre-explosion ZTF $g$-band image of SN~2024iss with a limiting apparent magnitude of $m_g \geq 21.0$~mag at time $t_0$. {\it Middle:} First detection image of SN~2024iss in ZTF $i$-band at $\delta t = 1.3$~hrs after the last non-detection, indicating a $\gtrsim 50\times$ increase in flux at a rate of $>3.2$~mag/hr. {\it Right:} Second detection image of SN~2024iss in ZTF $r$-band at 1~day after the last non-detection.  }.
\label{fig:discovery}
\end{figure*}

SN 2024iss was first reported to the Transient Name Server (TNS) by the Gravitational-wave Optical Transient Observer (GOTO; \citealt{2018SPIE10704E..0CD, 2020SPIE11452E..1QD}) on 2024-05-13 (MJD 60443.03, \citealt{Godson24}). SN~2024iss was classified as a Type IIb SN \citep{Srivastav24} and is located at $\alpha = 12^{\textrm{h}}59^{\textrm{m}}06.130^{\textrm{s}}$, $\delta = +28^{\circ}48'42.62^{\prime \prime}$ in the host galaxy DESI J194.7752+28.8122. \citet{yamanaka_sn_2025} presented estimates of physical parameters of SN 2024iss from its bolometric light curve and blackbody properties, while \citet{chen_sn_2025} presented estimates of physical properties through analysis and modeling of optical and UV photometry, optical spectroscopy, and X-ray observations. We present, analyze, and model a range of multi-wavelength data including UV/Optical/NIR photometry, UV/Optical/NIR spectroscopy, and X-ray observations.

The last pre-explosion image of SN~2024iss was by ZTF on MJD 60442.23 ($m > 21.037$~mag in $g$-band), followed by an intra-night ZTF $i$-band detection ($m = 16.8$~mag) 1.3~hours later on MJD 60442.28 (see Figure \ref{fig:discovery}). Given the depth of the last non-detection, we constrain the time of first light to be MJD $60442.254 \pm 0.0271$, which is later and more constrained than the explosion epoch reported in \cite{chen_sn_2025}. We define $\delta t$ throughout as the time since first light in rest-frame days. For SN~2024iss, we use a redshift of $z = 0.003238 \pm 0.000033$ \citep{DESI24}, which corresponds to a luminosity distance of $14.04 \pm 0.14$~Mpc for standard $\Lambda$CDM cosmology ($H_{0}$ = 70 km s$^{-1}$ Mpc$^{-1}$, $\Omega_M = 0.27$, $\Omega_{\Lambda} = 0.73$). Unfortunately, no redshift-independent distance is available. The main parameters of SN~2024iss and its host galaxy are displayed in Table \ref{tbl:params24iss}.


\begin{table}[ht]
\begin{center}
\caption{Observational properties of SN\,2024iss and its host galaxy \label{tbl:params24iss}}
\begin{tabular}{lcccccc}
\hline
\hline
Host Galaxy &  &  & &  & &   DESI J194.7752+28.8122 \\ 
Redshift &  &  & &  & &  $0.003238 \pm 0.000033$\\  
Distance &  &  & &  & &  $14.04 \pm 0.14$~Mpc\\ 
Distance Modulus, $\mu$ &  &  & &  & & $30.74 \pm 0.02 $~mag\\ 
$\textrm{RA}_{\textrm{SN}}$ &  &  & &  & &  $12^{\textrm{h}}59^{\textrm{m}}06.130^{\textrm{s}}$\\
$\textrm{Dec}_{\textrm{SN}}$ &  &  & &  & & $+28^{\circ}48'42.62^{\prime \prime}$\\
Time of First Light (MJD) &  &  & &  & & 60442.254 $\pm$ 0.0271\\ 
$E(B-V)_{\textrm{MW}}$ &  &  & &  & & 0.01~mag\\
$E(B-V)_{\textrm{host}}$ &  &  & &  & & $0.019 \pm 0.004$ mag\\
$m_{g}^{\mathrm{peak}}$ &  &  & &  & & $13.302 \pm 0.001 $~mag\footnote{No extinction correction applied.}\\
$M_{g}^{\mathrm{peak}}$ &  &  & &  & & $-17.469 \pm 0.022$~mag\footnote{Extinction correction applied.}\footnote{Second $g$-band light curve peak}\\
\hline
\end{tabular}
\end{center}
\label{table:Observations}
\end{table}

Section \ref{sec:observations} describes the observations used in this paper. In Section \ref{sec:photometry} we present an analysis of photometric data, including light curve comparisons, shock breakout constraints, blackbody evolution, derivation of the $^{56}$Ni mass, and shock cooling emission modeling. In Section \ref{sec:spectroscopy}, we analyze ultraviolet, optical, and NIR spectroscopy with comparisons to model and other SNe IIb spectra. In Section \ref{sec:csm}, we derive the CSM and progenitor mass-loss history through X-ray observations. Finally, our discussion and conclusions are presented in Sections \ref{sec:discussion} and \ref{sec:conclusions}.


\section{Observations} \label{sec:observations}

\begin{figure*}[ht]
\centering
\includegraphics[width=\textwidth]{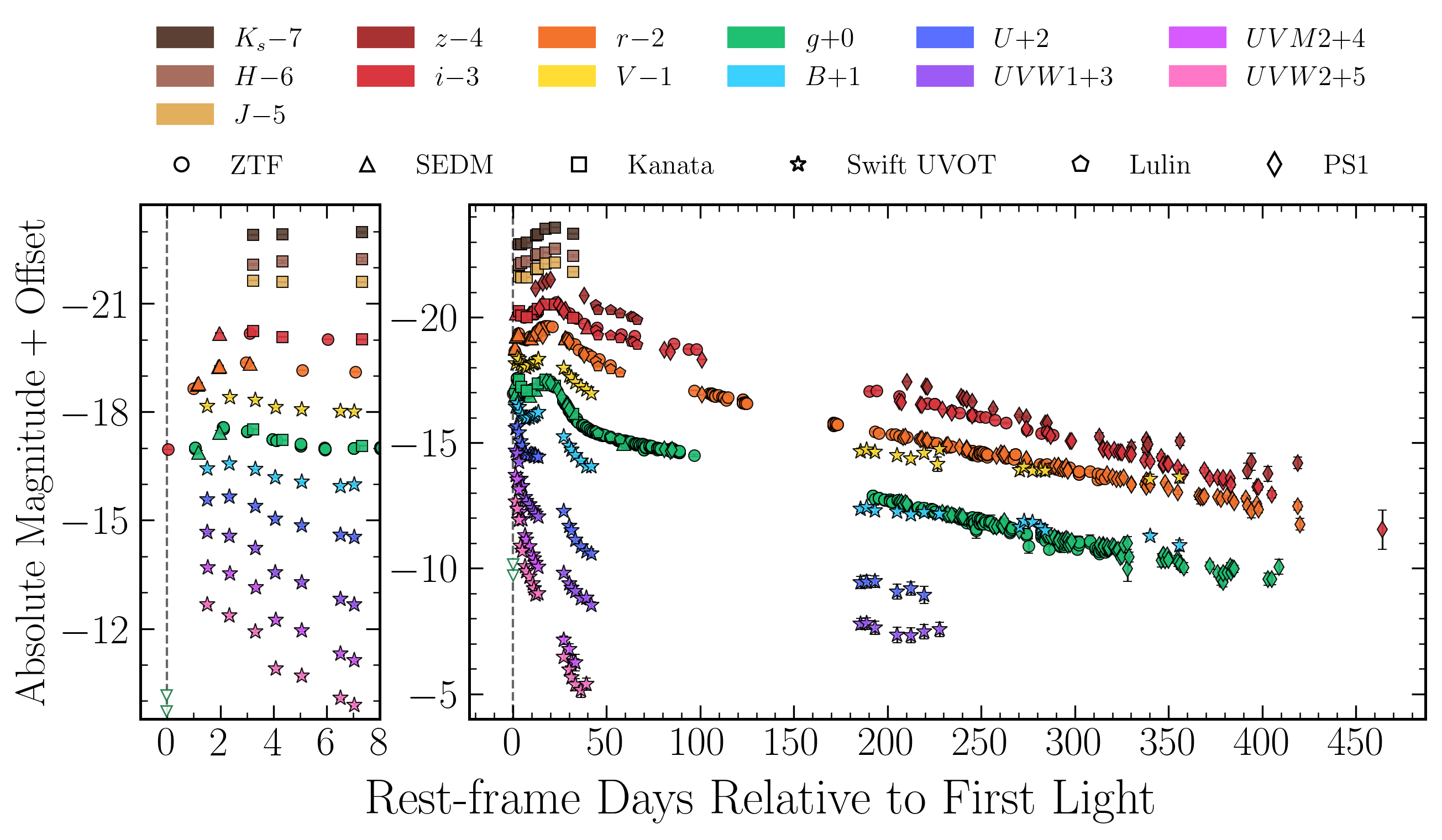}
\caption{Multi-band photometry of SN 2024iss. The left panel shows the first peak from shock breakout and cooling, while the right panel shows the full light curve. The downward triangle points are the last ZTF $g$-band non-detection upper limits at 0.027 days before first light. All data points are corrected for Milky Way and host galaxy extinction.}.
\label{fig:lightcurve}
\end{figure*}

\subsection{UV/Optical/NIR Photometry} \label{subsec:phot}
The complete, multi-band light curve of SN~2024iss is presented in Figure \ref{fig:lightcurve}. SN~2024iss was first detected by the Zwicky Transient Facility (ZTF; \citealt{bellm_zwicky_2019}) in $i$-band and was subsequently monitored in $gri$ bands. ZTF photometry comes from Public, Partnership and Caltech streams and was generated through the ZTF forced-photometry service \citep{Masci19}. Additional $ugri$ band photometry was obtained with the Spectral Energy Distribution Machine (SEDM; \citealt{blagorodnova_sed_2018}) on the 60-in telescope at Palomar Observatory. Data storage and observation follow-up was performed through the SkyPortal interface and the Fritz broker \citep{skyportal, Coughlin23}. 

SN~2024iss was observed with the Pan-STARRS telescope \citep[PS1;][]{Kaiser2002, 2016arXiv161205560C} through the Young Supernova Experiment (YSE) \citep{Jones2021} between 2024-05-24 and 2025-08-19 ($\delta t= 12.1-464.0$~days since first light). Data storage/visualization and follow-up coordination was done through the YSE-PZ web broker \citep{Coulter22, Coulter23}. The YSE photometric pipeline is based on {\tt photpipe} \citep{Rest+05}, which relies on calibrations from \cite{Magnier20a} and \cite{waters20}. Each image template was taken from stacked PS1 exposures, with most of the input data from the PS1 3$\pi$ survey \citep{2016arXiv161205560C}. All images and templates were resampled and astrometrically aligned to match a skycell in the PS1 sky tessellation. An image zero-point is determined by comparing PSF photometry of the stars to updated stellar catalogs of PS1 observations \citep{flewelling16}. The PS1 templates are convolved with a three-Gaussian kernel to match the PSF of the nightly images, and the convolved templates are subtracted from the nightly images with {\tt HOTPANTS} \citep{becker15}. Finally, a flux-weighted centroid is found for the position of the SN in each image and PSF photometry is performed using ``forced photometry": the centroid of the PSF is forced to be at the SN position. The nightly zero-point is applied to the photometry to determine the brightness of the SN for that epoch. 

We also observed SN~2024iss with the 1-m Lulin telescope in $griz$ bands from 2024-06-25 to 2024-07-17. Standard calibrations for bias and flat-fielding were performed on the images using {\tt IRAF} \citep{1986SPIE..627..733T}, and we reduced the calibrated frames in {\tt photpipe} using the same methods described in \citet{kilpatrick18}.


The Ultraviolet Optical Telescope (UVOT; \citealt{roming_swift_2005}) onboard the Neil Gehrels \emph{Swift} Observatory \citep{gehrels_swift_2004} observed SN~2024iss from 2024-05-13 to 2025-05-03 ($\delta t = 1.51 - 356.0$~days). We performed aperture photometry with a 5$\arcsec$ region radius with \texttt{uvotsource} within HEAsoft v6.33 \citep{HEAsoft}\footnote{We used the most recent calibration database (CALDB) version.}, following the standard guidelines from \cite{Brown14}\footnote{\url{https://github.com/gterreran/Swift_host_subtraction}}. In order to remove contamination from the host galaxy, we employed pre-explosion images to subtract the measured count rate at the location of the SN from the count rates in the SN images and corrected for point-spread-function (PSF) losses following the prescriptions of \cite{Brown14}. 

Near-infrared (NIR) data of SN~2024iss was obtained with the Hiroshima Optical and Near-InfraRed camera (HONIR) instrument of the 1.0-m Kanata Telescope, Hiroshima. The sky-background subtraction was done using a template sky image obtained by dithering individual frames at different positions. We performed PSF photometry and calibrated the SN magnitudes using comparison stars in the 2MASS catalog \citep{1998AJ....116.2475P}. 


\subsection{Reddening and Host Galaxy Properties} \label{subsec:host}

The Milky Way (MW) $V$-band extinction and color excess along the SN line of sight is $A_{V} = 0.031$~mag and \textit{E(B-V)} = 0.01~mag \citep{schlegel98, schlafly11} respectively, which we correct for using a standard \cite{fitzpatrick_correcting_1999} reddening law (assuming \textit{$R_V$} = 3.1). In addition to the MW color excess, we estimate the contribution of host galaxy extinction in the local SN environment by measuring the equivalent widths (EWs) of each component of the \ion{Na}{i} doublet in the medium resolution spectrum from the Palomar Double Spectrograph (DBSP; \citealt{oke_efficient_1982}). We use $A_V^{\rm host} = (0.78\pm0.15)~{\rm mag} \times ({\rm EW_{NaID}}$/\AA) from \cite{Stritzinger18} to convert these EWs into to a host-galaxy $E(B-V)$ and derive a host galaxy extinction of $E(B-V)_{\textrm{host}} = 0.019 \pm 0.004$~mag, also corrected for using the \cite{fitzpatrick_correcting_1999} reddening law.

We utilize a Sloan Digital Sky Survey (SDSS) spectrum to constrain the properties of the dwarf host galaxy of SN~2024iss. In the SDSS spectrum, we find strong H$\alpha$ emission but no significant detections of other typical galaxy lines. We derive upper limits on the line fluxes of H$\beta$, [\ion{O}{iii}] $\lambda 5007$, and \ion{N}{ii} $\lambda 6584$ by simulating a synthetic Gaussian profile with the Full-Width-Half-Maximum (FWHM) set to the spectral resolution until a 3$\sigma$ detection is recovered above the continuum. However, because only flux limits can be determined for these emission lines, we cannot estimate a reliable limit on the $O3N2$ ratio. Nonetheless, we find $N2 = \rm log_{10}(\rm [NII]/H\alpha) < -0.662$, which corresponds to a limit on the oxygen abundance of $\rm 12 + log(O/H) < 8.44$ using the relation in \cite{Marino13}. This oxygen abundance corresponds to a likely sub-solar host galaxy metallicity of $Z < 0.6 ~ \rm Z_{\odot}$.


\subsection{UV/Optical/NIR Spectroscopy} \label{subsec:spec}

We present our full range of UV, Optical, and NIR spectroscopic observations in Table \ref{tab:spec_all}. We obtained 33 optical spectra covering $
\delta t = 1.1$ to 411.3 days, an $HST$ UV spectrum at $
\delta t = 7.0$ days, and NIR spectra in both photospheric and nebular phases. We obtained an early-time ultraviolet spectrum of SN 2024iss with the {\it Hubble Space Telescope (HST)} Space Telescope Imaging Spectrograph (STIS; \cite{woodgate_space_1998}) on 19 May 2024 ($\delta t = 7$~days). The SN was observed through {\it HST} Cycle 31 ``Flex Thursday'' Target of Opportunity Program 17507 (PI: Jacobson-Galán). Spectra were obtained in the G230LB (1680 – 3060~\AA), G430L (2900 – 5700~\AA), and G750L (5236 – 10266~\AA) gratings, resulting in a $
\delta t = 7$~day spectrum ranging from 1680 – 10266~\AA. We reduced the spectra using the \texttt{stistools} package by performing cosmic ray rejection, matching the extraction aperture size across the gratings, increasing the local background size, and placing the local background closer to the spectral trace.

We obtained optical spectra of SN~2024iss with the Kast spectrograph on the 3-m Shane telescope at Lick Observatory \citep{KAST}. For all of these spectroscopic observations, standard CCD processing and spectrum extraction were accomplished with \textsc{IRAF}\footnote{\url{https://github.com/msiebert1/UCSC\_spectral\_pipeline}}. The data was extracted using the optimal algorithm of \citet{1986PASP...98..609H}.  Low-order polynomial fits to calibration-lamp spectra were used to establish the wavelength scale and apply small adjustments derived from night-sky lines in the object frames. Additional optical spectra of SN~2024iss were also obtained with the Palomar Spectral Energy Distribution Machine (SEDM; \citealt{blagorodnova_sed_2018, Kim22}), the Alhambra Faint Object Spectrograph and Camera (ALFOSC)\footnote{\href{http://www.not.iac.es/instruments/alfosc}{{http://www.not.iac.es/instruments/alfosc}}}, the Spectrograph for the Rapid Acquisition of Transients (SPRAT; \citealt{piascik_sprat_2014}), and DBSP. Late-time optical spectra were obtained with the Low Resolution Imaging Spectrometer (LRIS; \citealt{oke_keck_1995}) on the 10-m Keck I telescope and reduced with {\tt Lpipe} \citep{Perley19}. Near-infrared (NIR) spectra were obtained with the $R\approx2700$ Near-Infrared Echelle Spectrograph (NIRES; \citealt{2004SPIE.5492.1295W}) located on the 10-m Keck II telescope. The data were reduced using a custom version of the IDL based reduction package {\tt Spextool} \citep{2004PASP..116..362C} modified for use with NIRES as well as the {\tt Pypeit} spectral reduction pipeline \citep{2020JOSS....5.2308P}. For the {\tt Spextool} reduction, we used {\tt xtellcor} \citep{2003PASP..115..389V} to correct for telluric features in our spectrum using an A0 standard star observed close in airmass and time to our target. NIRES data were obtained through the Keck Infrared Transient Survey (KITS; \citealt{2024PASP..136a4201T}). A complete log of spectroscopic observations is presented in Table \ref{tab:spec_all}. 

\subsection{X-ray Observations \& Comparisons} \label{subsec:xray}

The X-Ray Telescope (XRT, \citealt{2005SSRv..120..165B}) on board the \emph{Swift} spacecraft \citep{gehrels_swift_2004} observed the field of SN~2024iss from 2024-05-13 to 2025-05-03 ($\delta t = 1.51 - 356.0$~days, with a total exposure time of 60 ks). We analyzed the data using \texttt{XSNAP} \citep{Ferdinand25} and followed standard filtering and screening prescriptions. A bright source of X-ray emission is clearly detected in each individual observation with significance of $>3\sigma$ against the background in the first seven epochs ($\delta t = 1.51 - 7.10$~days; total exposure time of 10.7 ks). In order to increase the detection significance, we chose to merge event files in 2 day bins and use the combined epochs for analysis of the X-ray spectrum. Similar to the X-ray evolution reported in \cite{chen_sn_2025}, we find a significant decrease in X-ray luminosity at $\delta t > 8.88$~days. A complete log of Swift-XRT observations is presented in Table \ref{tab:xrt} and \ref{tab:xray_obs}.

From each merged event file, we extracted a spectrum using a 25$\arcsec$ region centered at the location of SN~2024iss and corrected for background emission with a 100$\arcsec$ source-free region. We use {\tt Xspec} to model each 0.3-10~keV spectra with an absorbed thermal bremsstrahlung model ({\tt tbabs*ztbabs*bremss}), which includes solar abundances \citep{2009ARA&A..47..481A}, and a line-of-sight hydrogen column density of $N_{\rm H,MW} = 1.1 \times 10^{20}$~cm$^{-2}$. We are unable to constrain the temperature in our model fits so we adopt a forward shock temperature evolution following the self-similar solutions by \cite{1982ApJ...258..790C} and \cite{Chevalier2017}. Here, temperature goes as $T_{\rm FS}(t) \propto t^{2(s-3)(n-s)}$, and we adopt $s=2$ for a ``wind-like'' CSM and outer ejecta profile index of $n=10$. This is based on what is commonly adopted for SNe~IIb that are thought to arise from more extended progenitors (e.g., see \citealt{Fransson96, Nymark09, Chandra09, 2016ApJ...818..111K}). We normalize the X-ray temperature evolution to be $T = 33 \ \rm keV \ (t/5d)^{-0.25}$ based on the temperature measurement presented by \cite{chen_sn_2025} using {\it NuSTAR} observations \citep{Margutti24}. 

\begin{figure}[t!]
\centering
\includegraphics[width=\columnwidth]{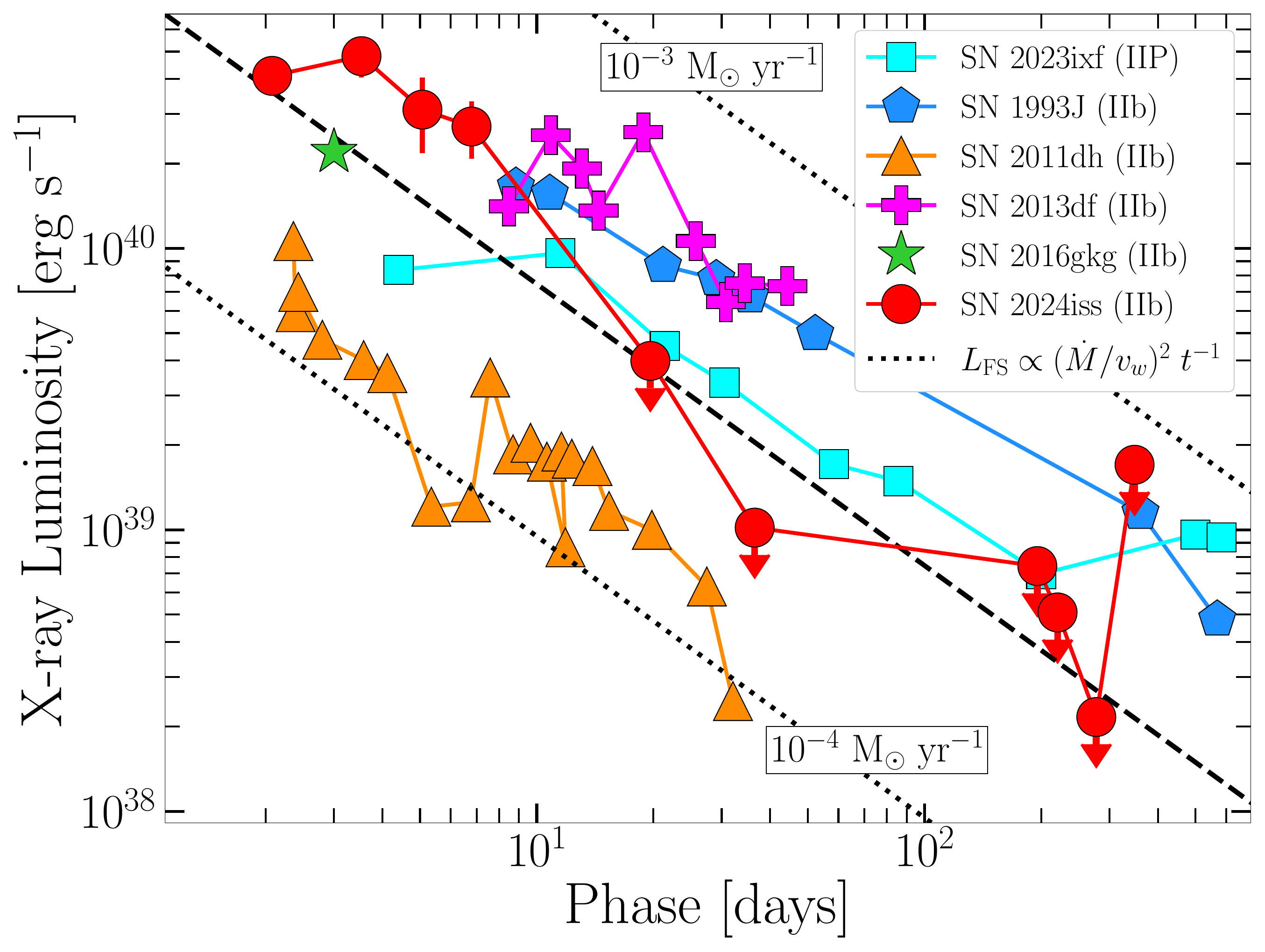}
\caption{Unabsorbed 0.3-10~keV light curve of SN~2024iss (red circles) compared to SNe IIb 1993J (blue polygons), 2011dh (orange triangles), and 2013df (magenta plus signs), as well as type II SN~2023ixf (cyan squares). Black dotted lines represent the analytic prediction for free-free emission from forward shock luminosity \citep{Chevalier03, Chevalier2017} produced at different mass-loss rates and assuming a wind velocity of $v_w = 100~\kms$.\label{fig:XrayLC} }
\end{figure}

We present the 0.3-10~keV luminosities and derived upper limits in Figure \ref{fig:XrayLC}, with all best-fit model parameters and flux measurements being presented in Table \ref{tab:xray_obs}. Compared to other SNe~IIb, SN~2024iss is more luminous than SNe~2011dh \citep{2012ApJ...752...78S} and 2016gkg \citep{Margutti16} at $\delta t < 7$~days. Similarly, SN~2024iss is also brighter in X-rays than CSM-interacting SNe~II-P 2023ixf \citep{Grefenstette23, 2024ApJ...963L...4C, Nayana25} and 2024ggi \citep{Ferdinand25}. However, the X-ray light curve of SN~2024iss fades significantly more dramatically than SNe~IIb 1993J \citep{Chandra09} and 2013df \citep{2016ApJ...818..111K} at $\delta t > 7$~days. The earliest detections of SN~2024iss are also fainter than H-free SNe e.g., SNe~2008D, 2019ehk, 2021gno \citep{Soderberg08, Modjaz09, wjg20, wjg22}. The X-ray light curve evolution and its connection to confined CSM in SN~2024iss is discussed in Section \ref{sec:csm}.


\section{Photometric Analysis} \label{sec:photometry}

\subsection{Light Curve} \label{subsec:lightcurve}



\begin{figure*}[t]
\centering
\subfigure{\includegraphics[width=0.49\textwidth]{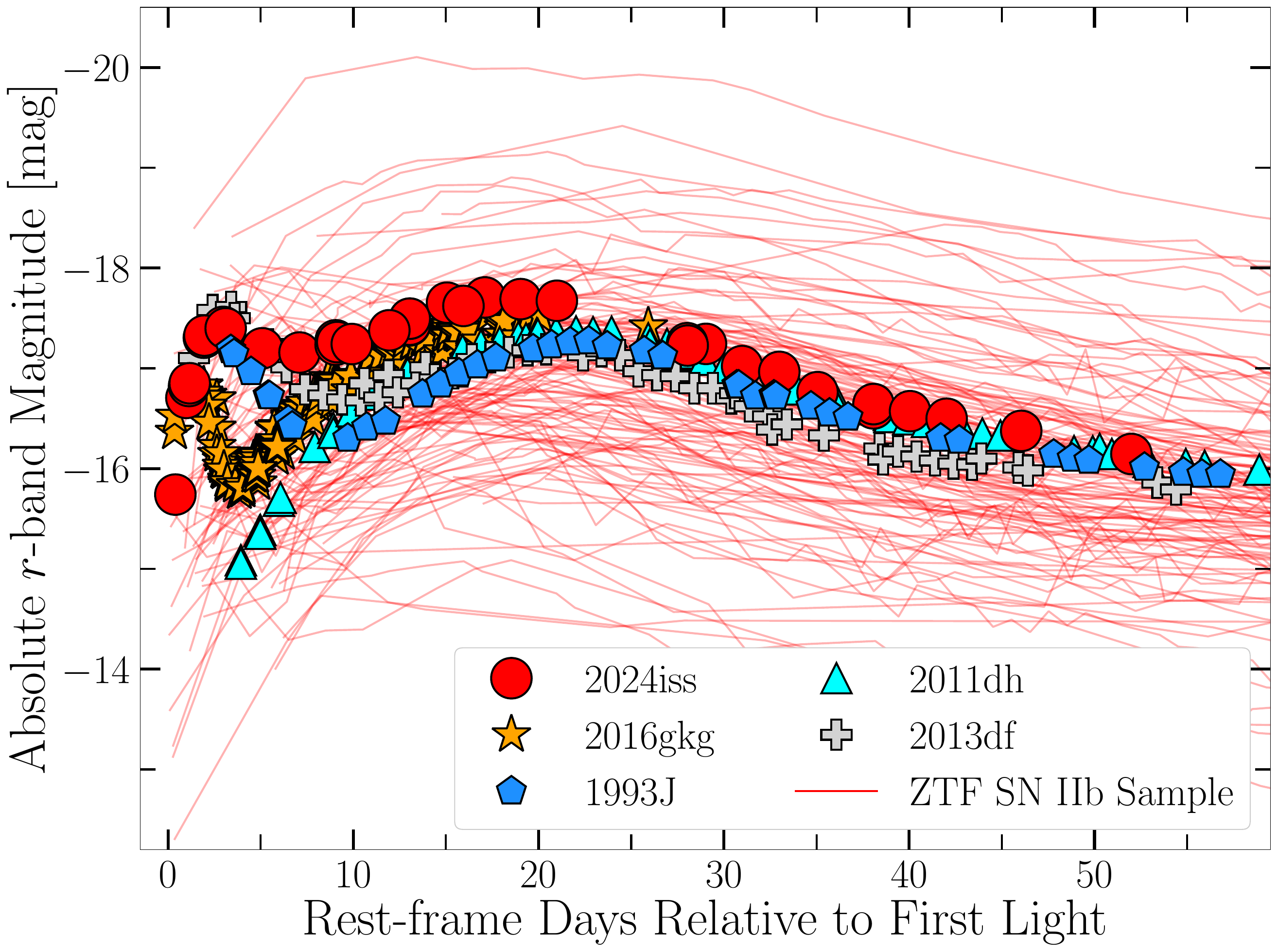}}
\subfigure{\includegraphics[width=0.49\textwidth]{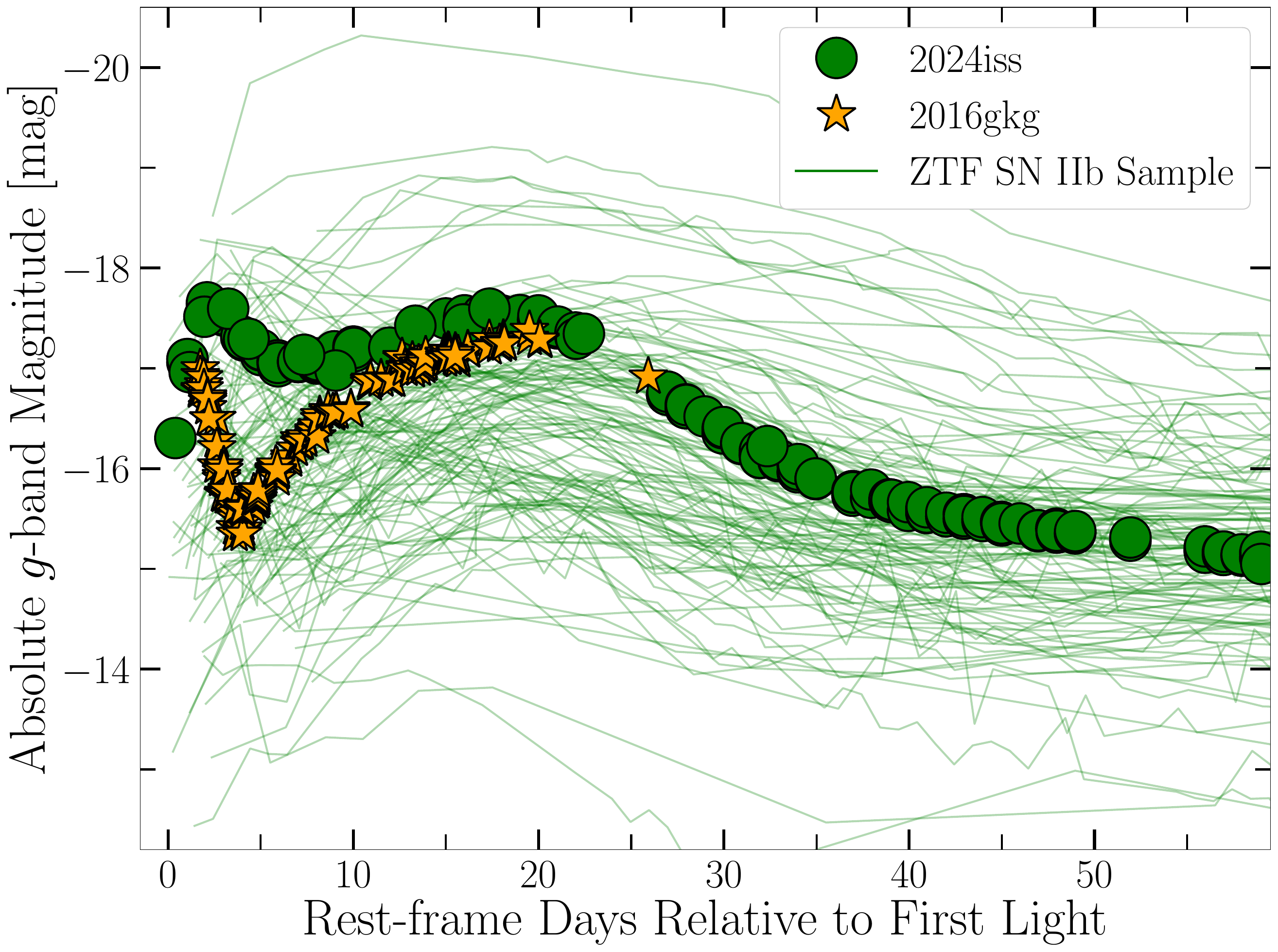}}
\caption{ {\it Left:} Early-time $r$-band photometry of SNe IIb 2016gkg \citep{tartaglia_progenitor_2017, kilpatrick_progenitor_2017, Bersten18}, 2013df \citep{morales-garoffolo_sn_2014}, 1993J \citep{woosley_sn_1994}, 2011dh \citep{bersten_type_2012}, and 2024iss. Complete ZTF SN IIb sample (Hinds et al., in prep) shown as red lines (Hinds et al., in prep). All data points are corrected for extinction. {\it Right:} SN~2024iss $g$-band light curve (green circles) compared to 2016gkg (orange stars) complete ZTF SN IIb sample shown as green lines.}\label{fig:lc comparison}
\end{figure*}

The multiband light curve of SN 2024iss (see Figure \ref{fig:lightcurve}) shows two prominent peaks (as seen in Figure \ref{fig:lc comparison}). We refer to the primary as the shock cooling peak, which can be seen in all optical and ultraviolet bands, with a peak $g$ absolute magnitude \footnote{All absolute magnitudes include extinction correction (see Section \ref{subsec:phot})} of $-17.584 \pm 0.022$ mag at $\delta t = 2.24$ days. We calculated this peak by fitting a third-order polynomial to $g$-band photometry between $\delta t = 1-4$ days. Similarly, we fit a third-order polynomial to $g$-band photometry between $\delta t = 3-10$ days to find the minimum between the two peaks, which has an absolute magnitude of $-16.972 \pm 0.022$ mag at $\delta t =7.01$ days. The second peak, ranging from $\delta t = \sim7 - 40$ days in the $g$-band, results from radiation during the decay chain of $^{56}\mathrm{Ni}\rightarrow^{56}\mathrm{Co}\rightarrow^{56}\mathrm{Fe}$, synthesized in the inner ejecta of the supernova \citep{1980ApJ...237..541A}. The $^{56}\mathrm{Ni}$ peak is seen most prominently in the optical bands, with a peak $g$ absolute magnitude of $-17.469 \pm 0.022$ mag at $\delta t =16.77$ days. We calculated this peak by fitting a third-order polynomial to $g$-band photometry between $\delta t = 10-30$ days. Since the host galaxy is nearby, we note that there may be additional uncertainty in the redshift measurement and the resulting redshift-dependent distance calculation. To account for this, we calculate the following $g$-band peak absolute magnitudes for the second light curve peak for a distance uncertainty of $\pm5$ Mpc: at a distance of 9 Mpc, $M^{\textrm{peak}}_g = -16.504$ mag, and at a distance of 19 Mpc, $M^{\textrm{peak}}_g = -18.127$ mag. This uncertainty may propagate into the physical parameters derived from photometric data in Section \ref{sec:photometry}. However, due to the lack of constraints on this uncertainty, we adopt the redshift-dependent distance of $14.04 \pm 0.14$ Mpc for further analysis.

We compare the light curve of SN 2024iss with other prototypical SNe IIb (1993J, 2011dh, 2013df, 2016gkg) and the upcoming complete ZTF SNe IIb sample from Hinds et al., in prep (see Figure \ref{fig:lc comparison}). Photometry in the $r$-band (1993J: \citealt{woosley_sn_1994}; 2011dh: \citealt{bersten_type_2012}; 2013df: \citealt{morales-garoffolo_sn_2014}; 2016gkg: \citealt{tartaglia_progenitor_2017, kilpatrick_progenitor_2017, Bersten18}) reveals SN 2024iss to have a similar magnitude shock cooling peak but a brighter second peak from $^{56}$Ni decay. Additionally, the transition between the first and second peaks of the light curve is less pronounced in SN 2024iss compared to the other SNe (see Section \ref{sec:csm} discussion on how CSM interaction impacts this observable).

\subsection{Shock Breakout Constraints} \label{subsec:SBO}

As shown in Figures \ref{fig:discovery} \& \ref{fig:SBO}, the time of first light in SN~2024iss is tightly bracketed between the last ZTF $g$-band non-detection on MJD 60442.23 and first ZTF $i$-band detection on MJD 60442.28. Consequently, the uncertainty on time of first light is empirically constrained to $\pm 40$~min, assuming first light is when the SN is brighter than $M_g > -9.8$~mag, which reflects the limiting magnitude of the last $g$-band non-detection. As shown in the right panel of Figure \ref{fig:SBO}, we jointly fit the $gri$-band photometry at $\delta t < 2$~days using the following power-law formalism:

\begin{equation}
F_\nu(t) =
\begin{cases}
a (t - t_{\mathrm{fl}})^b & \text{if } t \ge t_{\mathrm{fl}} \\
0 & \text{if } t < t_{\mathrm{fl}}
\end{cases}
\end{equation}

\noindent
where $a$ and $b$ can vary across filters but $t_{fl}$ remains fixed (e.g., see \citealt{Miller20, Rehemtulla25}). We first fit the light curves with the earliest $i$-band point excluded, which unsurprisingly leads to a time of first light that is too late in time to be consistent with the last non-detection and first detection of SN~2024iss. Second, we fit with first light fixed to be the phase of the last $g$-band non-detection and force a consistent temperature evolution wherein $F_{\nu, \ g} > F_{\nu, \ r} > F_{\nu, \ i}$ (shown as dashed lines in Fig. \ref{fig:SBO}). Overall, the power law indices across filters are consistent with thermal SN emission where bluer filters are more sensitive to decreasing temperature and rise slower than redder filters i.e., $a_g < a_r$. However, even when first light is fixed to the last non-detection date, this single index power law model cannot reproduce the flux observed in the earliest $i$-band detection, which deviates from the model at a $\sim44~\sigma$ level. A similar early light curve evolution was observed in type II SNe~2023ixf and 2024ggi, both of which showed evidence for delayed shock breakout from dense CSM given the need for multiple powerlaw components when fitting their observed photometry in the first $\sim$day after first light \citep{Hosseinzadeh23, Li24, Shrestha24, WJG24ggi}. Therefore, by using the last non-detection in $g$-band and first photometry points consistent with the power law model, we can constrain the regime of shock breakout emission to be $\delta t < 0.42$~day, where the first $i$-band detection is likely associated with shock breakout.

We can further constrain the shock breakout emission in SN~2024iss by considering the expected energy, temperature and various timescales that are at play for massive star breakout emission. For a typical SN~IIb progenitor with envelope mass and radius of $M_e = 0.1~\Msun$ and $R_e = 300~\Rsun$, the amount of energy that travels through this envelope at breakout can be expressed as:

\begin{equation}
    E_e \approx 1.3\times 10^{50} \Bigg(\frac{E_{k}}{10^{51} \rm \ erg} \Bigg ) \Bigg(\frac{M_{c}}{2 \ \rm \Msun} \Bigg )^{-0.7} \Bigg(\frac{M_{e}}{0.1  \ \rm \Msun} \Bigg )^{0.7} \rm erg  
\end{equation}

\noindent
where $E_{k}$ is the SN kinetic energy and $M_{c}$ is the core mass \citep{2014ApJ...788..193N}. Treating this as a low energy SN in an extended envelope we can use the relations for a $n=3/2$ polytrope from \cite{Matzner99} to express the shock breakout radiation temperature as: 

\begin{equation}
\begin{aligned}
    T_{\rm SBO} = 6.2 \times 10^{5} \left(\frac{E_e}{1.3 \times 10^{50}\ {\rm erg}}\right)^{0.20} \\ \left(\frac{M_e}{0.1\,M_\odot}\right)^{-0.052} \left(\frac{R_e}{300\,R_\odot}\right)^{-0.54}\ {\rm K}
\end{aligned}
\end{equation}

\noindent
where we assume $\kappa = 0.34$~cm$^{2}$~g$^{-1}$ and $\rho / \rho_{\star} = 1$. Then the total energy in shock breakout goes as: 

\begin{equation}
\begin{aligned}
    E_{\rm SBO} =
1.7 \times 10^{48}
\left(\frac{E_e}{1.3 \times 10^{50}\ {\rm erg}}\right)^{0.56}\\
\left(\frac{M_e}{0.1\,M_\odot}\right)^{-0.44}
\left(\frac{R_e}{300\,R_\odot}\right)^{1.74}
\ {\rm erg}
\end{aligned}
\end{equation}

The total luminosity depends on the timescales over which photons are observed. According to \cite{Matzner99}, the diffusion timescale goes as:

\begin{equation}
\begin{aligned}
t_{\rm diff} \approx
500
\left(\frac{E_e}{1.3 \times 10^{50}\ {\rm erg}}\right)^{-0.79}\\
\left(\frac{M_e}{0.1\,M_\odot}\right)^{0.21}
\left(\frac{R_e}{300\,R_\odot}\right)^{2.16}
\ {\rm s}
\end{aligned}
\end{equation}

\noindent 
Additionally, the light travel timescale can be expressed as:

\begin{equation}
\begin{aligned}
t_{\rm lt} = \frac{R_e}{c}
= 700
\left(\frac{R_e}{300\,R_\odot}\right)
\ {\rm s}
\end{aligned}
\end{equation}

\noindent
Finally, for the non-spherical effect timescale ($t_{\rm nsph} \approx \frac{R_e}{v_{\rm sh}}$), we can express the shock velocity as: 

\begin{equation}
\begin{aligned}
    v_{\rm sh} \approx
3.5 \times 10^{9}
\left(\frac{E_e}{1.3 \times 10^{50}\ {\rm erg}}\right)^{0.57}\\
\left(\frac{M_e}{0.1\,M_\odot}\right)^{-0.44}
\left(\frac{R_e}{300\,R_\odot}\right)^{-0.26}
\ {\rm cm\ s^{-1}}
\end{aligned}
\end{equation}

\noindent
This then allows us to describe the non-spherical timescale as:

\begin{equation}
\begin{aligned}
t_{\rm nsph} \approx
\frac{R_e}{v_{\rm sh}}
\approx
6 \times 10^{3}
\left(\frac{E_e}{1.3 \times 10^{50}\ {\rm erg}}\right)^{-0.57}\\
\left(\frac{M_e}{0.1\,M_\odot}\right)^{0.44}
\left(\frac{R_e}{300\,R_\odot}\right)^{1.26}
\ {\rm s}
\end{aligned}
\end{equation}

\noindent
Given our fiducial SN IIb values, the diffusion timescale is likely insignificant and shock breakout is defined by light travel time and non-spherical effects i.e., $t_{\rm SBO} \approx 12 \rm  \min - 2 \ hrs$. Assuming that shock breakout occurs directly after the last ZTF non-detection, the first $i$-band detection at 1.3~hrs later does lie within this shock breakout time window, and would be more consistent with non-spherical effects, which are known to elongate the shock breakout signal in massive stars when calcuated in 3-D simulations (e.g., see \citealt{Goldberg22}).


\subsection{Blackbody Evolution} \label{subsec:bolometric}

\begin{figure*}[ht]
\centering
\subfigure{\includegraphics[width=0.46\textwidth]{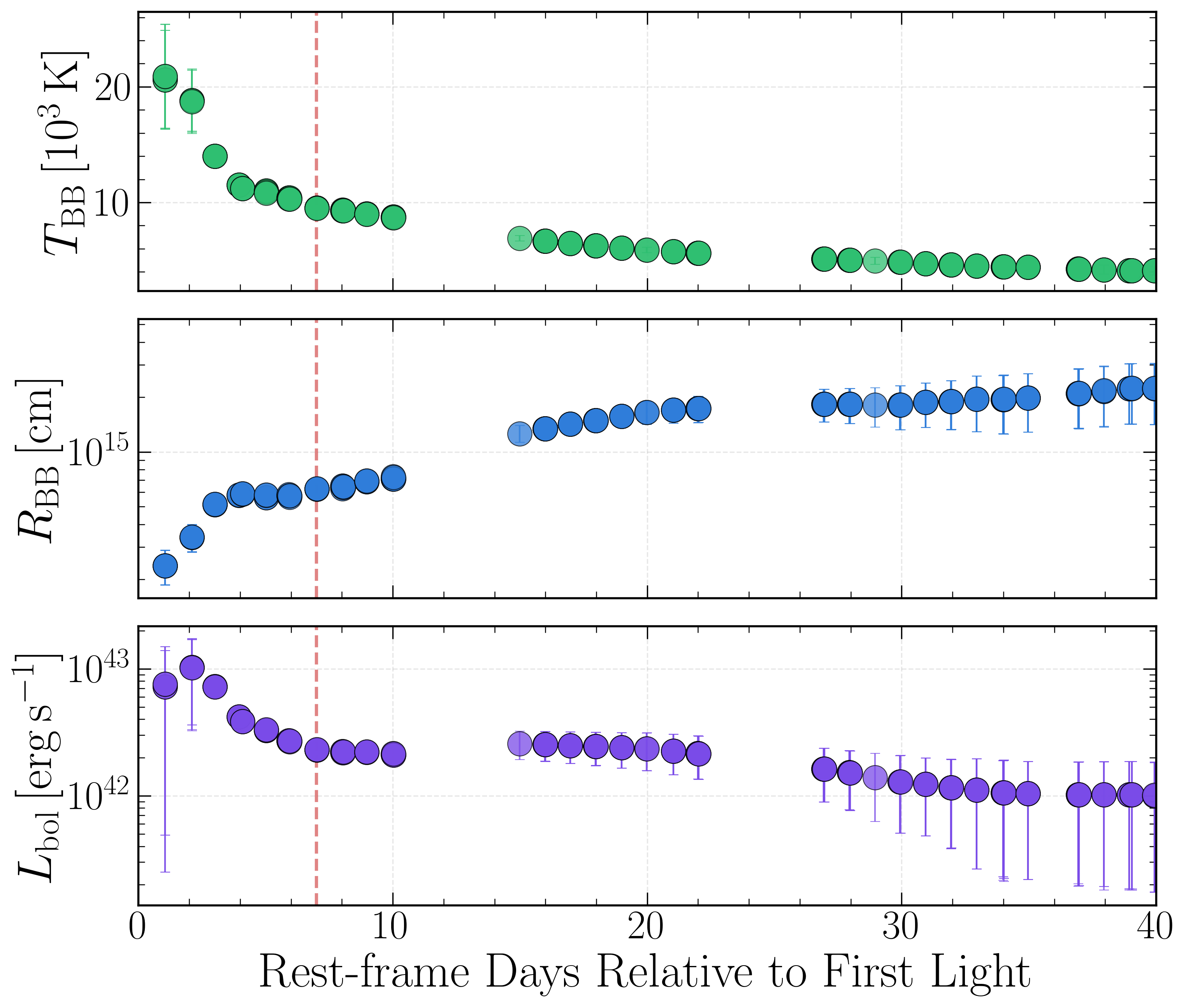}}
\subfigure{\includegraphics[width=0.52\textwidth]{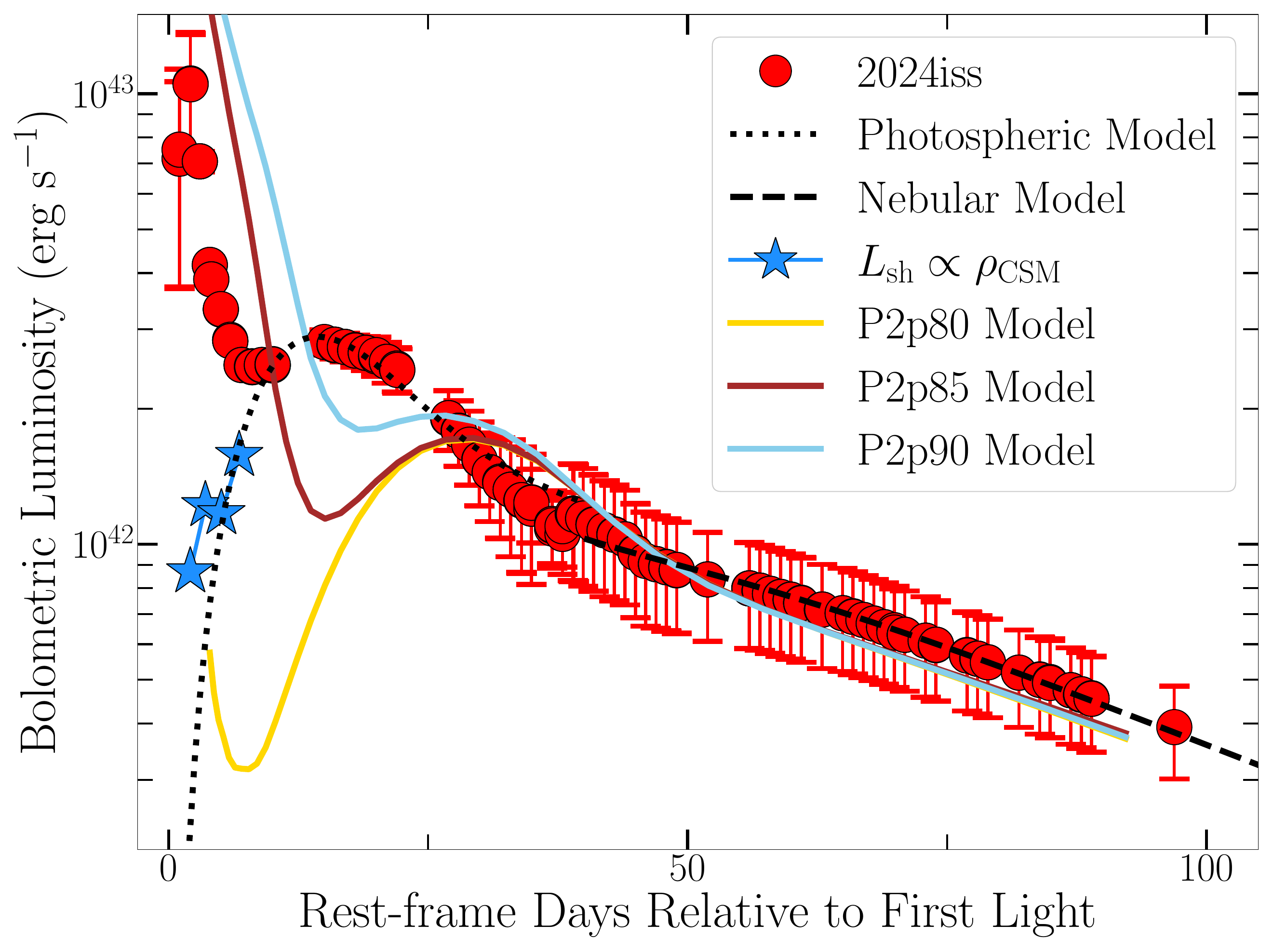}}
\caption{ {\it Left:} Blackbody temperature (top), blackbody radius (middle), and bolometric luminosity (bottom) of SN 2024iss at early times, as calculated with \texttt{superbol} \citep{nicholl_superbol_2018} for $\delta t < 40$ days. The dashed vertical line shows values at $\delta t = 7$ days. {\it Right:} Bolometric light curve of SN~2024iss (red circles) with best-fit photospheric and nebular light curve models (see Section \ref{subsec:ni56}) shown as dotted and dashed lines, respectively. Radiated luminosity from CSM-interaction power ($\epsilon_{\rm rad} = 30$\%) using CSM densities inferred from X-ray detections (see Section \ref{sec:csm}) shown as blue stars. Binary progenitor model light curves from \cite{dessart_sequence_2024} shown as solid blue, yellow and brown lines (Section \ref{subsec:binary progenitor}) and represent a range of ejecta and H-rich envelope masses of $M_{\rm ej} = 2.3 - 2.5~\Msun$ and $M_{\rm env} = 0.02 - 0.12~\Msun$. \label{fig:bolometric} }
\end{figure*}

We calculated the bolometric light curve of SN 2024iss based on photometry interpolated from the \textit{UVW2, UVM2, UVW1}, and $UBgVriz$ bands up to $\delta t = 50$ days using the \texttt{superbol} package \citep{nicholl_superbol_2018}. A blackbody spectral energy distribution was fit to the interpolated observations, resulting in calculations for the bolometric luminosity, blackbody temperature, and blackbody radius for each epoch (see Figure \ref{fig:bolometric} left panel). The bolometric luminosity shows the same double-peaked shape as the observed photometry. The highest temperature of $(2.09 \pm 0.45) \times 10^4$ K occurs at $\delta t = 1.05$ days and the peak bolometric luminosity is $(1.03 \pm 0.70) \times 10^{43}$ erg s$^{-1} \ \ (\textrm{log}_{L_{\textrm{bol}}}=43.01)$ at $\delta t = 2.10$ days. The peak temperature is higher than that of SN 2016gkg ( $\sim 10,000$ K at $\delta t = 1.70$ days; \citealt{tartaglia_progenitor_2017}) and lower than that of SN 1993J ($\sim10^5$ K; \citealt{woosley_sn_1994}), while the bolometric luminosity is comparable to other IIb SNe \citep{bersten_type_2012,morales-garoffolo_sn_2014, tartaglia_progenitor_2017}.

\begin{figure}[h!]
\includegraphics[width=\columnwidth]{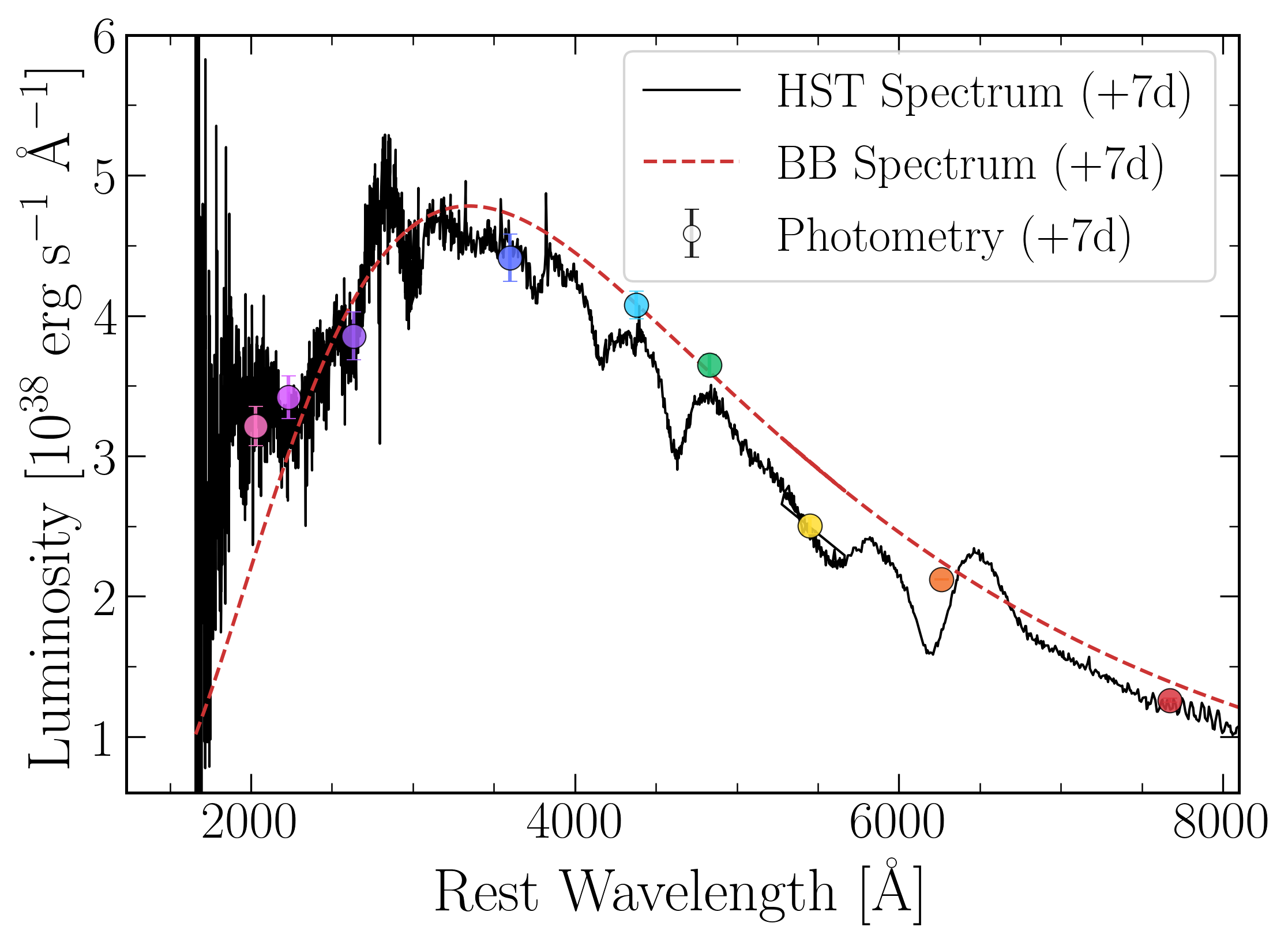}
\caption{The observed \textit{HST} $\delta t = 7$ days spectrum with the calculated blackbody spectrum and observed photometry at $\delta t = 7$ days. The observed spectrum and photometry are corrected for dust extinction.}
\label{fig:bb spectrum}
\end{figure}

From these parameters, we calculated a blackbody spectrum at $\delta t = 7$ days, and compared it to the observed $\delta t = 7$ days \textit{HST} spectrum and observed photometry at similar phases (see Figure \ref{fig:bb spectrum}). For this phase, we only consider interpolated photometry from the $i$ to \textit{UVW2} bands since there are no $z$-band observations in the earliest phases. The overall shape of both spectra match well, indicating a $\delta t = 7$ days blackbody temperature of $8,660 \pm 130$ K and a blackbody radius of $(8.3 \pm 0.2) \times 10^{14}$ cm. To quantify the departure from a thermal continuum, we compute the root mean square (RMS) fractional deviation between the observed spectrum and the best-fit blackbody model over various wavelength ranges. At optical wavelengths ($3000 - 9000$~\AA), the flux uncertainty weighted RMS deviation is 9\%, which increases to 17\% across near-UV wavelengths ($1600 - 3000$~\AA). Interestingly, the RMS deviation from $1600 - 2200$~\AA\ reaches 47\%, which can be observed visually in Figure \ref{fig:bb spectrum} as the blackbody continuum drops sharply at $<2200$~\AA\ compared to emission/absorption profiles from Fe-group elements SN~2024iss.

\subsection{$^{56}$Ni Mass Estimation}\label{subsec:ni56}

In order to infer ejecta mass ($M_{\rm ej}$), kinetic energy ($E_{\rm k}$), and ${}^{56}\textrm{Ni}$ mass ($M_{\textrm{Ni}}$), we fit the SN~2024iss bolometric light curve with a two-component model presented in Appendix A of \cite{valenti08} and in \cite{wheeler15} (see Figure \ref{fig:bolometric} right panel). This analytic formalism distinguishes between photospheric ($\delta t < 30$~days; \citealt{arnett82}) and nebular ($\delta t > 40$~days; \citealt{sutherland84,Cappellaro97}) phases of the light curve and implements the possibility of incomplete $\gamma$-ray trapping and a typical opacity of $\kappa = 0.1$ cm$^2$ g$^{-1}$. Given the known degeneracy between $E_{\rm k}$ and $M_{\rm ej}$, we adopt a photospheric velocity of $v_{\rm ph} \approx 6500~\kms$ based on the \ion{Fe}{ii} absorption minimum at peak brightness. For the photospheric model ($10 < \delta t < 30$~days), we fit the secondary light curve peak and find an ejecta mass of $M_{\rm ej} = 1.1 \pm 0.1~\Msun$, kinetic energy of $E_{\rm k} = (2.6 \pm 0.3) \times 10^{50}$~erg, and ${}^{56}\textrm{Ni}$ mass of $M_{\textrm{Ni}} = 0.11 \pm 0.01~\Msun$. However, this $E_k$ value is nonphysical given the large ejecta velocities ($\sim15{,}000-18{,}000~\kms$) of the fastest moving SN ejecta observed in the SN spectra at $\delta t\sim3$~days (e.g., see Fig. \ref{fig:phase_vel}). We note that using these ``Arnett-like'' formalisms fit H-poor SNe at peak can over-estimate the ${}^{56}\textrm{Ni}$ mass by up to $\sim$50\% (e.g., see \citealt{Dessart15, Dessart16, Afsariardchi21, Haynie23}). However, we compare the rise-time and luminosity at peak in SN~2024iss to radioactive decay model light curves calibrated with numerical radiation transport simulations presented in \cite{Khatami19} and find that $M_{\textrm{Ni}}$ is likely $\sim 0.09-0.1~\Msun$. 

For the nebular phase model ($\delta t > 30$~days), we find a larger kinetic energy of $E_{\rm k} = (6.3 \pm 0.1) \times 10^{50}$~erg and, consequently, a larger ejecta mass of $M_{\rm ej} = 2.5 \pm 0.03~\Msun$. We find a ${}^{56}\textrm{Ni}$ mass estimated from the nebular phase model to be $M_{\textrm{Ni}} = 0.11 \pm 0.01~\Msun$, similar to the photospheric model estimate. Additionally, we fit the bolometric light curve from $\delta t = 60 - 300$~days with a radioactive decay model that has the timescale of $\gamma$-ray escape during energy disposition from ${}^{56}\textrm{Co}$ decay ($t_{\gamma}$) and $M_{\textrm{Ni}}$ as free parameters \citep{sutherland84,clocchiatti97, wheeler15, wjg21}. We find a similar ${}^{56}\textrm{Ni}$ mass to the other model formalisms ($M_{\textrm{Ni}} = 0.091 \pm 0.002~\Msun$) and a $\gamma$-ray leakage timescale of $t_{\gamma} = 145.7 \pm 2.7$~days, which is similar to other SNe~IIb \citep{wheeler15, Haynie23}. Modeling of the light curve ``tail'' also provides a consistency check for the parameters derived above using the photospheric model. 

\subsection{Shock Cooling Emission Modeling} \label{subsec:shock cooling}

The first peak in the light curve corresponds to cooling of the extended stellar envelope after the shock from the supernova propagates through it \citep{woosley_sn_1994, Dessart11, bersten_type_2012, 2014ApJ...788..193N, Bersten18}. The effects from this phenomenon are visible until emission from the $^{56}\mathrm{Ni}$ decay becomes brighter than the fading emission from shocked envelope. In the $g$-band, the first peak decreases until $\delta t = 7$ days, after which the flux begins to increase again. This can also be seen in the bolometric luminosity light curve, in which the second peak begins to rise around $\delta t = 8$ days.
We modeled photometry in the $i$, $r$, $V$, $g$, $B$, $U$, \textit{UVW1}, \textit{UVM2}, and \textit{UVW2} bands with the \citealt{sapir_uvoptical_2017} (SW17), \citealt{morag_shock_2023} (MSW23), and \citealt{piro_shock_2021} (P21) shock cooling emission models. 

SW17 is a physically calibrated model for red and blue supergiant progenitors based on luminosity from shock cooling emission. In the model, the assumed value of the polytropic index $n$ (related to the density profile of the stellar envelope) differentiates between the red and blue supergiant, with $n=1.5$ taken to represent a red supergiant progenitor with a convective envelope, and $n=3$ a blue supergiant progenitor with a fully radiative envelope. Since type IIb supernovae are especially sensitive to the value of $n$ due to their low envelope masses \citep{farah_shock-cooling_2025} and the inefficient convection in the progenitor envelopes \citep{2025arXiv250812486G}, we modeled the shock cooling emission with both versions of SW17 for each $n$ value. MSW23 is an analytic model that builds from the SW17 $n=1.5$ model with a two-phase expansion of the supernova ejecta. Both SW17 and MSW23 approximate a constant opacity during shock cooling emission, which is only valid during times when the supernova temperature is above $8{,}120$ K, according to the model setup described in \cite{sapir_uvoptical_2017}. From our bolometric fits (see Section~\ref{subsec:bolometric}), the blackbody temperature is above this threshold until $\delta t = 9$ days. However, due to the rise of emission from $^{56}\mathrm{Ni}$ decay, we limit our shock cooling emission fitting to observations to $\delta t < 7$ days. 

\begin{figure*}[h!t]
\centering
\subfigure{\includegraphics[width=0.48\textwidth]{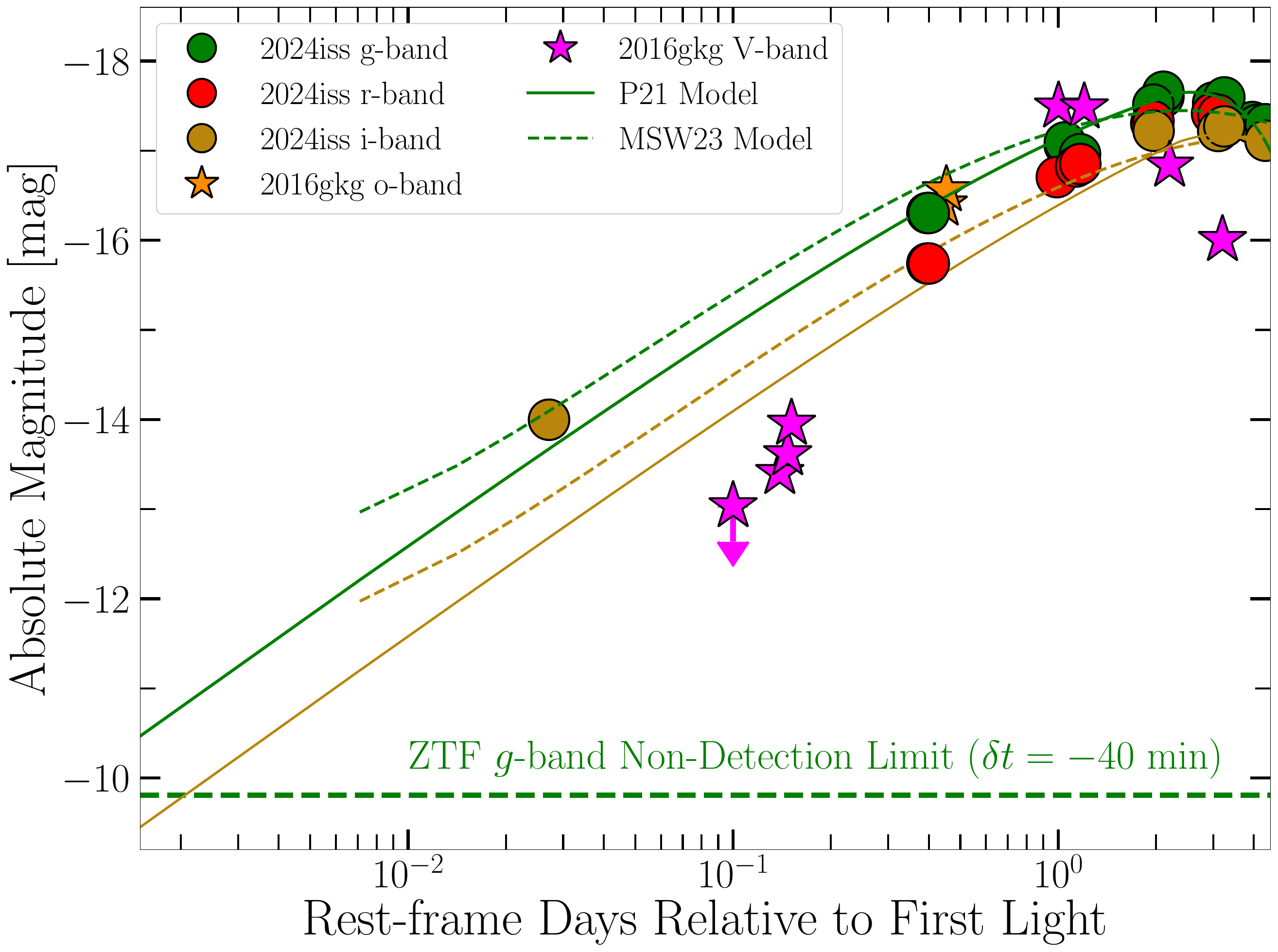}}
\subfigure{\includegraphics[width=0.48\textwidth]{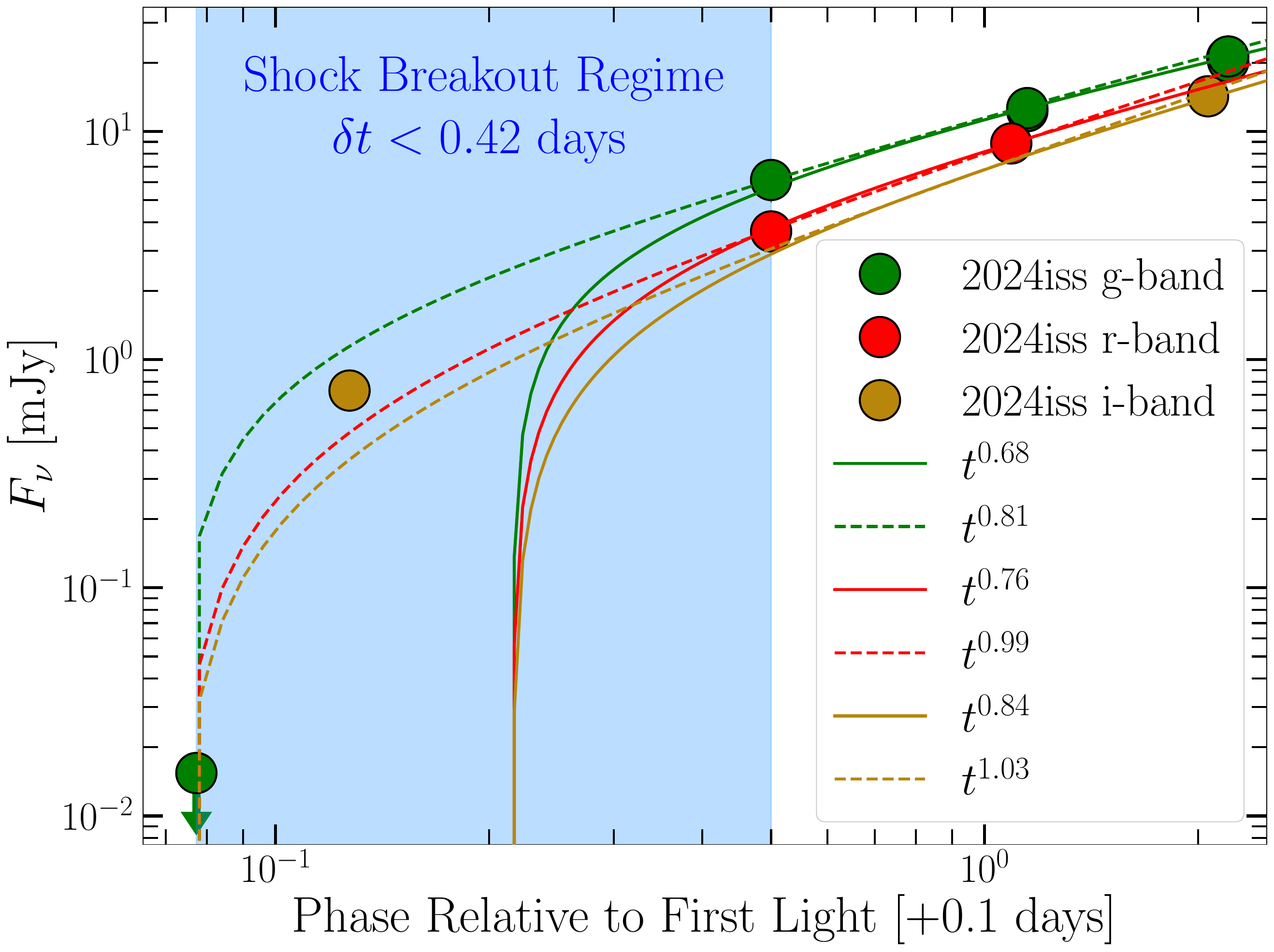}}
\caption{ {\it Left:} Multi-band photometry of SN~2024iss (circles) compared to first detections of SN~IIb 2016gkg in $V$-band (stars) from \cite{Bersten18}. Horizontal dashed line represents last non-detection limit in $g$-band at 1.33~hours before first $i$-band detection. Solid and dashed lines represent best-fit shock cooling models from \cite{piro_shock_2021} and \cite{morag_shock_2023}. There is a significant deviation from models predictions at $\delta t<0.1$~days, likely connected to shock breakout emission. {\it Right:} Power-law model fits to early multi-band photometry with (dashed lines) and without (solid lines) the inclusion of the earliest $i$-band detection. Dashed lines represent fits that force the last $g$-band non-detection as the time of first light and $F_{\nu, \ g} > F_{\nu, \ r} > F_{\nu, \ i}$ show that the $i$-band point cannot be fit with a single power-law index.  }
\label{fig:SBO}
\end{figure*}

P21 is an analytic model that builds on the low-mass extended material model of \cite{piro_using_2015} by incorporating two-zones with varying densities and velocities. We used the Python packages \texttt{shock-cooling-curve} \citep{venkatraman_shock_cooling_curve_2024} to fit P21 and SW17, and \texttt{lightcurve-fitting} \citep{hosseinzadeh_light_2023} to fit MSW23. We tested fits for each model for photometry up to $\delta t = 5$, 6, and 7 days, and chose the best statistical fit for each model (see Appendix \ref{appendix shock cooling}). As a result, we fit SW17 and MSW23 with photometry up to $\delta t = 7$ days and P21 with photometry up to $\delta t = 5$ days. Additionally, MSW23 contains a factor for UV flux suppression, $A$, which we set equal to $1$ (indicating no UV flux suppression) due to the model underestimating flux in the UV bands (see Figure \ref{fig:shock cooling}). We found best fit parameters using MCMC fitting with 1000 steps and 25 walkers for each model. The results are shown in Table~\ref{tab:shock cooling results}. The 4 models result in a progenitor radius ($R_{\star}$) range of $100-320 \ \Rsun$, a hydrogen-rich envelope mass ($M_{\textrm{env}}$) range of $0.07-5.16\ \Msun$, a shock velocity ($v_s$) range of $(0.84-1.60)\times10^9\ \textrm{cm s}^{-1}$, and a first light time offset (estimated difference from the time of first light given in Table \ref{table:Observations}, $t_{\textrm{offset}}$) of $0.000 - 0.003$ days. These parameter ranges likely reflect the variety of assumptions made across the different models, which we discuss in Section \ref{sec:discussion}. The first ZTF $i$-band detection point at $\delta t = 0.027$ days likely captures emission from the initial shock breakout (as calculated in Section \ref{subsec:SBO}). At this phase, the shock cooling emission models significantly deviate from the observed value (as seen in Figure \ref{fig:SBO}). We explain further model constraints and our final selection of parameter ranges in Section \ref{sec:discussion}.

\section{Spectroscopic Analysis} \label{sec:spectroscopy}

\subsection{Optical} \label{subsec:optical spectra}

The evolution of the optical spectrum of SN 2024iss in the photospheric phase shows several key features (Figure \ref{fig:spec}). Early spectra up to $\delta t \sim 3$ days are featureless blackbody curves \citep{filippenko_optical_1997} reflecting a blackbody temperature peak in ultraviolet wavelengths. After 3 days, features from hydrogen begin to emerge, such as the characteristic P-Cygni profile (the combined absorption and emission features) of H$\alpha$ around 6500~\AA. Prominent helium features, such as \ion{He}{i} at 5800~\AA, appear in early spectra after $\sim$5 days. As the photospheric velocity decreases, the H and \ion{He}{i} features become narrower, with redshifted absorption features approaching the rest wavelength emission line. \ion{Fe}{ii}, \ion{O}{i}, and \ion{Ca}{ii} features also appear in optical spectra after $\sim20$ days. 

We measured the spectral line velocities of prominent P-Cygni profiles that can be traced throughout the photospheric phase optical spectra ranging from $\delta t = 4.0 - 94.9$ days (Figure \ref{fig:phase_vel}). We calculated velocities using the absorption component minimum of the P-Cygni profile seen in each spectrum. H$\alpha$ appears to probe the highest velocities, starting at $\sim18{,}000 \ \kms$ and decreasing to $\sim10{,}000 \ \kms$ by $\delta t = 90.4$ days. H$\beta$ and H$\gamma$ velocities follow a similar pattern,  with H$\beta$ decreasing from $\sim15{,}000 \ \kms$ to $\sim8{,}000 \ \kms$ and  H$\gamma$ from $\sim12{,}000 \ \kms$ to $\sim7{,}000 \ \kms$. The velocity of \ion{He}{i} evolves like that of H$\beta$ and H$\gamma$, starting at $\sim15{,}000 \ \kms$ and decreasing to $\sim7{,}000 \ \kms$. \ion{Fe}{ii} displays a noticeably lower velocity than the elements listed above, starting at $\sim8{,}000 \ \kms$ at $\delta t = 15.6$ days and decreasing to $\sim4{,}000 \ \kms$ at $\delta t = 77.0$ days. This difference highlights various layers in the ejecta, with \ion{Fe}{ii} $\lambda5169$ probing the velocity just outside the photosphere \citep{Kasen09, Paxton18, Goldberg19}.

Optical spectra of SN 2024iss appear similar to optical spectra of other SNe IIb at similar phases (see Figure \ref{fig:IIb_specs}). The SNe 2016gkg, 2011dh, 2013df, 1993J, and 2024iss all show prominent features from H, \ion{He}{i}, \ion{Fe}{ii}, and \ion{Ca}{ii} at $\delta t \sim 27$ days. SN 2024iss appears to have less pronounced \ion{Ca}{ii} features at $\sim4000$~\AA~ and $\sim8500$~\AA~, and more prominent H$\beta$ absorption compared to the other SNe IIb. No other distinguishable differences are seen between SN 2024iss and the other optical spectra of SNe IIb.

\begin{figure*}[ht]
\centering
\includegraphics[width=\textwidth]{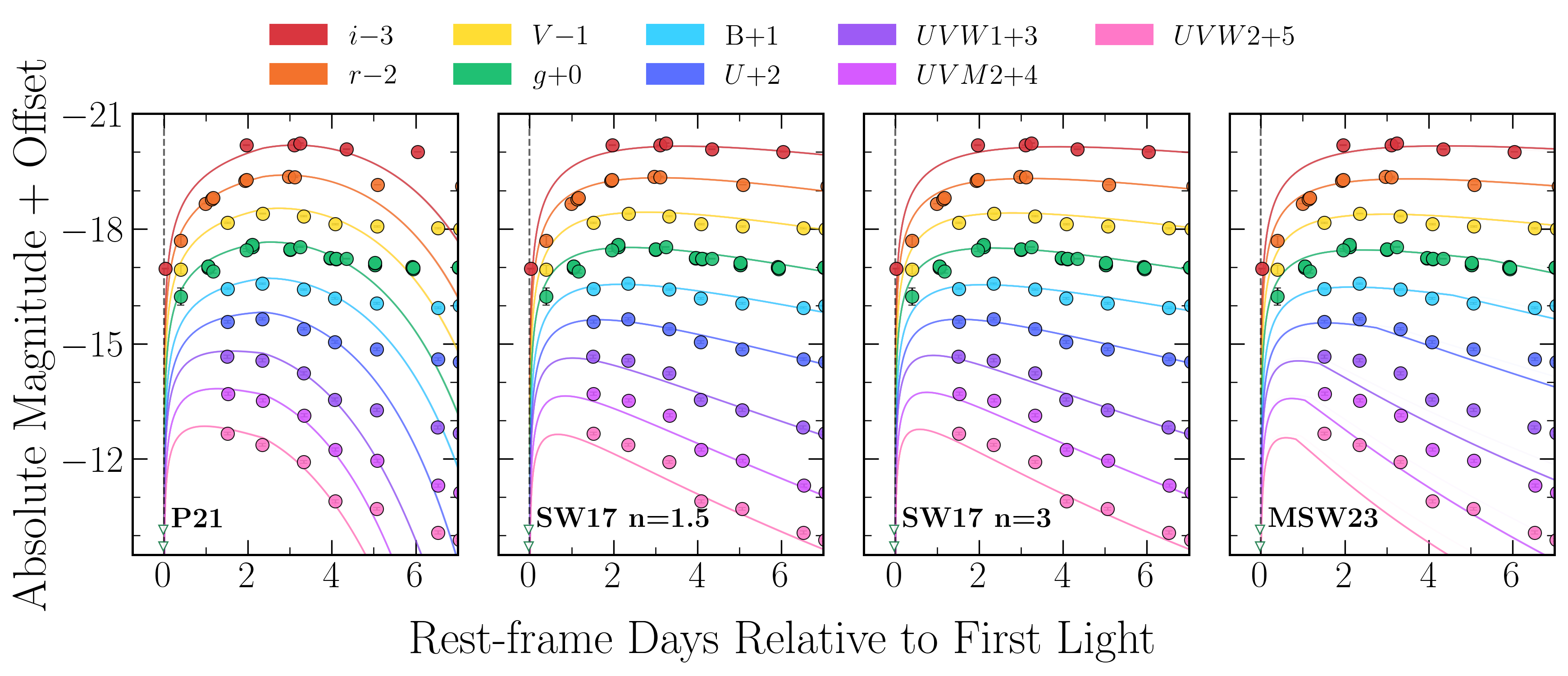}
\caption{Shock cooling model fits calculated from the best fit parameters listed in Table \ref{tab:shock cooling results}. P21 was fit to photometry until $\delta t < 5$ days, while SW17 and MSW23 were fit to photometry until $\delta t < 7$ days (see Appendix \ref{appendix shock cooling}). The $r$, $V$, and $g$-band points at $\delta t = 0.396$ days are from \cite{chen_sn_2025}. }
\label{fig:shock cooling}
\end{figure*}

\begin{figure*}[h]
\centering
\includegraphics[width=\textwidth]{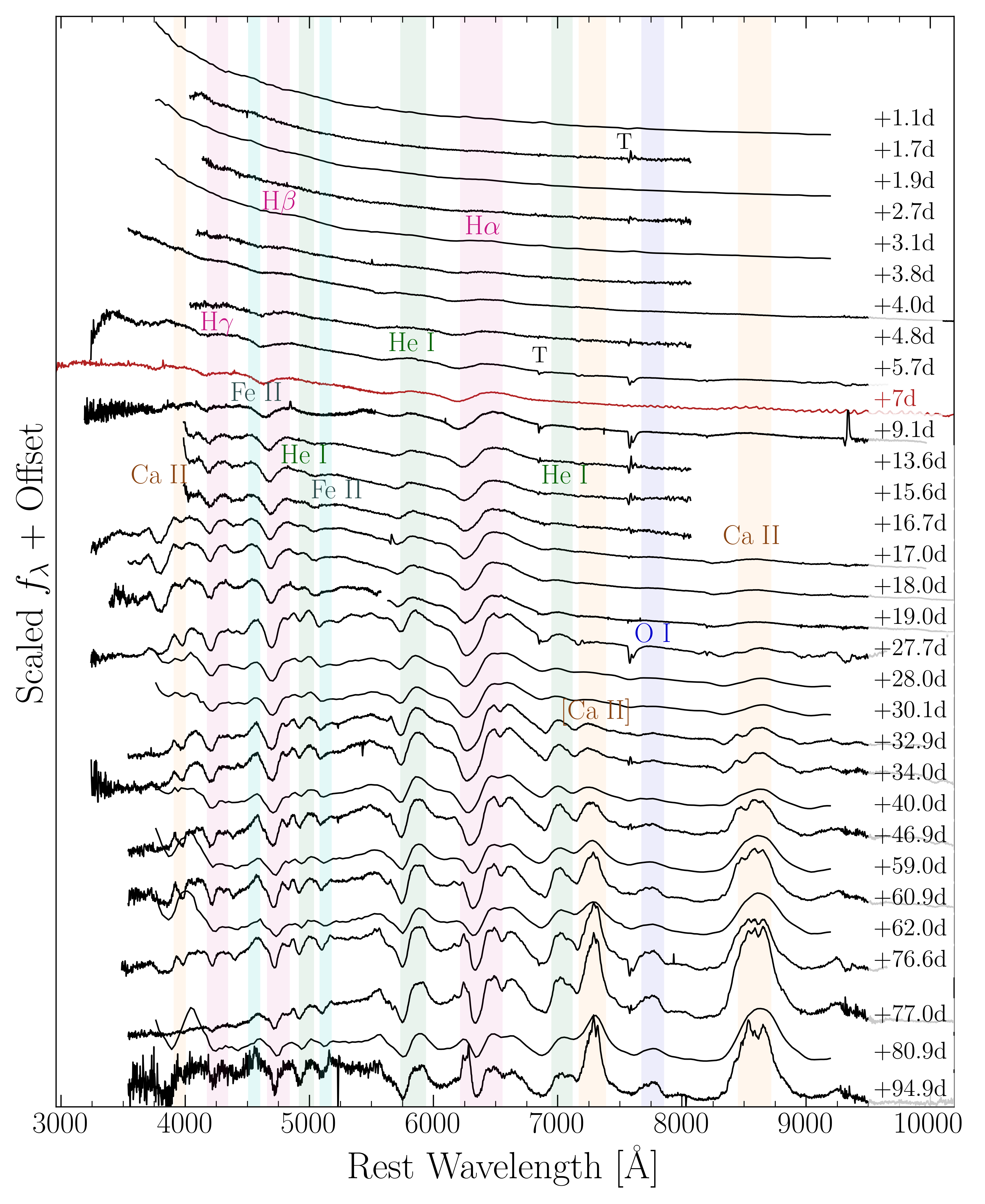}
\caption{The evolution of the optical spectrum of SN 2024iss in the photospheric phase, ranging from $\delta t = 1.1-94.9$ days. Prominent spectral features and their corresponding elements are highlighted, and telluric absorption features are marked with 'T'. The \textit{HST} $\delta t = 7$ days spectrum is shown in red. All spectra are corrected for redshift and dust extinction, and spectra are scaled relative to mean flux and offset.}
\label{fig:spec}
\end{figure*}

\begin{figure}[h!]
    \centering
    \includegraphics[width=\columnwidth]{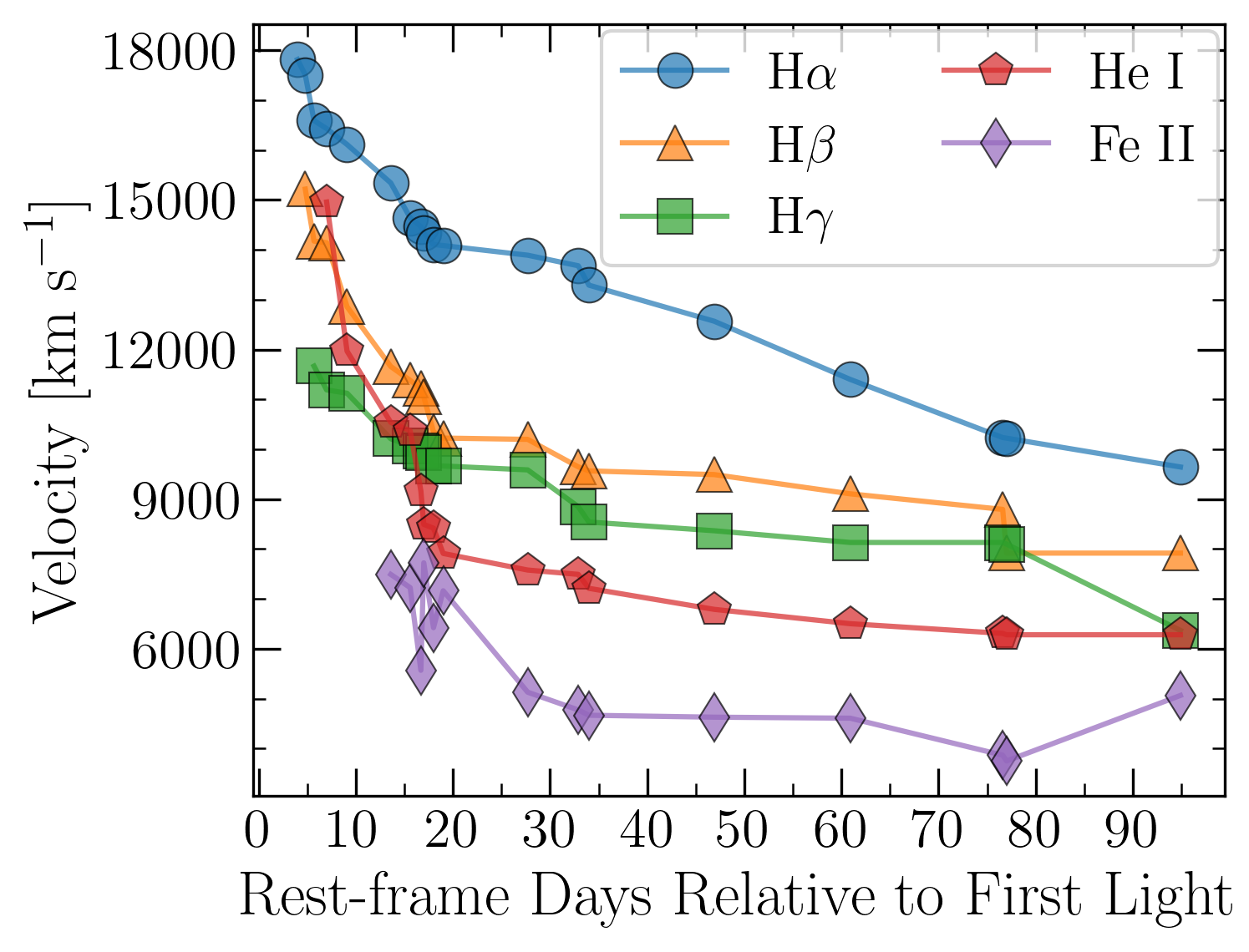}
    \caption{Velocities of P-Cygni profiles (H$\alpha$: 6563 Å, H$\beta$: 4861 Å, H$\gamma$: 4341 Å, \ion{He}{i}: 5876 Å, \ion{Fe}{ii}: 5169 Å) in SN 2024iss optical spectra ranging from $\delta t = 4.0-94.9$ days.}
    \label{fig:phase_vel}
\end{figure}

\subsection{Ultraviolet} \label{subsec:uv spectrum}

The $\delta t = 7$ days UV spectrum contains blended features from several iron-group elements, causing ultraviolet flux suppression from line blanketing. This also makes it difficult to identify features from individual ions, since many of their signatures blend together \citep{pun_ultraviolet_1995, dessart_using_2008, gezari_probing_2008, dessart_supernova_2010, bostroem_sn_2023}. We have identified the features shown in Figure \ref{fig:UVspec} from similar features in previous early-time SNe II UV spectra \citep{ben-ami_ultraviolet_2015, vasylyev_early-time_2022, vasylyev_early-time_2023, bostroem_sn_2023} and comparison to known prominent spectral lines. The SN 2024iss spectrum is mostly dominated by iron absorption from \ion{Fe}{ii} and \ion{Fe}{iii}. Additional iron-group metals with possible broad features include \ion{Mg}{ii} ($\sim2800$~\AA), \ion{Al}{iii} ($\sim1820$~\AA), \ion{Ni}{iii} ($\sim1770$~\AA), and \ion{Ti}{iii} ($\sim2500$~\AA). Narrow absorption features from the host galaxy interstellar medium can also be seen for \ion{Fe}{ii} (2344~\AA, 2383~\AA) and \ion{Mg}{ii} (2796~\AA). We compared the $\delta t = 7$ days UV spectrum to other early-time SNe II UV spectra at similar phases (SN 2022acko $\delta t = 7$ days: \citealt{bostroem_sn_2023}; SN 2023ixf $\delta t = 8$ days: \citealt{zimmerman_complex_2024}; SN 1993J $\delta t = 9$ days: \citealt{1993A&A...280L..15D, 1993AAS...182.5503S}; SN 2021yja $\delta t = 9$ days: \citealt{vasylyev_early-time_2022}; SN 2022wsp $\delta t = 10$ days: \citealt{vasylyev_early-time_2023}; SN 2013df $\delta t = 13$ days: \citealt{ben-ami_ultraviolet_2015}; see Figure \ref{fig:UVspec}). We found that the SN 2024iss UV spectrum shows less prominent features than the other UV spectra. The SNe IIb (2024iss, 1993J, and 2013df) all lack distinctive absorption/emission features that are seen in other SNe II in the 2000 – 2500~\AA~ range.

We measured the velocities of P-Cygni profiles present in the $\delta t = 7$ days \textit{HST} spectrum and compared the velocities from the UV wavelengths to those in the optical. We calculated the velocities using the minimum of the P-Cygni profile as seen in Figure \ref{fig:line velocities}, except for the velocity of the \ion{Mg}{ii} emission line. Since the \ion{Mg}{ii} profile is surrounded by other blended absorption features, we used the width of a fitted gaussian profile to measure the velocity. This resulted in an \ion{Mg}{ii} velocity of $\sim12{,}000\ \kms$. From absorption features identified in the UV spectrum, we calculated velocities that may correspond to P-Cygni profiles of \ion{Ti}{iii} ($\sim6{,}000\ \kms$), \ion{Al}{iii} ($\sim6{,}000 \ \kms$), and \ion{Ni}{iii} ($\sim7{,}000\ \kms$) as seen in Figure \ref{fig:line velocities}. The UV velocities are consistent with H$\alpha$, H$\beta$, and \ion{He}{i} velocties (identified from the optical spectrum) of $\sim16{,}000$, $\sim14{,}000$, and $\sim12{,}000\  \kms$ respectively. However, due to the many overlapping transitions in the UV, the identifications of UV P-Cygni profiles are not definite.

\subsection{Comparison to Early-Time Binary Progenitor Models} \label{subsec:binary progenitor}

The hydrogen-rich envelope present at the time of explosion is a key differentiator between various subtypes of SNe II, and can be defined by the orbital period of the binary system \citep{Ercolino24, dessart_sequence_2024}. We compared spectral models developed by \cite{dessart_sequence_2024}, in which the mass of the hydrogen-rich envelope ($M_{\textrm{env}}$) present at the time of explosion is a direct function of the orbital period of the progenitor’s binary system, to observed UV and optical spectra. The models assume a progenitor mass of $12.6 \ \Msun$, a companion mass of $11.97 \ \Msun$, a He core mass of $3.8 \ \Msun$, and a $^{56}\mathrm{Ni}$ mass of $0.09 \ \Msun$. More model specifics can be found in \cite{dessart_sequence_2024} and \cite{Ercolino24}, and the full grid of 10 models with corresponding parameters are listed in Tables 1 and 2 of \cite{dessart_sequence_2024}. A shorter orbital period corresponds to more interaction between the progenitor star and its companion, stripping the progenitor's hydrogen envelope and leaving less $M_{\textrm{env}}$ at the time of explosion. For orbital periods ranging from 562 to 2818 days (corresponding to a $M_{\textrm{env}}$ range of $0.00 - 6.86 \ \Msun$), early-time model spectra were calculated using radiation hydrodynamics and non-local thermodynamic equilibrium time-dependent radiative transfer. We note that these models assume solar metallicity, which differs from the likely sub-solar ($Z < 0.6~\rm Z_{\odot}$) metallicity of SN~2024iss given its host environment. However, given the numerous metal lines in the UV and the high optical depths, significant effects in the model spectra would only be observed at extremely low metallicities of $<0.1~\rm Z_{\odot}$.

We compared model spectra to the full \textit{HST} $\delta t = 7$ days UV to NIR spectrum and a range of early-time optical spectra (see Figure \ref{fig:binary_models}). The best-matched UV to NIR spectral model was the 2p85 model, corresponding to $M_{\textrm{env}} = 0.19 \ \Msun$, $M_{\textrm{ej}} = 2.43 \ \Msun$, and $R_{\star}= 619.8 \ \Rsun$. We note that the 2p90 model, which has similar physical properties, is also a closely matched model to the data but at a later model phase (Fig. \ref{fig:binary_models}). The 2p85 model, while being the best-matched, is still highly discrepant in the UV wavelengths. The deviation in the UV is likely the result of the model $R_{\star}$ being too large, leading to the model spectrum being overly luminous compared to SN~2024iss. Furthermore, there is no additional contribution from CSM interaction included in the model spectra and the effects of line-blanketing could be treated with more accuracy. The best-matched model for the optical spectra was the 2p90 model, corresponding to $M_{\textrm{env}}=0.28 \ \Msun$, $M_{\textrm{ej}} = 2.49 \ \Msun$, and $R_{\star}=710.2 \ \Rsun$. In the right panel of Figure \ref{fig:bolometric}, we compare the bolometric light curve of SN 2024iss to the model bolometric light curves of 2p80 ($M_{\textrm{env}}=0.12\ \Msun, M_{\textrm{ej}}=2.25 \ \Msun, R_{\star}=360.7\ \Rsun$), 2p85, 2p90, and 2p95 ($M_{\textrm{env}} = 0.31 \ \Msun$, $M_{\textrm{ej}} = 2.56 \ \Msun$, $R_{\star}=740.9\ \Rsun$). The models do not appear to match the rise and peak phases for the observed $^{56}$Ni-powered curve in early times. This could be a result of the progenitor radius being too large in the model as well as the uncertain effects of $^{56}$Ni mixing/clumping in the outer ejecta \citep{Dessart18}. After $\gtrsim50$ days, the binary progenitor model light curves match the observed bolometric light curve and the nebular light curve model described in Section \ref{subsec:ni56}.

\begin{figure}[h!]
\centering
\includegraphics[width=\columnwidth]{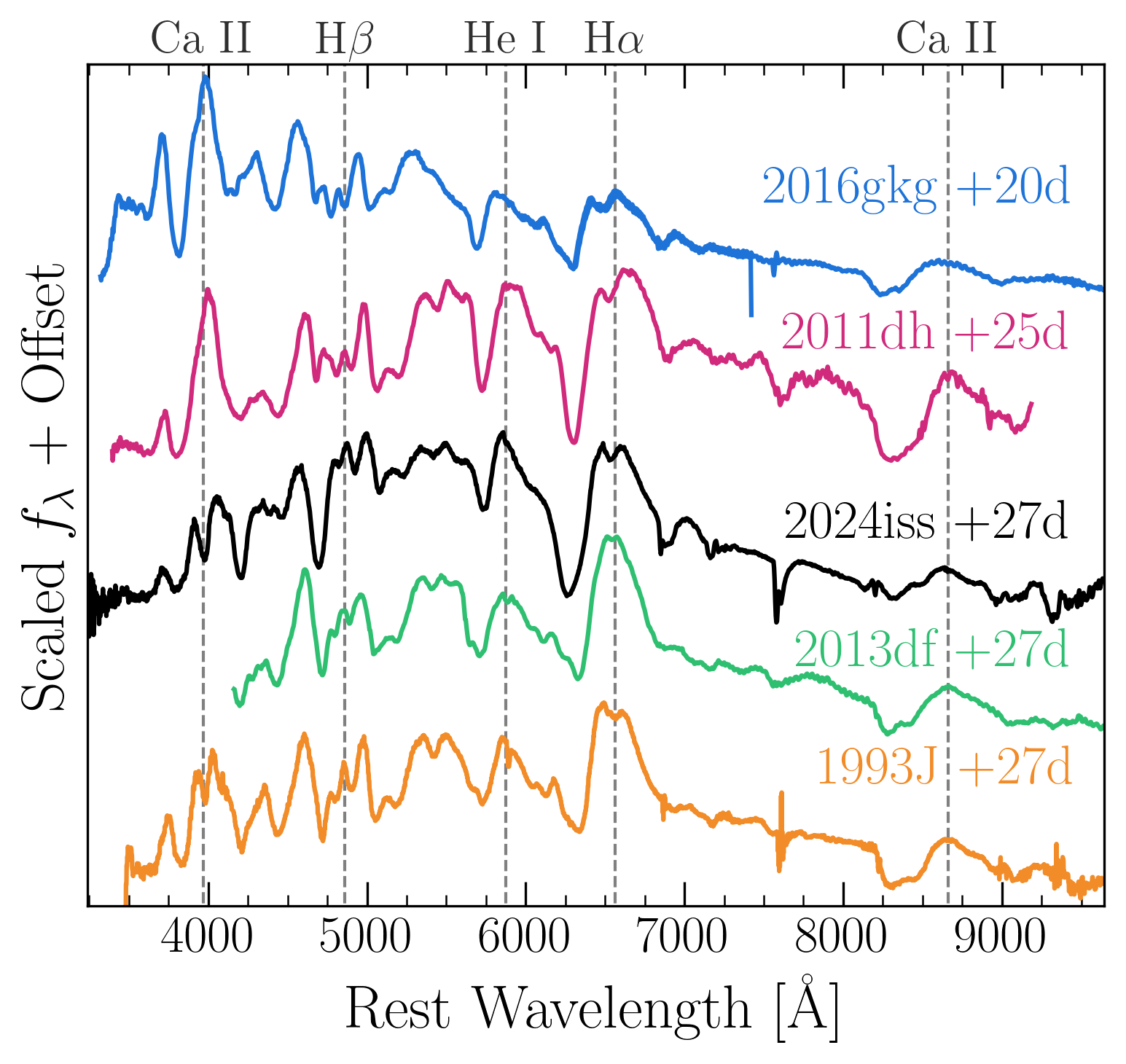}
\caption{Optical spectrum of SN 2024iss at $\delta t = 27$ days compared with optical spectra of IIb SNe 2016gkg, 2011dh, 2013df, and 1993J at similar phases (spectra obtained from WISeREP; \citealt{2012PASP..124..668Y}). All spectra are corrected for redshift and extinction.}
\label{fig:IIb_specs}
\end{figure}

\begin{figure*}[ht]
\centering
\subfigure{\includegraphics[width=0.62\textwidth]{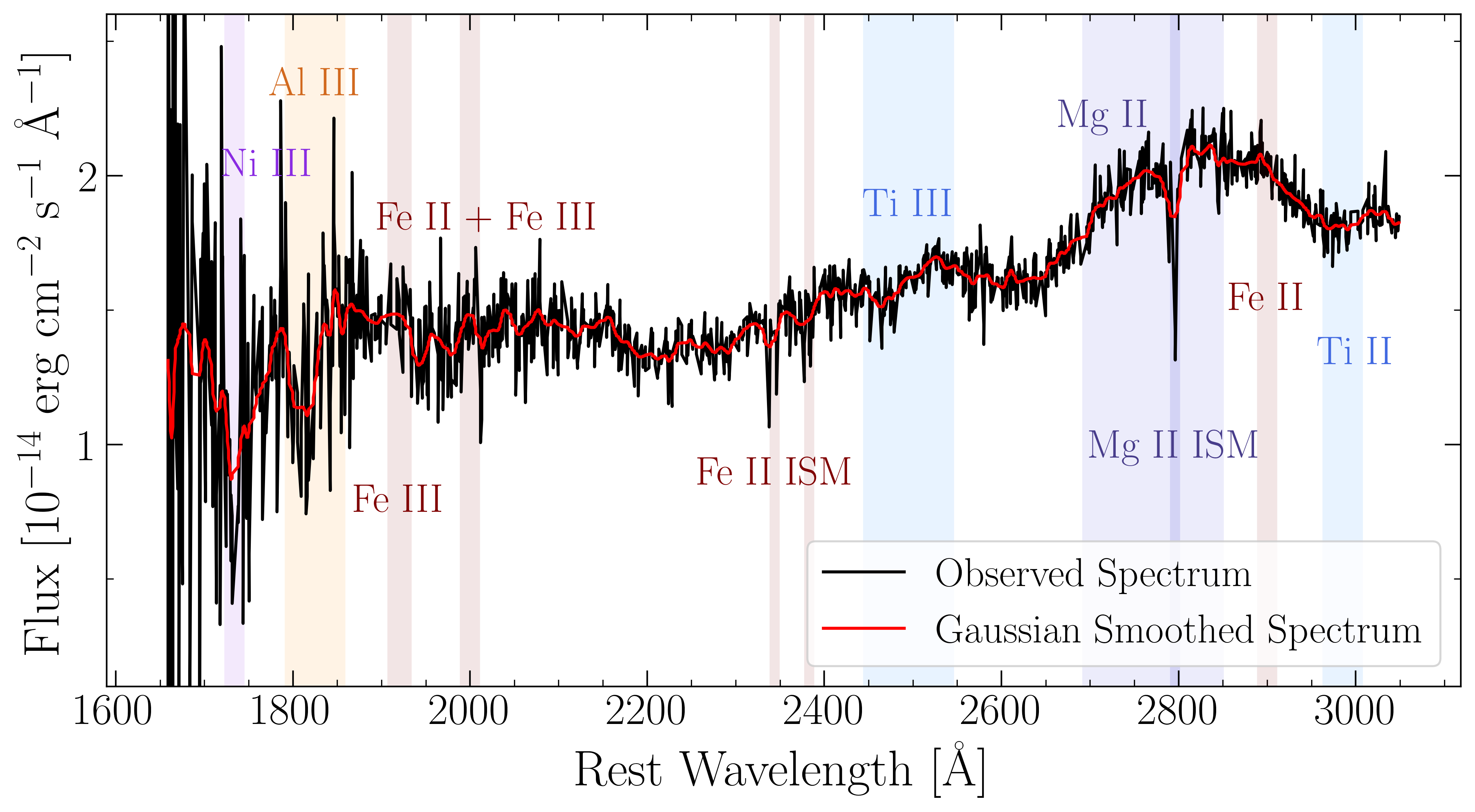}}
\subfigure{\includegraphics[width=0.36\textwidth]{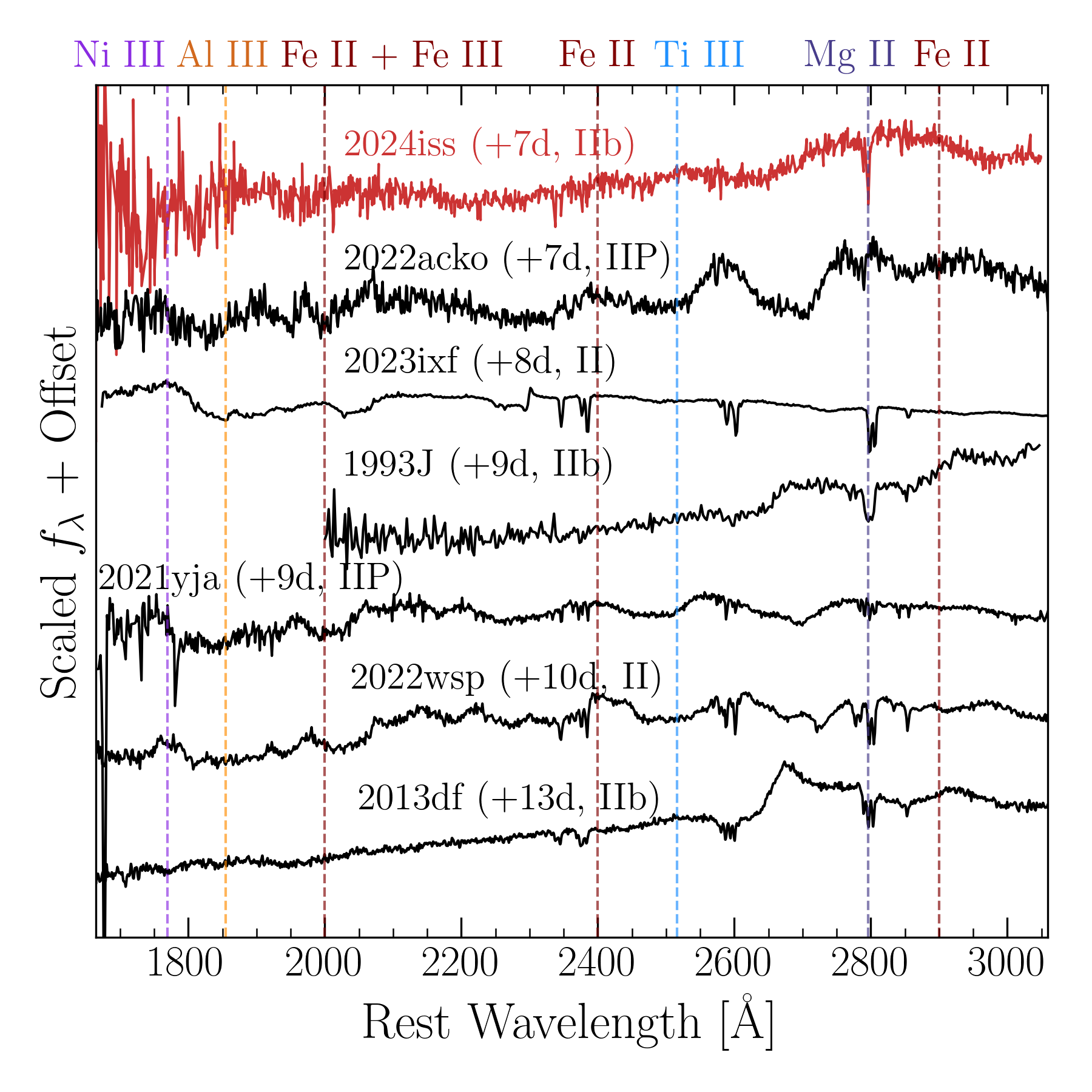}}
\caption{ {\it Left:} The $\delta t = 7$ days \textit{HST} UV spectrum of SN 2024iss, with prominent features from both the supernova and the host galaxy interstellar medium (ISM) highlighted. The spectrum is corrected for redshift and extinction. {\it Right:} The $\delta t = 7$ days \textit{HST} UV spectrum of SN 2024iss compared with the $\delta t = 7$ days spectrum of SN 2022acko \citep{bostroem_sn_2023}, $\delta t = 8$ days spectrum of SN 2023ixf \citep{zimmerman_complex_2024}, $\delta t = 9$ days spectrum of SN 1993J \citep{1993A&A...280L..15D, 1993AAS...182.5503S}, $\delta t = 9$ days spectrum of SN 2021yja \citep{vasylyev_early-time_2022}, $\delta t = 10$ days spectrum of SN 2022wsp \citep{vasylyev_early-time_2023}, and $\delta t = 13$ days spectrum of SN 2013df \citep{ben-ami_ultraviolet_2015}. All spectra are corrected for redshift and extinction. \label{fig:UVspec} }
\end{figure*}

\subsection{Nebular Phase Spectra} \label{subsec:nebularmodels}

We present the late-time optical ($\delta t = 259$~days) and NIR ($\delta t = 290$~days) spectra of SN~2024iss in Figure \ref{fig:nebular}. In the optical, we identify prominent emission lines of [\ion{Ca}{ii}], [\ion{O}{i}], \ion{N}{ii}, \ion{Mg}{i}] and \ion{Fe}{ii}, all consistent with expected emission from SNe~IIb once the ejecta has become optically thin (i.e., ``nebular''). In the NIR, we observe strong \ion{He}{i} $\lambda$1.083$\mu$m emission as well as emission lines from intermediate-mass elements such as \ion{Na}{i} and \ion{Mg}{i} in addition to iron-group elements such as [\ion{Co}{ii}], [\ion{Fe}{i/ii}] and [\ion{Ni}{ii}]. Given the lack of H emission at late-time phases, we compare the SN~2024iss nebular spectra to a grid of He-star models by \cite{Dessart21} computed at similar late-time phases. As shown in Figure \ref{fig:nebular}, the he6p0 ($M_{\rm ej} = 2.82~\Msun$) and he7p0 ($M_{\rm ej} = 3.33~\Msun$) model spectra provide a reasonable match to most emission line profiles at optical and NIR wavelengths. In the optical, we note that the model spectra over-predict the flux from Mg and Fe between 4000-5500~\AA\ and under-predicts the \ion{He}{i} $\lambda$1.083$\mu$m emission strength; however, the former can be remedied by additional clumping of the inner ejecta as shown in \cite{Dessart21}. In the NIR, the model under-predicts the line strengths of \ion{Mg}{i} $\lambda$1.50~$\mu$m and [\ion{Fe}{ii}] $\lambda$1.64~$\mu$m, but over-predicts the strength of [\ion{Ni}{ii}] $\lambda$1.94~$\mu$m.


In Figure \ref{fig:OIvels}, we present forbidden [\ion{O}{i}] $\lambda$6300 and semi-forbidden \ion{Mg}{i}] line velocities from low and medium resolution spectroscopy of SN~2024iss at $\delta t = 259\ \& \ 412$~days. Overall, both profiles show significant structure e.g., the [\ion{O}{i}] profile is double-peaked, with the bulk of the emission being offset to redder wavelengths by $\sim$580~km~s$^{-1}$ and can be well described by a Gaussian profile with a FHWM of $\sim$2600~km~s$^{-1}$. However, redward emission extends out to $\sim$6000~km~s$^{-1}$ and is not a monotonic function. Blueward of line center, the [\ion{O}{i}] profile has a narrow peak centered at 660~km~s$^{-1}$, with FWHM of 1100~km~s$^{-1}$. Comparison to the \ion{Mg}{i}] profile shows an identical line profile structure, but the redward emission does not extend to as high of velocities as in [\ion{O}{i}] ($\sim$4000-6000~km~s$^{-1}$). This line profile structure is suggestive of an aspherical ejecta distribution and/or ejecta clumping, as shown by \cite{Dessart21} (see their section 7).


\section{Circumstellar Environment \& Progenitor Mass-loss History}\label{sec:csm}

The detection of luminous, fast-fading X-ray emission in SN~2024iss indicates SN ejecta interaction with confined CSM. In order to quantify the CSM densities that the shock samples at $\delta t < 7$~days, we use the normalization of the X-ray spectrum to calculate the total emission measure (EM) at each epoch. Emission measure is defined as $EM = \int n_e n_I dV$, where $n_e$ and $n_I$ are the electron and ion number densities in the emitting volume $V$. We follow the procedure outlined in \cite{Brethauer22} for deriving the unshocked CSM density, which is described by the following relation: 

\begin{equation}
	\rho_{\rm CSM}(r) = \frac{m_p}{4} \left(\frac{2 \times \rm EM(r)\mu_e \mu_I}{V_{\rm FS}(r)}\right)^{1/2}
\end{equation}

\noindent
where $\mu_e$ and $\mu_I$ are the mean molecular weight of electron and ion, respectively; $m_p$ is the proton mass; $ \mathrm{V_{FS}} = \frac{4\pi}{3}f \left(R_{\rm out}^3 - R_{\rm in}^3\right)$ is the emitting forward shock (FS) volume, with $R_{\rm in}$ and $R_{\rm out}$ are the inner and outer radius of the shell respectively and $f$ is the filling factor. We assume H-rich CSM with sub-solar composition ($\mu_e=1.15$, $\mu_I = 1.24$), spherical CSM geometry ($f=1$), and outer CSM radius of $R_{\rm out} = 1.2~ R_{\rm in}$ \citep{1982ApJ...258..790C}, where $R_{\rm in} = v_{\rm sh} t_{\rm SN}$. Based on the H$\alpha$ velocities observed in early-time spectra, we adopt a range of shock velocities of $v_{\rm sh} = (1-3) \times 10^4~\kms$ when deriving CSM density estimates shown with respect to shock radius in Figure \ref{fig:CSM}. Using these CSM densities, we make a rough estimate of the radiated luminosity from CSM interaction (i.e., $L = 2 \pi r² \rho v_{\rm sh}^3 \epsilon$) in Figure \ref{fig:bolometric}, assuming a radiative efficiency of $\epsilon = 30$\% \citep{Khatami24}. Compared to the bolometric light curve of SN~2024iss, this shock powered emission could account for $\sim 10~\%$ of the luminosity budget in SN~2024iss at the phase of the primary light curve peak and $\sim 50\%$ during the minimum between both light curve peaks. This additional emission mechanism from CSM interaction at early-times likely influences the accuracy of physical parameters such as envelope mass and radius that are derived from shock cooling model fits (e.g., Fig. \ref{fig:shock cooling}). 

For $v_{\rm sh} = 10^4~\kms$, the CSM density profile can be fit with a steady-state ($\rho \propto r^{-2}$) mass-loss rate of $\dot M = 5 \times 10^{-4}~$\mdot, for a wind velocity of $v_w = 100~\kms$. This value is chosen based on the velocities of unbound material found in 3D hydrodynamical simulations of SN IIb progenitor stars (e.g., \citealt{2025arXiv250812486G}) and the likely compact nature of the SN~2024iss progenitor star based on the envelope radius and mass derived from shock cooling and the ejecta mass derived from bolometric light curve modeling. Nonetheless, it should be noted that much of the CSM in the 3-D simulations by \cite{2025arXiv250812486G} remains bound and is not a steady-state wind-like profile. As shown in Figure \ref{fig:CSM}, we compare CSM densities estimated from X-ray spectral modeling to the the pre-explosion density profile predicted by 3D simulations. While the simulations from \cite{2025arXiv250812486G} do show elevated amounts of bound and unbound stellar material created by the star directly before collapse, the envelope densities derived from the range of shock cooling emission model parameters are higher than the average densities in the spherically-averaged 3D progenitor model snapshots for the YSG1L4.7 model. Furthermore, this simulation only traces CSM to $<10^{14}$~cm, which is not probed by the X-ray observations of SN~2024iss given that they occur at $\delta t > 1$~day. However, as shown in Figure \ref{fig:CSM}, the pre-explosion density profile of the larger 3-D progenitor model, YSG2L5.1, is consistent with the densities inferred from shock cooling emission and X-ray spectra modeling.



Also using the EM derived from modeling of the X-ray spectrum, \cite{chen_sn_2025} find a lower mass loss rate of $\dot M = 6 \times 10^{-5}~$\mdot, but they assume a wind velocity of $v_w = 10~\kms$. Intriguingly, the deviation from a $t^{-1}$ decline rate in the X-ray light curve at $\delta t > 7$~days indicates that these higher CSM densities, and enhanced $\dot M$, do not extend beyond $R_{\rm sh} \approx (0.6 - 2) \times 10^{15}$~cm for $v_{\rm sh} = (1-3) \times 10^4~\kms$. Assuming a progenitor wind velocity of $v_w = 100~\kms$, this corresponds to a phase of enhanced pre-explosion mass loss in the final $\sim 2-5$~years before core-collapse. However, steady-state mass loss may not be an appropriate descriptor of the circumstellar environment of SN~2024iss i.e., there could be CSM velocity gradients and/or not all the material is unbound. Overall, the CSM densities in SN~2024iss are too small for electron-scattering (``IIn-like'') emission line profiles observed in the early-time spectra of SN~IIb 2013cu \citep{gal-yam_wolf-rayet-like_2014, groh14, Grafener16, 2017A&A...605A..83D} and some SNe~IIP/L (e.g., \citealt{Yaron17, wjg23, jacobson-galan_final_2024}).

\begin{figure}[h!]
\centering
\subfigure{\includegraphics[width=0.52\columnwidth]{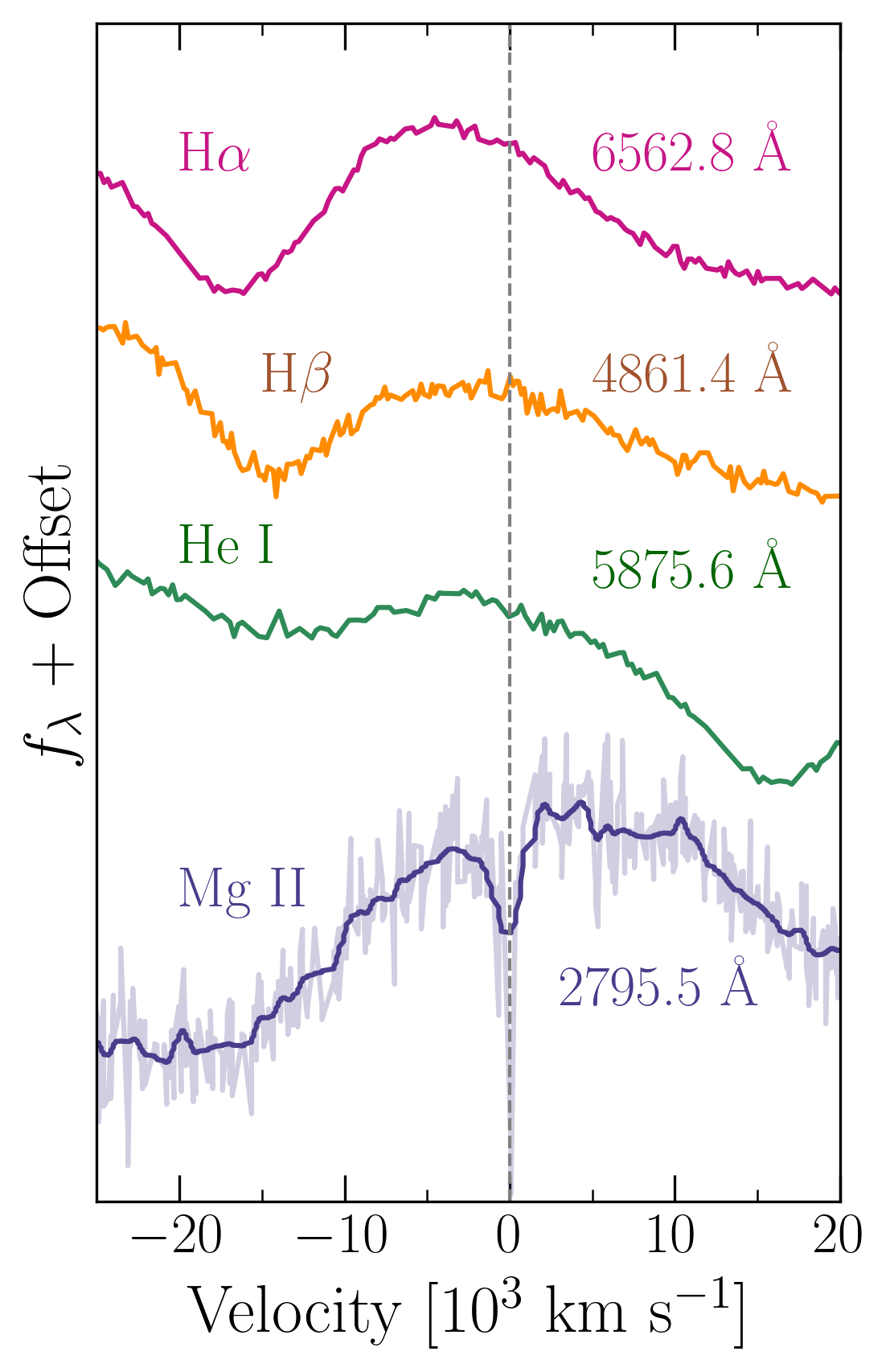}}
\subfigure{\includegraphics[width=0.45\columnwidth]{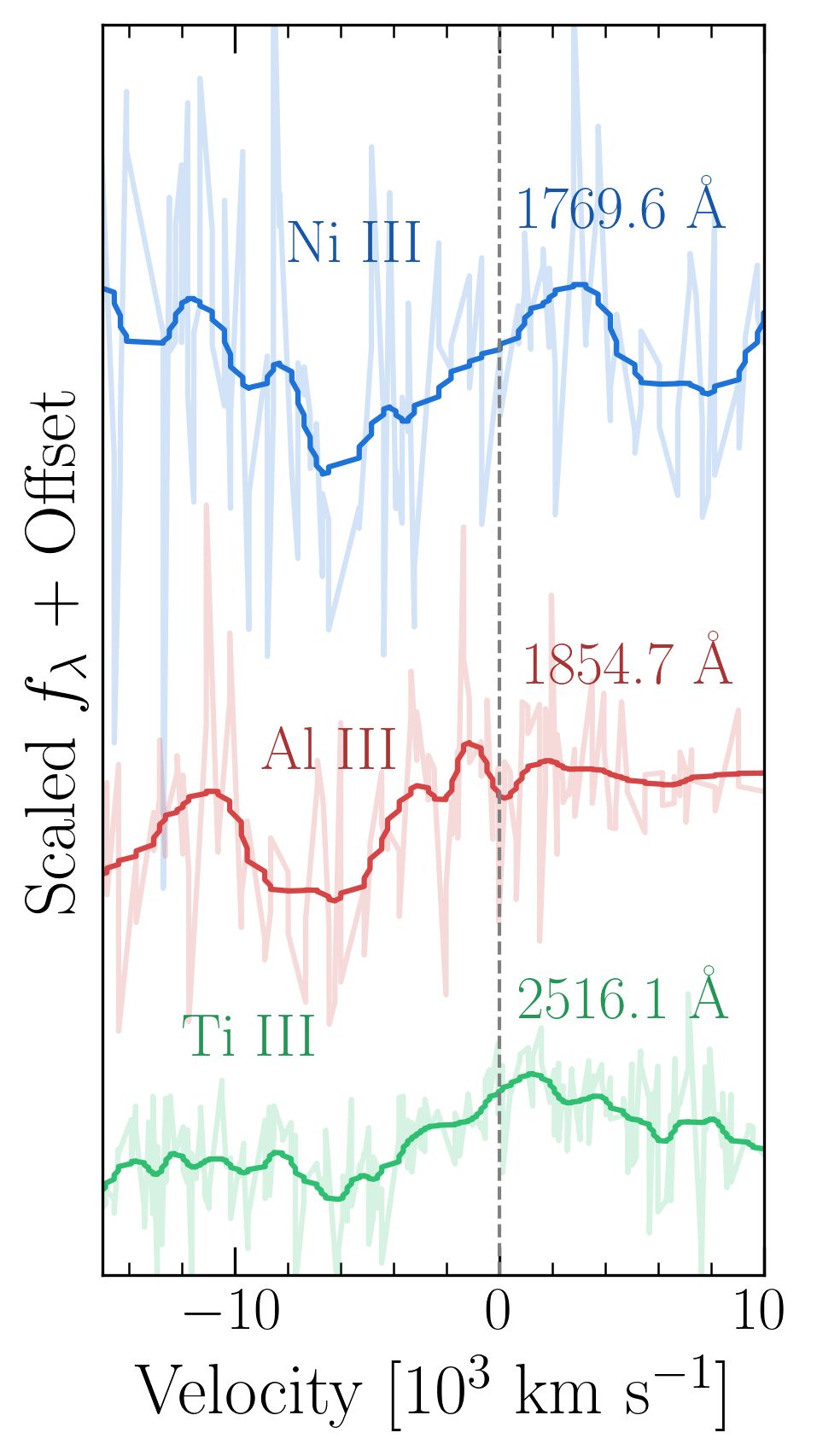}}
\caption{ {\it Left:} Velocities of identified P-Cygni profiles in the $\delta t = 7$ days \textit{HST} spectrum. {\it Right:} Velocities of possible P-Cygni profiles corresponding to ion signatures in the $\delta t = 7$ days \textit{HST} UV spectrum. All spectral features are shown after correcting for redshift and extinction.
\label{fig:line velocities} }
\end{figure}

\begin{figure*}
\centering
\subfigure{\includegraphics[width=0.49\textwidth]{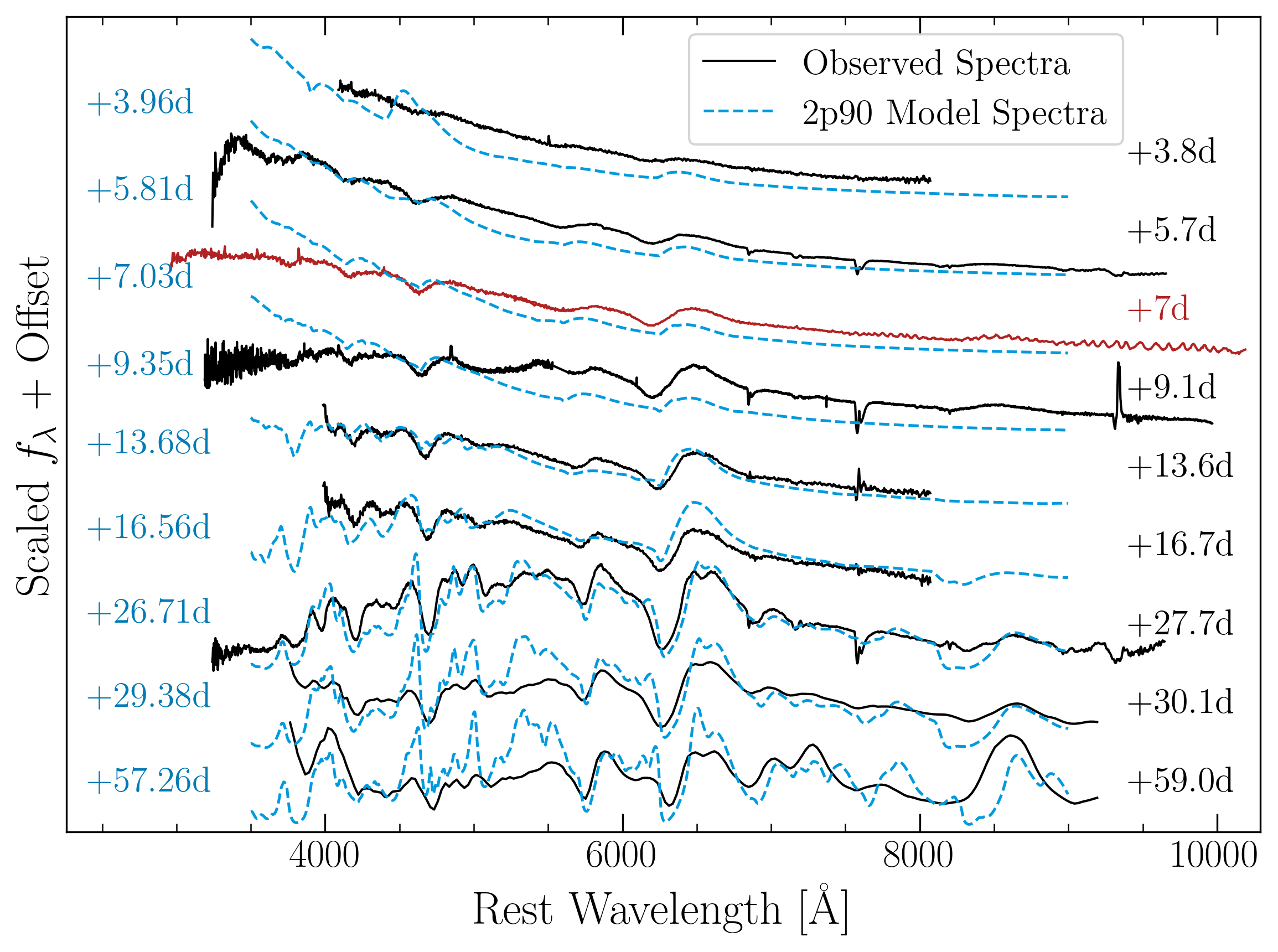}}
\subfigure{\includegraphics[width=0.495\textwidth]{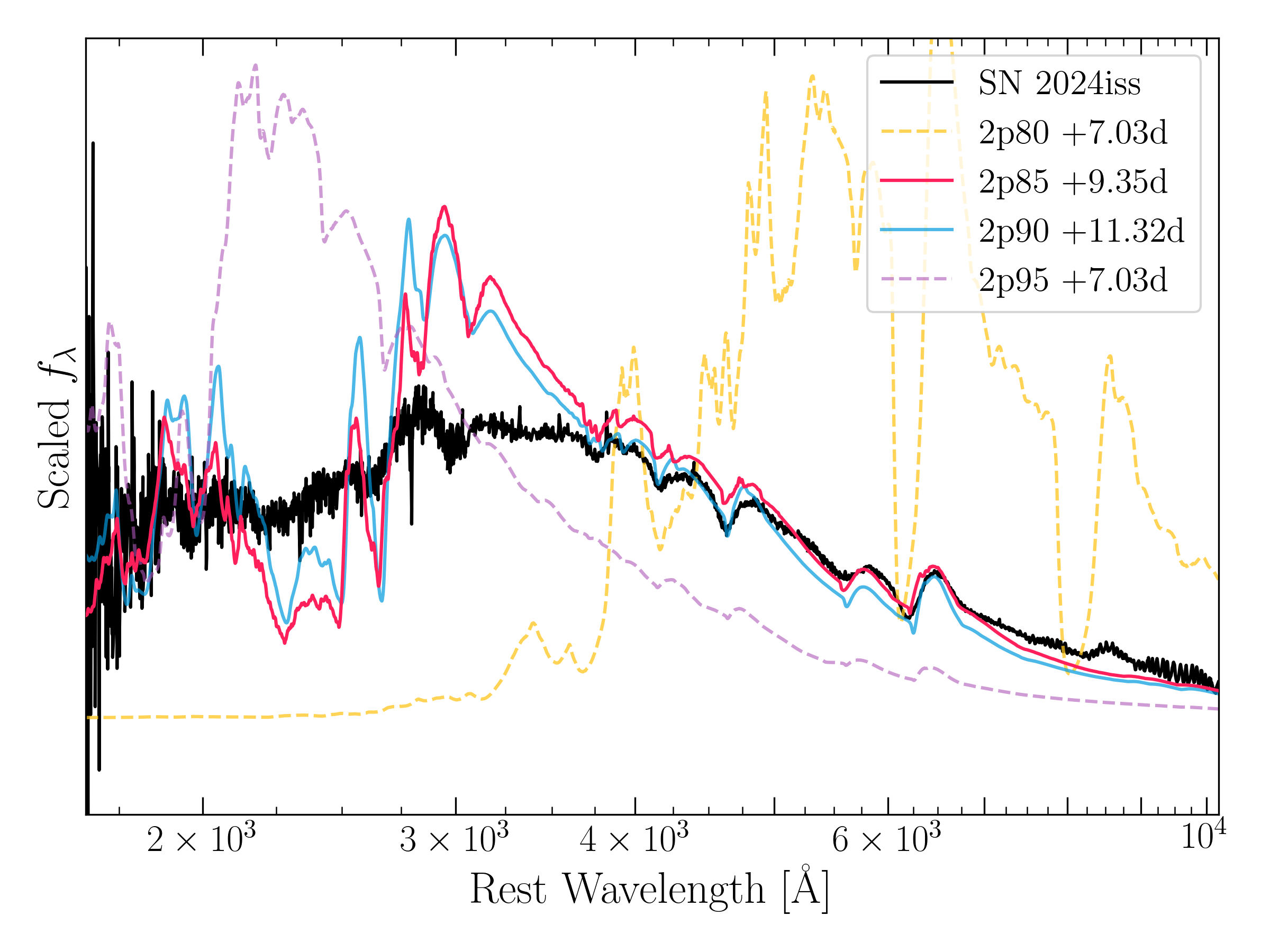}}
\caption{ {\it Left:} Optical spectra of SN 2024iss compared with the 2p90 ($M_{\textrm{env}} = 0.28 \ \Msun$) optical model spectra at similar phases from \citet{dessart_sequence_2024}. The $\delta t = 7$ days \textit{HST} spectrum is shown in red. {\it Right:} The full \textit{HST} UV to NIR $\delta t = 7$ days spectrum of SN 2024iss compared to a range of model spectra (2p80: $M_{\textrm{env}} = 0.12 \ \Msun$, 2p85: $M_{\textrm{env}} = 0.19 \ \Msun$, 2p90: $M_{\textrm{env}} = 0.28 \ \Msun$, 2p95: $M_{\textrm{env}} = 0.31 \ \Msun$) at similar phases from \citet{dessart_sequence_2024}. \label{fig:binary_models} }
\end{figure*}


\section{Discussion} 
\label{sec:discussion}

\subsection{Early Photometry}


The rich, multiwavelength dataset for SN~2024iss probes the explosions’s underlying physical properties across a broad range of physical scales and processes, from the outer envelope and CSM to the explosion dynamics and bulk ejecta properties. With an initial ZTF $i$-band detection at 1.3 hours after a $g$-band non-detection, we obtained one of the most well-constrained times of first light for an SN IIb. By constraining the shock breakout regime to $\delta t<0.42$ days, we find that the $i$-band point deviates from power-law models, making it consistent with shock breakout emission. Ultraviolet, optical, and near-infrared photometry displays the characteristic SN IIb double-peaked light curve. The shock cooling emission peak reaches a maximum at $\delta t = 2.24$ days, which falls within the average shock cooling peak rise time range of $2.07 \pm 1.0$ days for SNe IIb \citep{2025ApJ...989..192C}. The early detection and well-sampled shock cooling emission peak of SN 2024iss demonstrates the need for early-time, high-cadence photometry to better explore shock breakout emission in SNe.

\subsection{CSM Properties}

The shock cooling emission peak coincides with exceptionally luminous X-ray emission up to $\delta t = 7$ days, likely resulting from interaction between the SN shock and dense, confined CSM. With observed H$\alpha$ velocities, we consider a range of shock velocities from $(1-3) \times 10^4\ \kms$. Assuming a brief episode of constant, steady-state mass-loss, this corresponds to a high progenitor mass-loss rate of $5 \times 10^{-4} \ \Msun \ \textrm{yr}^{-1}$ with a wind velocity of $100 \ \kms$ in the $\sim 2-5$ years leading up to explosion. However, as noted previously, the mass-loss in the SN~2024iss progenitor may not have been ``steady-state'' given the known complexities of the local environment of SN~IIb progenitor stars \citep{2025arXiv250812486G}. Previous X-ray studies of SNe IIb 1993J, 2011dh, and 2013df reveal mass-loss rates of $\sim 10^{-5}$\mdot \citep{2001ApJ...561L.107I, 2012ApJ...752...78S, 2016ApJ...818..111K}, placing the mass-loss rate of SN 2024iss at an order of magnitude higher than other SNe IIb. SN II 2023ixf exhibited emission consistent with a comparable mass-loss rate ($10^{-4}$\mdot) to SN 2024iss, but with a lower assumed wind velocity ($20 \pm 5 \ \kms$) \citep{Grefenstette23, 2024ApJ...963L...4C, Nayana25, 2025ApJ...994L..14J}. In SN 2024iss, the high circumstellar density paired with a high wind velocity likely led to significant post-explosion CSM interaction. This resulted in excess emission that blurs the transition between the shock cooling emission peak and the $^{56}$Ni powered peak in the light curve (as seen in Figure \ref{fig:lc comparison}). Modeling of X-ray observations shows that $\sim50\%$ of the bolometric luminosity during this transitional phase can be attributed to emission from CSM interaction (see Figure \ref{fig:bolometric}). 

\subsection{Shock Cooling Emission \& Progenitor Star Properties}

The additional CSM interaction impacts the ability of models to explain the observations of SN 2024iss. Shock cooling models calibrated on supergiant progenitor characteristics (SW17 and MSW23) suggested a shock cooling emission curve lasting $\delta t = 7$ days, while the P21 analytical model suggested a curve lasting $\delta t = 5$ days. However, MSW23 was not calibrated for SNe IIb, a likely explanation for the inconsistent fits with UV photometry. Another likely culprit for the UV discrepancy is the shock power emergence from CSM interaction -- this UV excess has been seen in other SNe~IIb (e.g., 1993J, 2013df) and in UV spectra at later phases \citep{ben-ami_ultraviolet_2015}.

This study marks the second time MSW23 has been tested on an SN IIb, the first being on SN 2022hnt \citep{farah_shock-cooling_2025}. As seen with our results and in \citet{farah_shock-cooling_2025}, despite MSW23 not being specifically calibrated for IIb SNe, it produces similar results to those of SW17 n=1.5. The progenitor radii from these two models, which are based on similar red supergiant progenitor characteristics (e.g., adiabatic structure of a fully efficient convective envelope), are in agreement with each other, giving an $R_{\star}$ range of $100 - 110 \ \Rsun$. The SW17 n=3 model is based on blue supergiant progenitor characteristics (i.e., purely radiative envelope) and resulted in $R_{\star} = 174.93^{+0.75}_{-0.74} \ \Rsun$. However, the corresponding $M_{\textrm{env}}=5.12^{+0.04}_{-0.03} \ \Msun$ is not realistic, since SNe IIb are expected to have stripped envelope masses of $\lesssim1 \ \Msun$ \citep{1994AJ....107.1022R, 2000AJ....120.1499M}. This indicates that the progenitor of SN 2024iss is unlikely to be a blue supergiant star, so we disregard the SW17 n=3 results in further discussion. The P21 model, being fully analytical, is not calibrated on any specific progenitor type. Its resulting $R_{\star} = 317.55^{+2.83}_{-2.90} \ \Rsun$ and $M_{\textrm{env}} = 7.45^{+0.05}_{-0.04} \times 10^{-2} \ \Msun$ may not be specifically calibrated to a progenitor star type, but are the statistically best fit parameters to SN 2024iss photometry. As a result, we consider the final $R_{\star}$ range to be $101 - 320 \ \Rsun$ and the final $M_{\textrm{env}}$ range to be $0.07 - 0.46 \ \Msun$. 

\begin{figure*}
\centering
\subfigure{\includegraphics[width=0.49\textwidth]{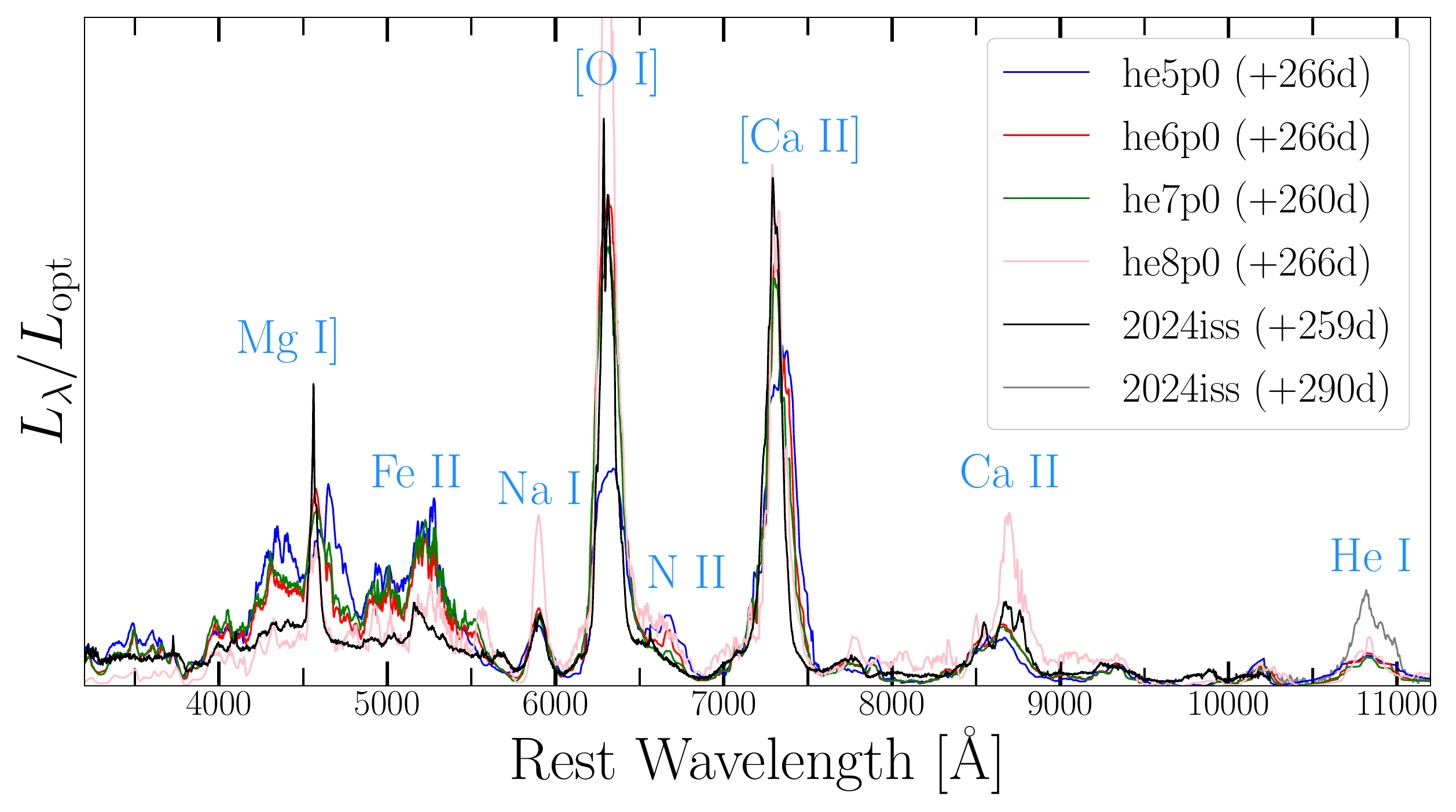}}
\subfigure{\includegraphics[width=0.49\textwidth]{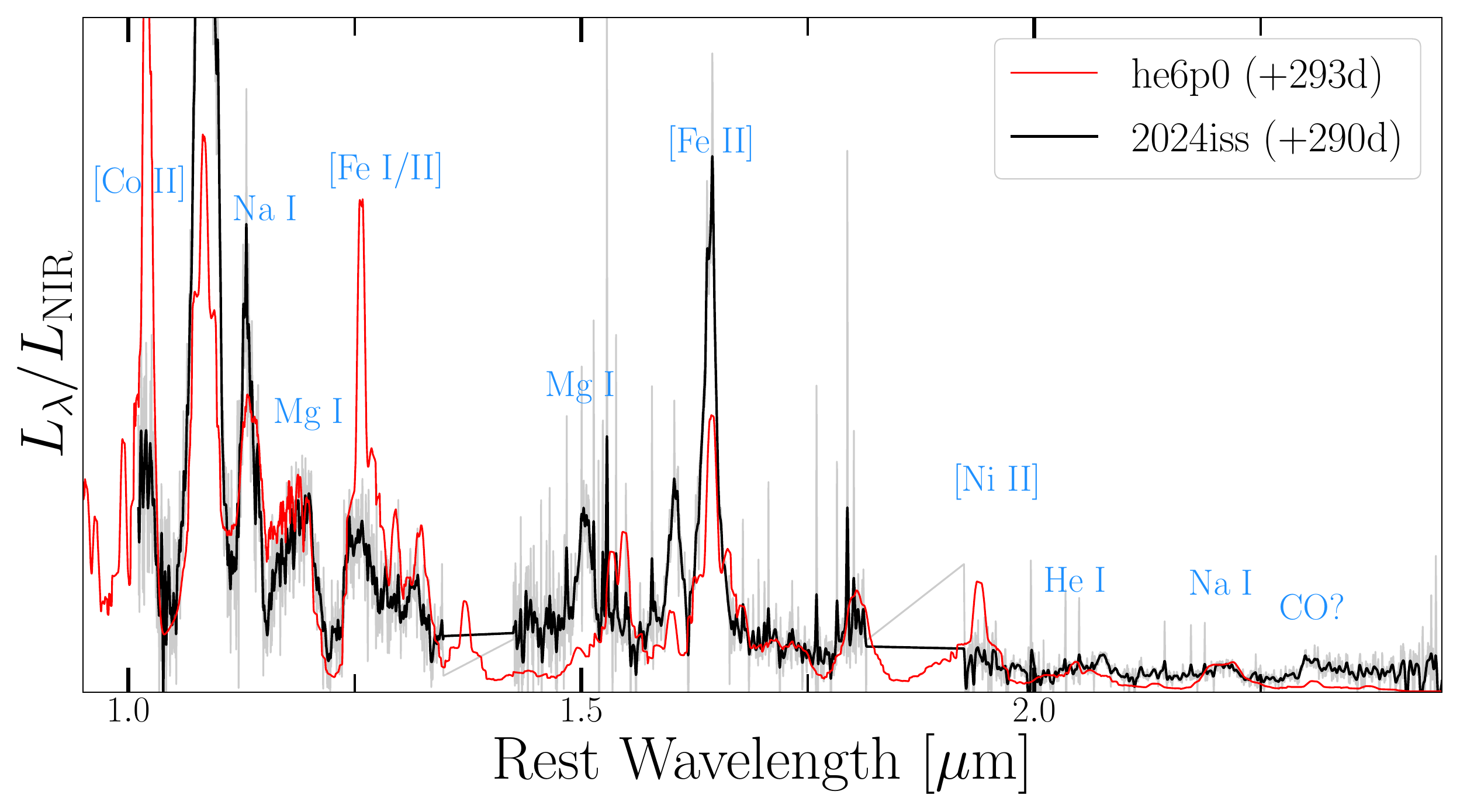}}
\caption{{\it Left:} Comparison of optical (black) and NIR (grey) spectra of SN~2024iss at nebular phases to He-star models ($M_{\rm preSN} = 3.8 - 5.6~\Msun$) from \cite{Dessart21}. Line identifications shown in blue. {\it Right:} Comparison of SN~2024iss NIR spectrum (black) to $M_{\rm preSN} = 4.4~\Msun$ He-star model (red) from \cite{Dessart25IR} \label{fig:nebular} }
\end{figure*}

\citet{yamanaka_sn_2025} derived an $R_{\star}$ range of $50 - 340\ \Rsun$ by comparing the photospheric temperature of SN 2024iss to analytic progenitor star models by \citet{milisavljevic_multi-wavelength_2013} and \citet{2010ApJ...725..904N}. We find a more constrained $R_{\star}$ range through shock cooling emission modeling that is in agreement with theirs. \citet{chen_sn_2025} obtained $R_{\star} = 244 \pm 43\ \Rsun$ by modeling the shock cooling emission to $\delta t < 5$ days with the SW17 n=1.5 model. While this fits within our overall progenitor radius range, we obtained $R_{\star}=106.31\pm0.50\ \Rsun$ with the SW17 n=1.5 model. One key difference is that we fit SW17 to $\delta t < 7$ days, which we found was a statistically better fit than to $\delta t < 5$ days (see Table \ref{tab:shock cooling phase fits}). Additionally, the offset from the time of first light ($t_{\textrm{offset}}$) is a free parameter in shock cooling emission models that impacts calculations of other best fit values. Due to our early ZTF detection, we have a much stronger constraint on this value, resulting in $t_{\textrm{offset}} <10^{-3}$ days, while the $t_{\textrm{offset}}$ from \citet{chen_sn_2025} is $0.20^{+0.04}_{-0.06}$ days. As explained in Appendix \ref{appendix shock cooling}, we also place a stricter constraint on $f_{p}M$ (a parameter describing the fraction of ejecta mass related to the progenitor's inner envelope structure) when using the \texttt{lightcurve-fitting} package. This allows for less uncertainty in estimated parameter values. Due to the tighter constraints and different phases for fitting, our hydrogen-rich envelope mass from SW17 n=1.5 ($M_{\textrm{env}}=0.457 \pm 0.003\ \Msun$) differs from that of \citet{chen_sn_2025} ($M_{\textrm{env}}=0.11 \pm 0.04\ \Msun$) as well.

Red supergiant progenitors are expected to have their hydrogen-rich envelopes mostly intact ($M_{\textrm{env}}\gtrsim 0.5\ \Msun$), but can evolve to yellow supergiants through mass-loss leading to a stripped hydrogen-rich envelope of $M_{\textrm{env}}\sim0.05 - 0.5 \ \Msun$ \citep{georgy_yellow_2012,2025arXiv250812486G}. $M_{\textrm{env}}$ for SN 2024iss is consistent with that of a yellow supergiant progenitor. As seen in Figure \ref{fig:parameter comparison}, the SN 2024iss parameters are most similar to those of SNe 2011dh, 2024abfo, and 2013df, whose yellow supergiant progenitors were confirmed through pre-explosion imaging \citep{bersten_type_2012, reguitti_sn_2025, van_dyk_type_2014, morales-garoffolo_sn_2014}. However, we caution the over-interpretation of this progenitor radius comparison given that this quantity is not derived through a consistent fitting procedure of the shock cooling peak across all of these works. 
\cite{chevalier_type_2010} introduced two sub-classes of SNe IIb: extended envelope (eIIb) and compact envelope (cIIb). SNe eIIb are estimated to have $R_{\star}\sim150\ \Rsun$ and $M_{\textrm{env}}>0.1 \Msun$, while SNe cIIb are estimated to have $R_{\star}\sim1.5\ \Rsun$ and $M_{\textrm{env}}<0.1 \ \Msun$. Additionally, X-ray emission from SNe cIIb is expected to be predominantly non-thermal, while SNe eIIb X-ray emission is expected to be thermal. The radius estimate for SN 2024iss is consistent with that of an eIIb, the envelope mass ranges from cIIb to eIIb, and the thermal X-ray emission is consistent with expectations for an eIIb. This suggests that SN 2024iss is likely an SN eIIb, though we can not definitely constrain $M_{\textrm{env}}$ to either sub-class.


\subsection{Progenitor Binarity}

The leading explanation for the stripped hydrogen envelope of SNe IIb is the interaction with a companion star in a binary system leading up to the progenitor's explosion \citep{1992ApJ...391..246P, sravan_progenitors_2019, Eldridge19, Ercolino24, dessart_sequence_2024}. In solar metallicity environments, the $R_{\star}$, $M_{\textrm{env}}$, and $M_{\textrm{ej}}$ of SN 2024iss are consistent with those of binary progenitor systems. The ejecta mass ($M_{\textrm{ej}}=1.1 - 3.3 \ \Msun$) is also inconsistent with that of a more massive single-star progenitor system ($M_{\textrm{He core, preSN}}\gtrsim9\ \Msun$) which would lose most of its hydrogen envelope naturally due to stronger winds \citep{2020ApJ...903...70S}. In low metallicity environments ($Z \approx 0.25 Z_{\odot}$), the $R_{\star}$, $M_{\textrm{env}}$, and $M_{\textrm{ej}}$ of SN 2024iss are all inconsistent with a single-star progenitor system \citep{2020ApJ...903...70S}. 
By comparing observed spectra of SN 2024iss with binary progenitor model spectra from \cite{dessart_sequence_2024}, we obtain a $M_{\textrm{env}}$ range of $0.19 - 0.28 \ \Msun$, which is consistent with the range derived from shock cooling emission modeling. However, as seen in the right panel of Figure \ref{fig:binary_models}, the model spectrum shows some inconsistencies with the observed spectrum in the UV. This is likely due to the additional emission from CSM interaction, which strongly impacts UV flux but is not included in the models. Additionally, the best-matched models suggest an $R_{\star}$ range of $619.8 - 710.2 \ \Rsun$. This does not agree with the $R_{\star}$ range from shock cooling emission modeling and likely leads to differences in the UV at early-time phases. As a result, we do not consider the values derived from these models to be definite constraints. \cite{dessart_sequence_2024} compared model spectra and light curves to observations of SN IIb 2011dh. For the yellow supergiant progenitor of SN 2011dh ($M_{\textrm{env}} = 0.1 \ \Msun, R_{\star} = 200 \ \Rsun$; \citealt{bersten_type_2012}), the best-matched model was 2p80 ($M_{\textrm{env}} = 0.12 \ \Msun, R_{\star}=360.7 \ \Rsun$). Like SN 2024iss, $M_{\textrm{env}}$ from the model is close to other estimates, but $R_{\star}$ appears overestimated. However, the models appear to match more closely to observed spectra of SN 2011dh than those of SN 2024iss, highlighting the impact of CSM interaction emission on observations of SN 2024iss.

\subsection{UV Spectroscopy \& Progenitor Environment}

\begin{figure*}
\centering
\subfigure{\includegraphics[width=0.49\textwidth]{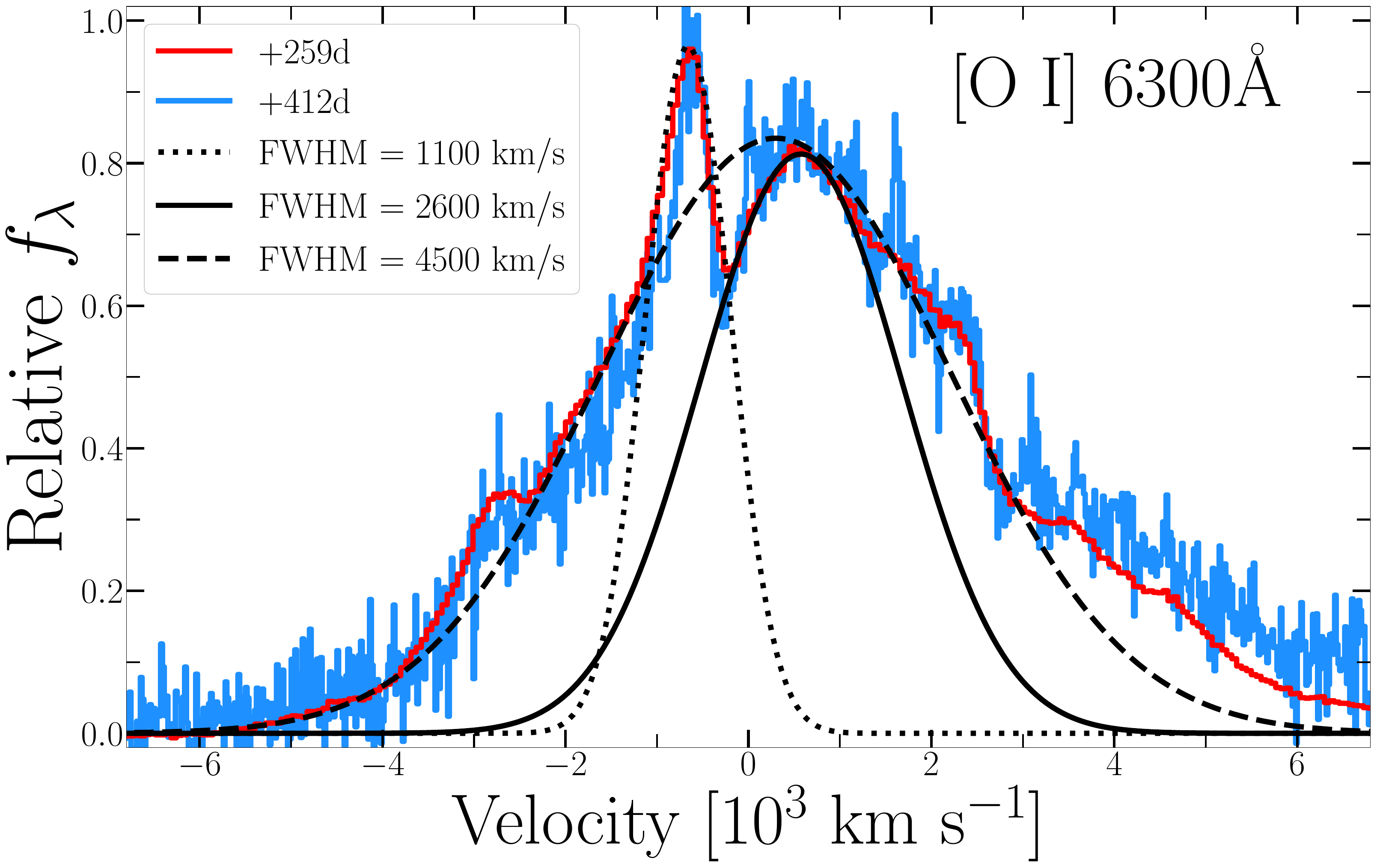}}
\subfigure{\includegraphics[width=0.49\textwidth]{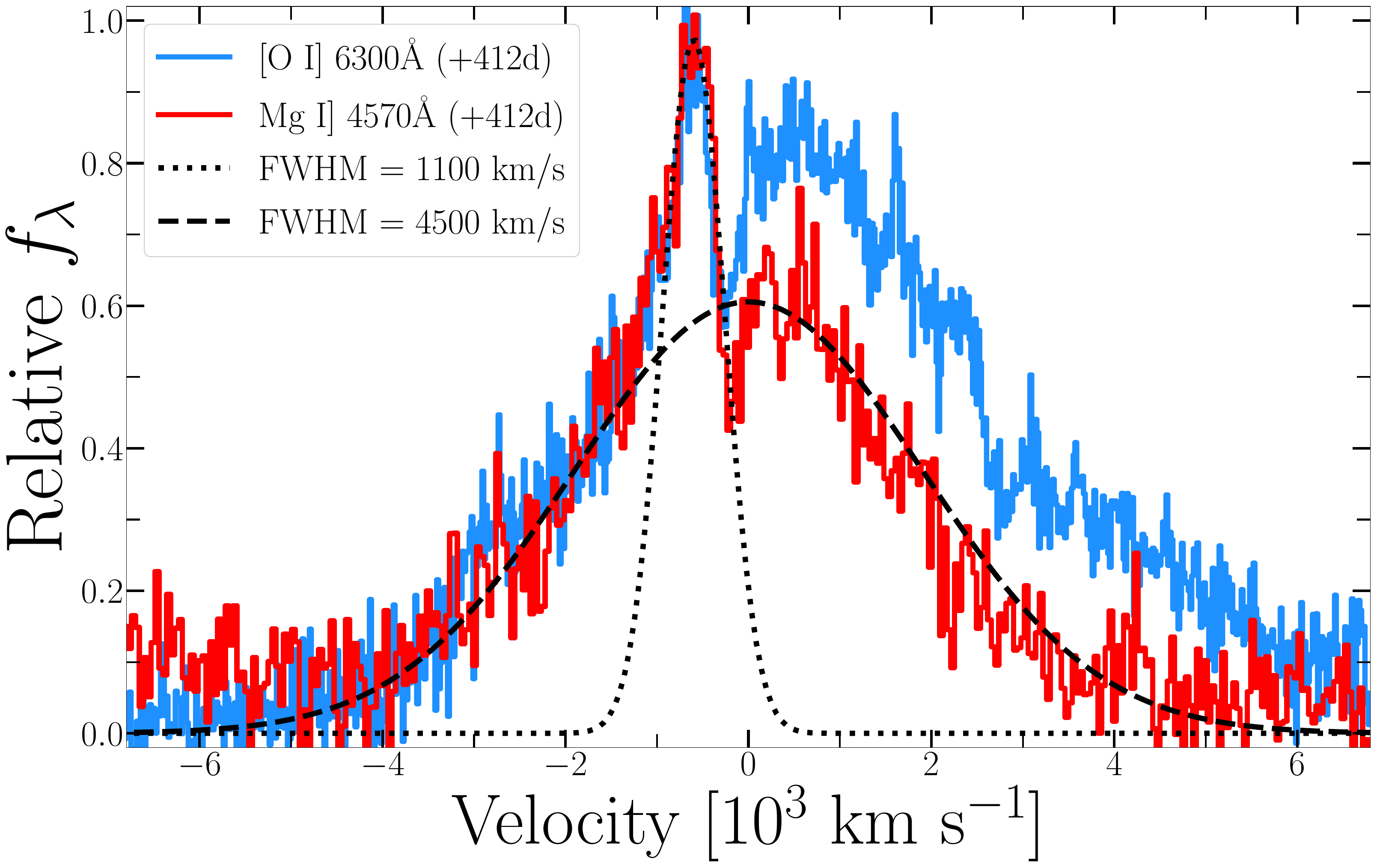}}
\caption{{\it Left:} Velocity profiles for [\ion{O}{i}] $\lambda$6300 at $\delta t =$ 259~days (red) and 412~days (blue). Gaussian profile fits shown in black demonstrate the potentially asymmetric distribution of O within the inner ejecta. {\it Right:} Comparison of the [\ion{O}{i}] $\lambda$6300 and \ion{Mg}{i}] $\lambda$4570 velocity profiles, both displaying double-peaked profiles suggestive of clumping and/or asymmetries with the inner ejecta. \label{fig:OIvels} }
\end{figure*}

The SN 2024iss $\delta t = 7$ days UV spectrum provides a unique snapshot of the outer layers of the supernova immediately following shock cooling emission, with likely contributions from additional CSM interaction emission. Numerous absorption features from iron-group elements result in heavy blanketing of UV flux and make it difficult to distinguish individual transition lines. We identified possible broad features corresponding to \ion{Ni}{iii}, \ion{Al}{iii}, \ion{Ti}{iii}. If these features are genuine, they suggest that these metals are traveling at about half the velocity of the H and \ion{He}{i} present in the outer ejecta. This is likely an optical depth effect that allows for slower moving hydrogen-rich ejecta with primordial Fe-group elements to be observed in the UV before these profiles emerge in the optical. Additionally, we detect a \ion{Mg}{ii} feature at $\sim2800$ ~\AA~ with a velocity similar to that of H and \ion{He}{i}. This places \ion{Mg}{ii} with fast-moving hydrogen-rich material in the outer ejecta. Multiple UV spectra showing the evolution of absorption features through $\sim$days/weeks in early phases would be useful for further investigating iron-group features in SNe II \citep[e.g.,][]{zimmerman_complex_2024, bostroem_circumstellar_2024}.


Early UV spectra for SNe 2024iss, 2013df, and 1993J (all SNe IIb) lack distinctive features in the $\sim2,000 - 2,500$ \AA ~ range. This contrasts with the other (non-IIb) SNe II in Figure \ref{fig:UVspec}, which contain several distinguishable emission and absorption features in this wavelength range. The lack of features could be attributed to additional line blanketing in SNe IIb UV spectra or the difference in hydrogen-rich envelope masses between SNe IIb and other SNe II. Since SNe IIb have less massive hydrogen-rich envelopes than other SNe II \citep{hachinger_how_2012, Ercolino24, dessart_sequence_2024}, they may have less prominent emission and absorption features due to lower relative abundances. Additionally, SN 2024iss could have weaker iron-group emission/absorption features due to the low-metallicity progenitor environment. The possible distinctions underscore the need for a larger sample of early-time SNe II UV spectra, including those of SNe IIb, to better understand the source of any differences and their implications for ejecta properties and the circumstellar environment. 

\begin{figure*}
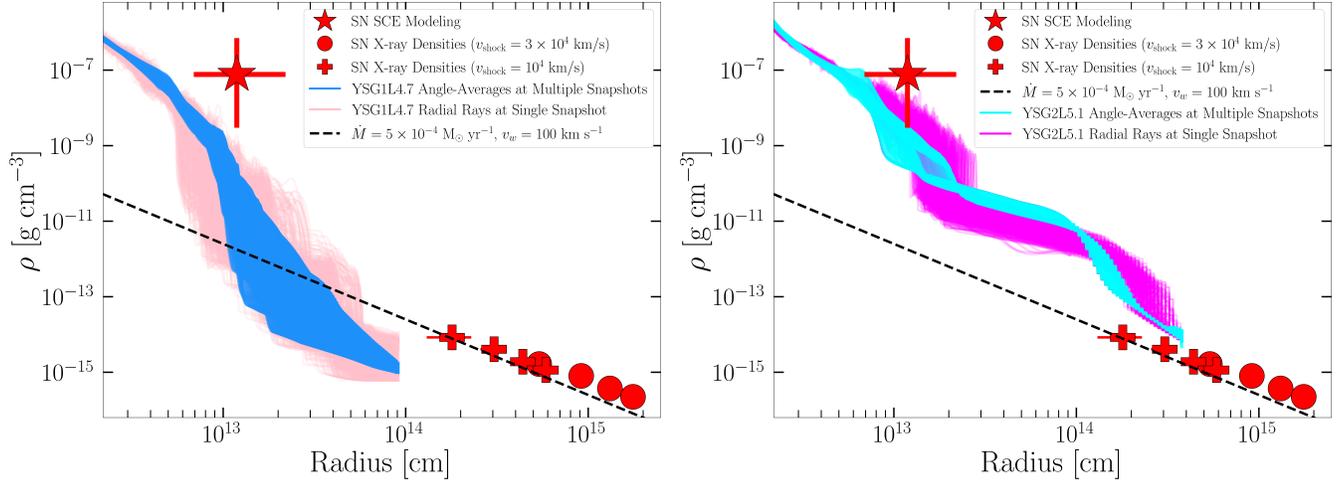

\centering
\subfigure{\includegraphics[width=0.49\textwidth]{rho_24iss_YSG1.png}}
\subfigure{\includegraphics[width=0.49\textwidth]{rho_24iss_YSG2.png}}
\caption{Progenitor envelope and CSM density profile derived from best-fit spherical shock cooling light curve models shown as the red star (Section \ref{subsec:shock cooling}) and the emission measures derived from modeling of the X-ray spectra shown as red plus signs and circles for different assumed shock velocities (Section \ref{sec:csm}). The spherically averaged density profiles for the YSG1L4.7 (left) and YSG1L5.1 (right) SN IIb progenitor models of \cite{2025arXiv250812486G} are shown in blue/cyan (angle-averages of individual model snapshots over a select time window) and pink/magenta (multiple view angles) lines. \label{fig:CSM} }
\end{figure*}

\begin{figure}[h]
\includegraphics[width=\columnwidth]{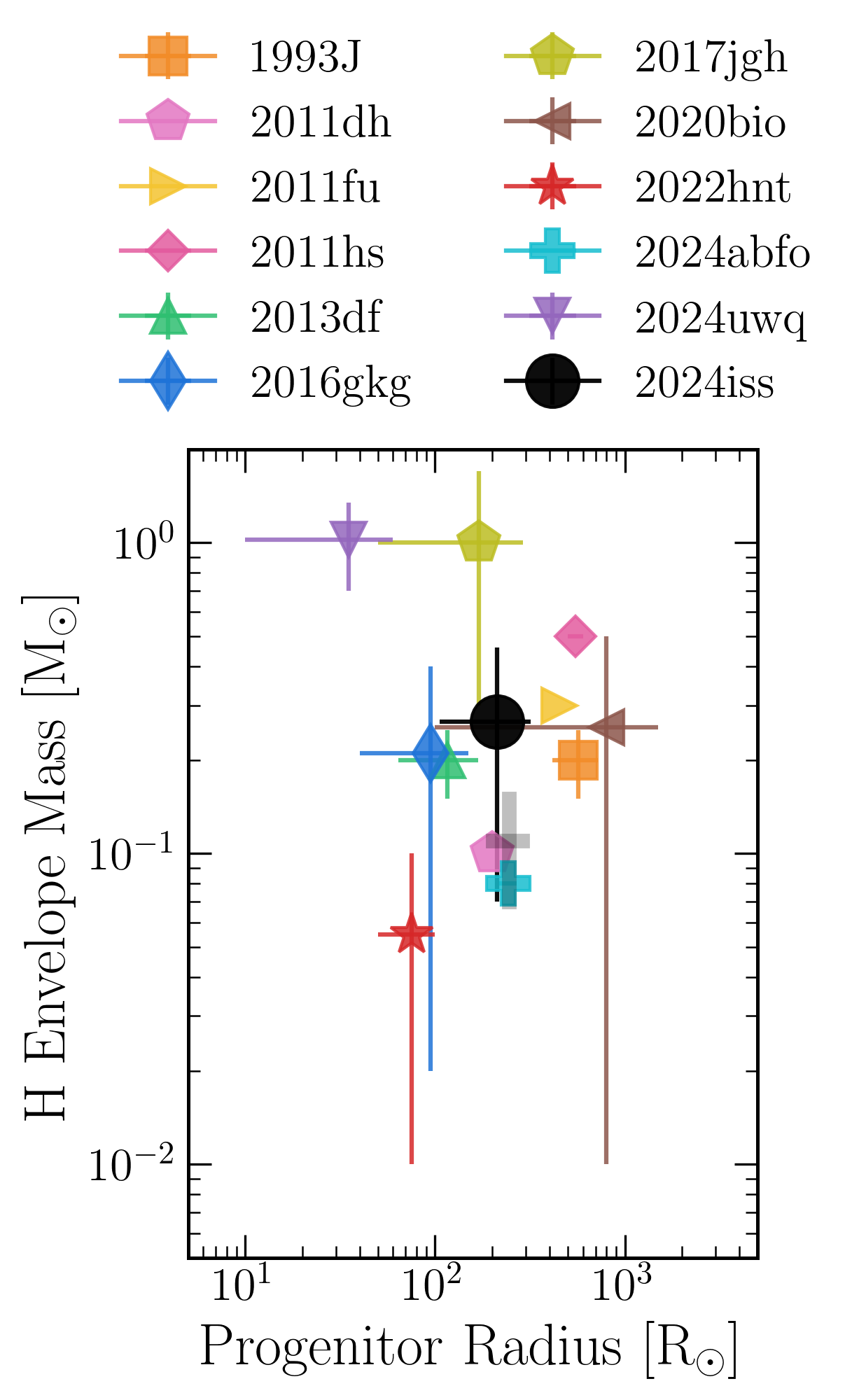}
\caption{Hydrogen-rich envelope mass vs. progenitor radius of SNe IIb. Values were obtained from: 1993J: \cite{woosley_sn_1994}; 2011dh: \cite{bersten_type_2012}; 2011fu: \cite{morales-garoffolo_sn_2015}, 2011hs: \cite{bufano_sn_2014}; 2013df: \cite{morales-garoffolo_sn_2014}; 2016gkg: \cite{arcavi_constraints_2017}; 2017jgh: \cite{armstrong_sn2017jgh_2021}; 2020bio: \cite{pellegrino_sn_2023}; 2022hnt: \cite{farah_shock-cooling_2025}; 2024abfo: \cite{reguitti_sn_2025}; 2024uwq: \cite{subrayan_early_2025}. SN 2024iss values from \cite{chen_sn_2025} are highlighted in gray.}.
\label{fig:parameter comparison}
\end{figure}

\section{Conclusions}
\label{sec:conclusions}

In this paper we have presented X-ray/UV/optical/NIR observations of the SN~IIb 2024iss located in a nearby dwarf host galaxy at $D\approx14$~Mpc. Below we summarize the primary observational findings of SN~2024iss:  

\begin{enumerate}
    \item We present one of the best constrained times of first light for a SN IIb, with a ZTF $i$-band detection at 0.027 days ($\sim$40~minutes) after last $g$-band non-detection. We find that a single power-law model fit to the multi-band photometry cannot reproduce the earliest $i$-band detection ($\sim$44$\sigma$ deviation), suggesting an association with shock breakout emission. We constrain the timescale of shock breakout in SN~2024iss to be $\delta t < 0.42$~days. 
    
    \item We obtained a UV spectrum of SN~2024iss at $\delta t = 7$~days, which represents the earliest {\it HST} UV spectrum obtained of a SN~IIb to date. In the UV spectrum, we detect possible iron-group features (\ion{Ni}{iii}, \ion{Al}{iii}, \ion{Ti}{iii}), which had not been previously identified in any SN IIb UV spectrum. We find a stratification in the velocities of these iron-group features when compared to velocities of H and \ion{He}{i} from optical spectra. We also identify a broad, blueshifted \ion{Mg}{ii} feature with a velocity consistent with that of fast-moving hydrogen-rich material, tracing the fastest moving ejecta. 

    \item We find that the early-time UV spectrum of SN~2024iss displays considerably less prominent P-Cygni profiles of Fe-group elements than observed in SNe~II-P observed in the UV at similar phases. This is likely related to increased line blanketing and/or lower metal abundances in the outer ejecta of SNe~IIb compared to SNe~II that arise from progenitor stars with less stripped H envelopes. Additionally, we observe an RMS deviation of $\sim47$~\% between SN~2024iss and a blackbody model in the wavelength range of $1600-2200$~\AA~ due to increased emission from Fe-group transitions.
    
    
    \item With shock cooling emission modeling, we constrain the progenitor radius to $100 - 320 \ \Rsun$ and the hydrogen-rich envelope mass to $0.07 - 0.46 \ \Msun$. These values are consistent with a yellow supergiant progenitor with a stripped hydrogen envelope.
    
    \item We find that the best-matched model for a binary progenitor system corresponds to a stripped hydrogen-rich envelope mass of $0.19 - 0.28 \ \Msun$ \citep{dessart_sequence_2024}, consistent with envelope mass estimates from shock cooling light curve model fits and other SNe IIb. The model is best-matched to observed optical spectra, but inconsistent with the observed UV spectrum.
    
    \item We derive an ejecta mass ranging from $1.1 - 3.3 \ \Msun$, a kinetic energy ranging from $(2.3 - 6.4) \times 10^{50}$ erg, and a $^{56}$Ni mass of $0.11 \pm 0.01 \ \Msun$ from modeling of the secondary peak and tail of the bolometric light curve.
    
    \item Using X-ray detections in the first week after first light, we estimate a CSM density of $8.4\times 10^{-15}$~g~cm$^{-3}$ at $R_{\rm sh} = 1.8\times 10^{14}$~cm, which decreases to $1.1\times 10^{-15}$~g~cm$^{-3}$ at $R_{\rm sh} = 5.9\times 10^{14}$~cm, assuming a shock velocity of $10^4~\kms$. These CSM densities are consistent with a steady-state progenitor mass-loss rate of $5 \times 10^{-4} \ \Msun \ \textrm{yr}^{-1}$ assuming a wind velocity of $100 \ \kms$ in the $\sim 2-5$ years before explosion. The X-ray light curve, combined with the indicates significant luminosity contribution from CSM interaction between the shock cooling emission and $^{56}$Ni powered light curve peaks.

    \item We obtained late-time optical/NIR spectra of SN~2024iss at $\delta t\sim$260-412~days, which are consistent with stripped-envelope SN models with $M_{\rm ej} = 2.8 - 3.3~\Msun$. Additionally, we identify multiple peaks in the forbidden emission lines of [\ion{O}{i}] and \ion{Mg}{i}], likely connected with inner ejecta asymmetry and/or clumping.
    
\end{enumerate}

This analysis highlights the importance of ultraviolet observations for constraining supernova properties. Early-time and high cadence ultraviolet photometry is fundamental for accurately modeling shock cooling emissions and blackbody properties, and early-time UV spectroscopy provides new insights into the supernova environment during and after shock breakout. Upcoming UV space missions such as NASA’s Ultraviolet Explorer (UVEX, \citealt{kulkarni_science_2021}) will be key to furthering supernova studies in the ultraviolet.

\begin{acknowledgments}
We thank Lars~Bildsten, Yan-Fei~Jiang, and Matteo~Cantiello for valuable discussions regarding the CSM profiles for the 3-D SN~IIb progenitor model shown in Figure~\ref{fig:CSM}. 


R.J.Y. and F.F.\ acknowledge support from the Caltech Summer Undergraduate Research Fellowship (SURF) program.
W.J.-G.\ is supported by NASA through Hubble Fellowship grant HSTHF2-51558.001-A awarded by the Space Telescope Science Institute, which is operated for NASA by the Association of Universities for Research in Astronomy, Inc., under contract NAS5-26555.
The UCSC team is supported in part by NASA grants 80NSSC23K0301 and 80NSSC24K1411; and a fellowship from the David and Lucile Packard Foundation to R.J.F.
I.A.\ is supported by the National Science Foundation award AST-2505775, NASA grant 24-ADAP24-0159, Scialog awards SA-LSST-2024-102a and SA-LSST-2025-112b, and the Discovery Alliance Catalyst Fellowship Mentors award 2025-62192-CM-19.
Parts of this research were supported by the Australian Research Council Centre of Excellence for Gravitational Wave Discovery (OzGrav), through project number CE230100016.
M.W.C.\ acknowledges support from the National Science Foundation with grant numbers PHY-2117997, PHY-2308862 and PHY-2409481.
C.G.\ is supported by a VILLUM FONDEN Villum Experiment grant (VIL69896).
A.G.\ acknowledges the support from the research project grant ``Understanding the Dynamic Universe'' funded by the Knut and Alice Wallenberg Foundation under Dnr KAW 2018.0067.
J.A.G.\ acknowledges financial support from NASA grant 23-ATP23-0070.
J.K.\ acknowledges support from the National Science Foundation Graduate Research Fellowship Program.
C.D.K.\ gratefully acknowledges support from the NSF through AST-2432037, the HST Guest Observer Program through HST-SNAP-17070 and HST-GO-17706, and from JWST Archival Research through JWST-AR-6241 and JWST-AR-5441.
R.M.\ acknowledges support by the National Science Foundation under award No.\ AST-2224255.
M.R.S.\ is supported by the STScI Postdoctoral Fellowship.
A.S.\ acknowledges support from the Knut and Alice Wallenberg Foundation through the ``Gravity Meets Light'' project.
S.T.\ is supported by the Fundamental Fund of Thailand Science Research and Innovation (TSRI) through the National Astronomical Research Institute of Thailand (Public Organization) (FFB690078/0269).
Q.W.\ is supported by the Sagol Weizmann-MIT Bridge Program. 

Based on observations obtained with the Samuel Oschin Telescope 48 inch and the 60inch Telescope at the Palomar Observatory as part of the Zwicky Transient Facility project. ZTF is supported by the National Science Foundation under grants Nos. AST-1440341 and AST-2034437 and a collaboration including current partners Caltech, IPAC, the Oskar Klein Center at Stockholm University, the University of Maryland, University of California, Berkeley, the University of Wisconsin–Milwaukee, University of Warwick, Ruhr University, Cornell University, Northwestern University, and Drexel University. Operations are conducted by COO, IPAC, and UW.

Zwicky Transient Facility access was supported by Northwestern University and the Center for Interdisciplinary Exploration and Research in Astrophysics (CIERA).

SED Machine is based upon work supported by the National Science Foundation under Grant No.\ 1106171. The ZTF forced-photometry service was funded under the Heising-Simons Foundation grant \#12540303 (PI: Graham). The Gordon and Betty Moore Foundation, through both the Data-Driven Investigator Program and a dedicated grant, provided critical funding for SkyPortal.

The Young Supernova Experiment (YSE) and its research infrastructure is supported by the European Research Council under the European Union's Horizon 2020 research and innovation programme (ERC Grant Agreement 101002652, PI K.\ Mandel), the Heising-Simons Foundation (2018-0913, PI R.\ Foley; 2018-0911, PI R.\ Margutti), NASA (NNG17PX03C, PI R.\ Foley), NSF (AST--1720756, AST--1815935, PI R.\ Foley; AST--1909796, AST-1944985, PI R.\ Margutti), the David \& Lucille Packard Foundation (PI R.\ Foley), VILLUM FONDEN (project 16599, PI J.\ Hjorth), and the Center for AstroPhysical Surveys (CAPS) at the National Center for Supercomputing Applications (NCSA) and the University of Illinois Urbana-Champaign.

Pan-STARRS is a project of the Institute for Astronomy of the University of Hawai‘i, and is supported by the NASA SSO Near Earth Observation Program under grants 80NSSC18K0971, NNX14AM74G, NNX12AR65G, NNX13AQ47G, NNX08AR22G, 80NSSC21K1572, and by the State of Hawai‘i. The Pan-STARRS1 Surveys (PS1) and the PS1 public science archive have been made possible through contributions by the Institute for Astronomy, the University of Hawai‘i, the Pan-STARRS Project Office, the Max-Planck Society and its participating institutes, the Max Planck Institute for Astronomy, Heidelberg and the Max Planck Institute for Extraterrestrial Physics, Garching, The Johns Hopkins University, Durham University, the University of Edinburgh, the Queen's University Belfast, the Harvard-Smithsonian Center for Astrophysics, the Las Cumbres Observatory Global Telescope Network Incorporated, the National Central University of Taiwan, STScI, NASA under grant NNX08AR22G issued through the Planetary Science Division of the NASA Science Mission Directorate, NSF grant AST-1238877, the University of Maryland, Eotvos Lorand University (ELTE), the Los Alamos National Laboratory, and the Gordon and Betty Moore Foundation.

This publication has made use of data collected at Lulin Observatory, partly supported by the TAOvA with the NSTC grant 112-2740-M-008-002.

We acknowledge the use of public data from the {\it Swift} data archive.


YSE-PZ was developed by the UC Santa Cruz Transients Team with support from The UCSC team is supported in part by NASA grants NNG17PX03C, 80NSSC19K1386, and 80NSSC20K0953; NSF grants AST-1518052, AST-1815935, and AST-1911206; the Gordon \& Betty Moore Foundation; the Heising-Simons Foundation; a fellowship from the David and Lucile Packard Foundation to R.\ J.\ Foley; Gordon and Betty Moore Foundation postdoctoral fellowships and a NASA Einstein fellowship, as administered through the NASA Hubble Fellowship program and grant HST-HF2-51462.001, to D.~O.~Jones; and a National Science Foundation Graduate Research Fellowship, administered through grant No.\ DGE-1339067, to D.~A.~Coulter.


Some/all of the data presented in this paper were obtained from the Mikulski Archive for Space Telescopes (MAST) at the Space Telescope Science Institute. The specific observations analyzed can be accessed via \dataset[https://doi.org/10.17909/wqem-3c51]{https://doi.org/10.17909/wqem-3c51}. STScI is operated by the Association of Universities for Research in Astronomy, Inc., under NASA contract NAS5–26555. Support to MAST for these data is provided by the NASA Office of Space Science via grant NAG5–7584 and by other grants and contracts.


This research is based on observations made with the NASA/ESA Hubble Space Telescope obtained from the Space Telescope Science Institute, which is operated by the Association of Universities for Research in Astronomy, Inc., under NASA contract NAS 5–26555. These observations are associated with programs HST-GO-17507 (PI Jacobson-Gal\'an).

A major upgrade of the Kast spectrograph on the Shane 3 m telescope at Lick Observatory, led by Brad Holden, was made possible through gifts from the Heising-Simons Foundation, William and Marina Kast, and the University of California Observatories. Research at Lick Observatory is partially supported by a generous gift from Google.

The data presented here were obtained in part with ALFOSC, which is provided by the Instituto de Astrofisica de Andalucia (IAA) under a joint agreement with the University of Copenhagen and NOT.

The Liverpool Telescope is operated on the island of La Palma by Liverpool John Moores University in the Spanish Observatorio del Roque de los Muchachos of the Instituto de Astrofisica de Canarias with financial support from the UK Science and Technology Facilities Council.

Some of the data presented herein were obtained at Keck Observatory, which is a private 501(c)3 non-profit organization operated as a scientific partnership among the California Institute of Technology, the University of California, and the National Aeronautics and Space Administration. The Observatory was made possible by the generous financial support of the W.\ M.\ Keck Foundation. 
The authors wish to recognize and acknowledge the very significant cultural role and reverence that the summit of Maunakea has always had within the Native Hawaiian community. We are most fortunate to have the opportunity to conduct observations from this mountain. 

\end{acknowledgments}




%
\facilities{{\it Swift} UVOT/XRT, Shane Telescope (Kast), Lulin Observatory, Pan-STARRS, Liverpool Telescope (SPRAT), P60 (SEDM), Nordic Optical Telescope (ALFOSC), Keck Observatory (LRIS/NIRES), Hale Telescope (DBSP), Zwicky Transient Facility (ZTF), {\it HST} (STIS)}

\software{IRAF (Tody 1986, Tody 1993), photpipe \citep{Rest+05}, DoPhot \citep{Schechter+93}, HOTPANTS \citep{becker15}, YSE-PZ \citep{Coulter22, Coulter23}, CMFGEN \citep{hillier12}, Lpipe \citep{Perley19}, HEAsoft \citep{HEAsoft} }


\bibliography{24iss_citations3}{}
\bibliographystyle{aasjournalv7}

    \appendix
\onecolumngrid

\begin{deluxetable}{ccccccc}[h]
\tablecaption{Ultraviolet/Optical/Infrared Spectroscopy \label{tab:spec_all}}
\tablecolumns{7}
\tablewidth{0.45\textwidth}
\tablehead{\colhead{UT Date} & \colhead{MJD} &
\colhead{Phase\tablenotemark{a}} &
\colhead{Telescope} & \colhead{Instrument} & \colhead{Wavelength Range} & \colhead{Data Source}\\
\colhead{} & \colhead{} & \colhead{[days]} & \colhead{} & \colhead{} & \colhead{[$\mu$m]} & \colhead{}
}
\startdata
    2024-05-13 9:21:52 & 60443.4 & 1.1 & P60 & SEDM & 0.38 - 0.92 & ZTF\\
    2024-05-13 23:15:32 & 60444.0 & 1.7 & LT & SPRAT & 0.41 - 0.81 & ZTF \\ 
    2024-05-14 4:32:40 & 60444.2 & 1.9 & P60 & SEDM & 0.38 - 0.92 & ZTF \\
    2024-05-14 23:25:13 & 60445.0 & 2.7 & LT & SPRAT & 0.42 - 0.81 & ZTF\\
    2024-05-15 6:18:57 & 60445.3 & 3.0 & Hale & DBSP & 0.34 - 1.05 & ZTF \\
    2024-05-15 8:07:30 & 60445.3 & 3.1 & P60 & SEDM & 0.38 - 0.92 & ZTF\\
    2024-05-16 0:44:41 & 60446.0 & 3.8 & LT & SPRAT & 0.41 - 0.81 & ZTF \\
    2024-05-16 07:01:17 & 60446.3 & 4.0 & Shane & Kast & 0.36 - 1.03 & TReX$^b$\\
    2024-05-17 1:01:05 & 60447.0 & 4.8 & LT & SPRAT & 0.41 - 0.81 & ZTF \\
    2024-05-17 23:25:17 & 60448.0 & 5.7 & NOT & ALFOSC & 0.33 - 0.97 & ZTF \\
    2024-05-19 05:55:12 & 60449.2 & 7.0 & \textit{HST} & STIS & 0.17 - 1.03 & {\it HST} GO-17507 \\
    2024-05-21 7:22:36 & 60451.3 & 9.1 & Hale & DBSP & 0.34 - 1.05 & ZTF \\
    2024-05-22 10:38:13 & 60452.4 & 10.2 & Keck & NIRES & 0.9 - 2.5 & KITS\\
    2024-05-25 21:20:06 & 60455.9 & 13.6 & LT & SPRAT & 0.40 - 0.81 & ZTF \\
    2024-05-27 21:20:17 & 60457.9 & 15.6 & LT & SPRAT & 0.40 - 0.81 & ZTF \\
    2024-05-28 23:51:27 & 60459.0 & 16.7 & LT & SPRAT & 0.40 - 0.81 & ZTF \\
    2024-05-29 05:29:18 & 60459.2 & 17.0 & Shane & Kast & 0.33 - 1.1 & YSE \\
    2024-05-30 07:08:30 & 60460.3 & 18.0 & Shane & Kast & 0.36 - 1.03 & TReX$^b$ \\
    2024-05-31 6:39:55 & 60461.3 & 19.0 & Hale & DBSP & 0.34 - 1.05 & ZTF \\
    2024-06-05 04:55:59 & 60466.21 & 24.0 & Shane & Kast & 0.33 - 1.1 & YSE \\
    2024-06-08 23:08:30 & 60470.0 & 27.7 & NOT & ALFOSC & 0.33 - 0.97 & ZTF \\
    2024-06-09 6:31:12 & 60470.3 & 28.0 & P60 & SEDM & 0.38 - 0.92 & ZTF \\
    2024-06-11 7:20:57 & 60472.3 & 30.1 & P60 & SEDM & 0.38 - 0.92 & ZTF \\
    2024-06-14 04:36:09 & 60475.2 & 32.9 & Shane & Kast & 0.36 - 1.03 & TReX$^b$ \\
    2024-06-15 05:34:17 & 60476.2 & 34.0 & Shane & Kast & 0.33 - 1.09 & YSE \\
    2024-06-21 5:36:23 & 60482.2 & 40.0 & P60 & SEDM & 0.38 - 0.92 & ZTF \\
    2024-06-28 04:29:03 & 60489.2 & 46.9 & Shane & Kast & 0.36 - 1.05 & TReX$^b$ \\
    2024-07-07 05:13:16 & 60498.2 & 56.0 & Shane & Kast & 0.33 - 1.1 & YSE \\
    2024-07-10 5:33:51 & 60501.2 & 59.0 & P60 & SEDM & 0.38 - 0.92 & ZTF \\
    2024-07-12 04:37:37 & 60503.2 & 60.9 & Shane & Kast & 0.33 - 1.09 & YSE \\
    2024-07-13 5:22:05 & 60504.2 & 62.0 & P60 & SEDM & 0.38 - 0.92 & ZTF \\
    2024-07-27 21:04:33 & 60518.9 & 76.6 & NOT & ALFOSC & 0.35 - 0.97 & ZTF \\
    2024-07-28 05:41:28 & 60519.2 & 77.0 & Shane & Kast & 0.36 - 1.05 & TReX$^b$ \\
    2024-08-01 4:21:32 & 60523.2 & 80.9 & P60 & SEDM & 0.38 - 0.92 & ZTF \\
    2024-08-06 04:25:48 & 60528.2 & 86.0 & Shane & Kast & 0.33 - 1.1 & YSE \\
    2024-08-15 04:18:21 & 60537.2 & 94.9 & Shane & Kast & 0.36 - 1.03 & TReX$^b$ \\
    2024-08-16T03:49:30 & 60538.2 & 95.0 & Shane & Kast & 0.33 - 1.1 & YSE \\
    2024-08-27 04:15:59 & 60549.2 & 107.0 & Shane & Kast & 0.33 - 1.1 & YSE \\
    2024-08-28 03:53:08 & 60550.2 & 107.9 & Shane & Kast & 0.36 - 1.00 & TReX$^b$ \\
    2024-11-11 12:43:52 & 60625.5 & 183.3 & Hale & DBSP & 0.34 - 1.05 & ZTF \\
    2024-12-15 15:20:42 & 60659.6 & 207.2 & Keck & NIRES & 0.9 - 2.5 & KITS\\
    2024-12-31 13:05:27 & 60675.5 & 233.3 & Shane & Kast & 0.33 - 1.1 & YSE \\
    2025-01-26 16:05:12 & 60701.7 & 258.7 & Keck & LRIS & 0.31 - 1.03 & ZTF \\
    2025-02-18 13:09:34 & 60724.5 & 272.1 & Keck & NIRES & 0.9 - 2.5 & KITS\\
    2025-06-28 10:08:57 & 60854.4 & 411.4 & Keck & LRIS & 0.31 - 1.03 & ZTF \\
\enddata
\tablenotetext{a}{Relative to time of first light.}
\tablenotetext{b}{TRansient EXtragalactic team at UC Berkeley (PIs Margutti and Chornock)}
\end{deluxetable}


{
\setlength{\tabcolsep}{12pt}
\begin{deluxetable}{l c c c c c c}[h]
\tablecaption{Best‐fitting Parameters of the Absorbed Thermal Bremsstrahlung Model for SN~2024iss\label{tab:xrt}}
\tablehead{
  \colhead{Time} &
  \colhead{Instrument} &
  \colhead{$\mathrm{N}_{\rm H,int}$} &
  \colhead{$T$\tablenotemark{a}} &
  \colhead{$\log_{10}$(Flux)\tablenotemark{b}} &
  \colhead{$\log_{10}$(Flux)\tablenotemark{b}} &
  \colhead{Norm\tablenotemark{c}} \\ [-6pt]
  \colhead{(days)} &
  \colhead{} &
  \colhead{$\left(10^{22} \ \rm cm^{-2}\right)$} &
  \colhead{(keV)} &
  \colhead{(Absorbed)} &
  \colhead{(Unabsorbed)} &
  \colhead{($10^{-4} \ \rm cm^{-5}$)}
}
\startdata
  2.08 & Swift-XRT & $0.11^{+0.10}_{–0.09}$ & 41.00 & $-11.80^{+0.06}_{–0.06}$ & $-11.77^{+0.06}_{–0.06}$ & $3.19^{+0.50}_{–0.45}$\\
  3.53 & Swift-XRT & $0.17^{+0.14}_{–0.10}$ & 35.91 & $-11.74^{+0.07}_{–0.07}$ & $-11.70^{+0.07}_{–0.07}$ & $3.71^{+0.73}_{–0.63}$ \\
  5.07 & Swift-XRT & $0.69^{+0.48}_{–0.40}$ & 32.80 & $-11.98^{+0.13}_{–0.11}$ & $-11.88^{+0.13}_{–0.11}$ & $2.40^{+0.84}_{–0.70}$ \\
  6.79 & Swift-XRT & $0.46^{+0.46}_{–0.38}$ & 30.50 & $-12.02^{+0.10}_{–0.09}$ & $-11.94^{+0.10}_{–0.09}$ & $2.09^{+0.62}_{–0.51}$ \\
\enddata
\tablecomments{All errors are reported at $1\sigma$ c.l.}
\tablenotetext{a}{Value is fixed in model fit.}
\tablenotetext{b}{$\log_{10}$(Flux) is from 0.3–10 keV. Flux units in $\rm erg \ cm^{-2} \ s^{-1}$}
\tablenotetext{c}{Normalization of the Bremsstrahlung model defined as $\rm Norm \equiv \frac{3.02 \times 10^{-15}}{4 \pi d^2} \int n_e n_I \ dV$, where $n_e$ and $n_I$ are the electron and ion number densities in $\rm cm^{-3}$ and $d$ is the distance to the source in $\rm cm$.}
\end{deluxetable}}


\begin{deluxetable}{cccccc}[h]
\tablecaption{X-ray Observation Log of SN~2024iss\label{tab:xray_obs}}
\tablehead{
\colhead{Instrument} &
\colhead{Observation Date} &
\colhead{Mid Time\tablenotemark{a}} &
\colhead{Observation ID} &
\colhead{Exposure} &
\colhead{PI} \\
&
\colhead{(yyyy/mm/dd)}
&
\colhead{(days)} & \colhead{00016638-}
&
\colhead{Time (ks)} &
}
\startdata
Swift-XRT & 2024-05-13 -- 2024-05-14 & 2.08  & 001, 002 & 3.5  & W.~Jacobson-Galan \\
Swift-XRT & 2024-05-15 -- 2024-05-16 & 3.53  & 004, 006 & 3.1  & W.~Jacobson-Galan \\
Swift-XRT & 2024-05-17              & 5.07  & 009              & 1.6  & W.~Jacobson-Galan \\
Swift-XRT & 2024-05-18 -- 2024-05-19 & 6.79  & 010, 012 & 2.5  & W.~Jacobson-Galan \\
Swift-XRT & 2024-05-21 -- 2024-06-11 & 19.64 & 013, 014, 015, 016, 017, 021, 022 & 10.9 & W.~Jacobson-Galan \\
Swift-XRT & 2024-06-12 -- 2024-06-23 & 36.49 & 023, 024, 025, 026, 027 & 17.1 & W.~Jacobson-Galan \\
Swift-XRT & 2024-11-13 -- 2024-12-03 & 195.46 & 029, 030, 031, 032 & 6.5  & W.~Jacobson-Galan \\
Swift-XRT & 2024-12-10 -- 2024-12-26 & 220.44 & 033, 034, 035, 036 & 5.5  & W.~Jacobson-Galan \\
Swift-XRT & 2025-02-06 -- 2025-02-20 & 277.42 & 037, 039, 040, 041, 042 & 6.5 & W.~Jacobson-Galan \\
Swift-XRT & 2025-04-17 -- 2025-05-03 & 348.20 & 043, 044, 045 & 2.8 & W.~Jacobson-Galan \\
\enddata
\tablenotetext{a}{With respect to time of explosion}
\end{deluxetable}

\section{Shock Cooling Models} \label{appendix shock cooling}

The priors used for each parameter in the MCMC fitting for the shock cooling models are listed in table \ref{tab:shock cooling priors}. The $f_{p}M$ parameter describes a fraction ($f_p$) of the ejecta mass ($M$) relating to the inner envelope structure of the progenitor \citep{sapir_uvoptical_2017}. Shock cooling emission is independent of $f_{p}M$ \citep{sapir_uvoptical_2017, morag_shock_2023}. As a result, it is a free parameter in the \texttt{lightcurve-fitting} package \citep{hosseinzadeh_light_2023} and a fixed value in the \texttt{shock-cooling-curve} package \citep{venkatraman_shock_cooling_curve_2024}. As seen in Figure 5 of \citet{sapir_uvoptical_2017}, possible values for $f_{p}$ when n$=1.5$ range from $0.1 - 3$. Since the expected ejecta mass is approximately $3 \ \Msun$, we limited our $f_{p}M$ MCMC prior to $0.3 - 10 \ \Msun$.

In order to determine the best phase to fit the shock cooling models to, we fit the SW17 n=1.5, SW17 n=3, and P21 models to photometry up to 5.0, 6.0, and 7.0 days after first light in the $i$, $r$, $V$, $g$, $B$, $U$, \textit{UVW1}, \textit{UVM2}, and \textit{UVW2} bands. We chose the fit with the lowest $\chi^2$ value as the best phase for each model. This resulted in the SW17 models being fit up to $\delta t = 7$ days, and the P21 model being fit up to $\delta t = 5$ days. We performed these fits using MCMC fitting with \texttt{shock-cooling-curve} \citep{venkatraman_shock_cooling_curve_2024}. The results can be seen in Table \ref{tab:shock cooling phase fits}. Since the MSW23 model is based on the SW17 n=1.5 model, we fit it to photometry up to $\delta t = 7$ days as well. Additionally, we compared our results for SW17 n=1.5 as derived from using \texttt{shock-cooling-curve} with the same model fit with \texttt{lightcurve-fitting} \citep{hosseinzadeh_light_2023}. We found the values to be very similar between the two, as seen in table \ref{tab:sw17 comparison}.

\begin{deluxetable*}{ccccccc}[h] 
\tabletypesize{\small}        
\tablecaption{Shock Cooling Model Results \label{tab:shock cooling results}}
\tablewidth{0pt}              
\tablehead{
\colhead{Model} & \colhead{$R_{\star}$ [$R_{\odot}$]} & \colhead{$M_{\textrm{env}}$ [$M_{\odot}$]} & 
\colhead{$v_s$ [$10^9 \textrm{cm s}^{-1}$]} & \colhead{$f_{p}M$ [$M_{\odot}$]} & \colhead{$t_{\rm{offset}}$ [days]} & \colhead{Reduced $\chi^2$}
}
\startdata
P21 & $317.55^{+2.83}_{-2.90}$ & $(7.45^{+0.05}_{-0.04}) \times10^{-2}$ & $1.604^{+0.005}_{-0.003}$ & - & $(1.002_{-0.002}^{+0.005})\times10^{-3}$ & 450.23 \\
SW17 n=1.5 & $106.31 \pm 0.50$ & $0.457 \pm 0.003$ & $1.025 \pm 0.001$ & - & $\sim0.000^a$ & 952.05 \\
SW17 n=3 & $174.93^{+0.75}_{-0.74}$ & $5.124^{+0.036}_{-0.033}$ & $0.836 \pm 0.001$ & - & $\sim0.000^a$ & 1100.34 \\
MSW23 (A=1) & $107.81^{+5.75}_{-7.19}$ & $0.275^{+0.013}_{-0.008}$ & $0.985^{+0.041}_{-0.028}$ & $3.9^{+0.6}_{-0.7}$ & $0.002 \pm 0.001$ & - \\
\enddata
\tablenotetext{a}{Resulting $t_{\textrm{offset}}$ value from model is less than $10^{-3}$ days.}
\end{deluxetable*}

\begin{deluxetable*}{cccccc}[h] 
\tablecaption{Shock Cooling MCMC Priors \label{tab:shock cooling priors}}
\tablehead{
\colhead{Model} & \colhead{$R_{\star}$ [$R_{\odot}$]} & \colhead{$M_{\textrm{env}}$ [$M_{\odot}$]} & 
\colhead{$v_s$ [$10^9 \textrm{cm s}^{-1}$]} & \colhead{$f_{p}M$ [$M_{\odot}$]} & \colhead{$t_{\rm{offset}}$ [days]}
}
\startdata
P21 & $0.1 - 1000$ & $0.01 - 1.0$ & $0.01 - 10.0$ & - & $0.001 - 0.03$ \\
SW17 n=1.5 & $1.44 - 1437.40$ & $0.01 - 10.0$ & $0.01 - 10.0$ & - & $0.00 - 0.03$ \\
SW17 n=3 & $1.44 - 1437.40$ & $0.01 - 10.0$ & $0.01 - 3.0$ & - & $0.00 - 0.03$ \\
MSW23 (A=1) & $1.44 - 1437.40$ & $0.01 - 10.0$ & $0 - 3.16$ & $0.3 - 10.0$ & $-0.03 - 0.03$ \\
\enddata
\end{deluxetable*}

\begin{deluxetable*}{ccc}[h] 
\tablecaption{Shock Cooling Model Fits \label{tab:shock cooling phase fits}}
\tablehead{
\colhead{Model} & \colhead{Phase Cutoff [days]} & \colhead{Reduced $\chi^2$} 
}
\startdata
SW17 n=1.5 & 5.0 & 1223.87 \\
SW17 n=1.5 & 6.0 & 1038.43 \\
\textbf{SW17 n=1.5} & \textbf{7.0} & \textbf{952.05} \\
\tableline
SW17 n=3 & 5.0 & 1313.58 \\
SW17 n=3 & 6.0 & 1201.57 \\
\textbf{SW17 n=3} & \textbf{7.0} & \textbf{1100.34} \\
\tableline
\textbf{P21} & \textbf{5.0} & \textbf{450.23} \\
P21 & 6.0 & 1475.76 \\
P21 & 7.0 & 1466.80 \\
\enddata
\end{deluxetable*}

\begin{deluxetable*}{cccccc}[h] 
\tablecaption{SW17 n=1.5 $\delta t < 7$ days Photometry Fits \label{tab:sw17 comparison}}
\tablehead{
\colhead{Package} & \colhead{$R_{\star}$ [$R_{\odot}$]} & \colhead{$M_{\textrm{env}}$ [$M_{\odot}$]} & 
\colhead{$v_s$ [$10^9 \textrm{cm s}^{-1}$]} & \colhead{$f_{p}M$ [$M_{\odot}$]} & \colhead{$t_{\rm{offset}}$ [days]}
}
\startdata
\texttt{shock-cooling-curve} & $106.31 \pm 0.50$ & $0.457 \pm 0.003$ & $1.025 \pm 0.001$ & - & $\sim0.000^a$ \\
\texttt{lightcurve-fitting} & $100.62^{+2.87}_{-4.31}$ & $0.52^{+0.02}_{-0.01}$ & $1.02 \pm 0.03$ & $0.8 \pm 0.2$ & $-0.001 \pm 0.001$ \\
\enddata
\tablenotetext{a}{Resulting $t_{\textrm{offset}}$ value from model is less than $10^{-3}$ days.}
\end{deluxetable*}





\end{document}